\documentclass[pdflatex,sn-mathphys-num]{sn-jnl}

\usepackage[a-1b]{pdfx}

\usepackage{graphicx}           
\usepackage[table]{xcolor}      
\usepackage{multirow}           
\usepackage{booktabs}           
\usepackage{threeparttable}     
\usepackage{longtable}
\usepackage{tabularx}
\usepackage{amsmath}
\usepackage{float}
\usepackage{subcaption}
\usepackage{tcolorbox}
\usepackage{mdframed}
\usepackage{pgf-pie}
\usepackage{comment}
\usepackage{lipsum}
\usepackage{algorithmic}
\usepackage{textcomp}
\usepackage{xcolor}            
\usepackage{hyperref}    

\begin{document}

\title[Ethics Practices in AI Development: An Empirical Study Across Roles and Regions]{Ethics Practices in AI Development: An Empirical Study Across Roles and Regions}


\author*[1]{\fnm{Wilder} \sur{Baldwin}}\email{wilder.baldwin@maine.edu}

\author*[1]{\fnm{Sepideh} \sur{Ghanavati}}\email{sepideh.ghanavati@maine.edu}

\author*[1,2]{\fnm{Manuel} \sur{Woersdoerfer}}\email{manuel.woersdoerfer@maine.edu}

\affil*[1]{\orgdiv{School of Computing and Information Science}, \orgname{University of Maine}, \orgaddress{\street{College Avenue}, \city{Orono}, \postcode{04469}, \state{ME}, \country{USA}}}

\affil[2]{\orgdiv{Maine Business School}, \orgname{University of Maine}, \orgaddress{\street{College Avenue}, \city{Orono}, \postcode{04469}, \state{ME}, \country{USA}}}


\abstract{Recent advances in AI applications have raised growing concerns about the need for ethical guidelines and regulations to mitigate the risks posed by these technologies. In this paper, we present a mixed-methods survey study - combining statistical and qualitative analyses - to examine the ethical perceptions, practices, and knowledge of individuals involved in various AI development roles. Our survey comprises 414 participants from 43 countries, representing various roles such as AI managers, analysts, developers, quality assurance professionals, and information security and privacy experts. The results reveal varying degrees of familiarity and experience with AI ethics principles, governance initiatives, and risk mitigation strategies across roles, regions, and other demographic factors. Our findings underscore the importance of a collaborative, role-sensitive approach that involves diverse stakeholders in ethical decision-making throughout the AI development lifecycle. We advocate for developing tailored, inclusive solutions to address ethical challenges in AI development, and we propose future research directions and educational strategies to promote ethics-aware AI practices.}




\maketitle

\section{Introduction} 
\label{sec:introduction}
The rise of generative AI has intensified debates over its socio-economic impacts. Critics argue that these technologies can amplify issues such as discrimination, disinformation, deception, fraud, manipulation, and mistrust~\cite{bender2021dangers,center2023examples,gebru2023statement,gebru2023considerations,guardian2023drake}. A central concern is the lack of accountability and transparency, as many algorithms operate as “black boxes,” obscuring how decisions are made. Other challenges stem from the use of non-representative datasets and insufficient diversity in development teams, both of which can contribute to cultural or ethnic biases. Some researchers further warn that AI may undermine democratic institutions and the rule of law \cite{nyt2023godfather,future2023pause,gebru2023statement,guardian2023laws,center2023examples,center2023statement,chowdhury2021developers,marcus2023world} and call for the establishment or reinforcement of (supra-)national governance frameworks to address these risks
\cite{chowdhury2021developers,marcus2023world,worsdorfer2023aibill,AIAct2024}.  
In response, governments worldwide have introduced regulations, initiatives, and ethical guidelines, such as the EU AI Act \cite{AIAct2024}, to mitigate emerging harms. Yet, despite progress in defining AI ethics principles~\cite{worsdorfer2023aibill,center2023examples,center2023statement} and implementing regulatory measures to reduce AI risks~\cite {AIAct2024,ayyamperumal2024current,dong2024building}, substantial obstacles remain in developing trustworthy AI systems, including: (a) the complexity and ambiguity of existing regulatory frameworks~\cite{MHG17}, 
(b) a persistent gap between ethical principles and real-world practice, worsened by the limited understanding of how to operationalize ethics~\cite{pant2024ethics,griffin2024ethical,morley2023operationalising,sanderson2023ai,Alomar2022,vakkuri2020aiethics,vakkuri2022how,khan2023aiethics,kawakami2024studyingup},
 and (c) the absence of comprehensive guidelines and tool-supported methods to facilitate the adoption of ethical and privacy-preserving solutions~\cite{morley2023operationalising,jain2023,prybylo2024evaluating,khan2023aiethics}. 
 
Recent research suggests that fostering collaboration through human-centered AI (HAI) approaches~\cite{hao2024idontknow,li2018ai,shneiderman2020bridging} can strengthen the development of AI ethics systems. Past studies, drawing on semi-structured interviews or systematic literature review, have examined how AI developers perceive and implement AI ethics principles~\cite{pant2024ethics,sanderson2023ai,morley2023operationalising,rakova2021where,vakkuri2022how,griffin2024ethical,holstein2019improving,vakkuri2019ethically,deng2022understanding,vakkuri2019implementing,vakkuri2019ethicallyaligned,agbese2023implementing}.
Khan et al.~\cite{khan2023aiethics} and Pant et al.~\cite{pant2024ethicsai} evaluated AI ethics across countries, roles, and organizations. 
In this paper, we extend these prior works by jointly analyzing familiarity with ethics principles \emph{and} governance initiatives across seven distinct SDLC role types, using both LASSO regression and linear mixed-effects modeling to identify predictive factors. Table~\ref{tab:comparison_prior_studies} shows a detailed comparison with prior studies. 



This paper presents a large-scale mixed-methods study involving 414 participants, 
from 43 countries. The study systematically examines their knowledge and perceptions of AI ethics and its associated principles, regulatory initiatives, and best practices for mitigating AI-related risks throughout their development activities (i.e., from AI ethics requirements specification to design, development, and testing). Participants were recruited through Prolific and online platforms such as X (formerly Twitter), Reddit AI-focused subreddits, Quora, LinkedIn, Hugging Face, and Kaggle. They represent a wide range of roles within the AI development lifecycle, including managerial positions (e.g., project and product managers, CEOs, CTOs), analysts, developers, quality assurance professionals, researchers, ethicists, and privacy and security experts. Of the 411 participants who reported their location, 201 were from North America, 118 from Europe (including 12 from outside the EU, UK, and EEA), 43 from Africa, 25 from Central and South America, 16 from Asia, four from Australia, three from the Middle East, and one from another region. 

Our survey comprises three sections: (1) questions about participants’ demographics and experience with AI development (e.g., education, role, organization size and type, and level of involvement with AI); (2) questions examining their perceptions and practices related to AI; and (3) questions focusing on their knowledge, experiences, and risk mitigation strategies regarding AI ethics principles, regulatory initiatives, and best practices. 
We aim to address the following research questions (RQs):

\noindent \textbf{-- RQ1:} What are the general perceptions around AI and its usage, and how do they vary based on AI development experience, roles, and other demographic factors? 

\noindent \textbf{-- RQ2:} To what extent are AI practitioners familiar with AI ethics principles, and how do the familiarity, practice, and perception differ across roles, and other demographics, and what are the predictive factors?  

\noindent \textbf{-- RQ3:} To what extent are AI practitioners familiar with AI governance initiatives, and how does familiarity and perception differ across roles, and other demographics? 

\noindent \textbf{-- RQ4:} What are the risk perceptions regarding AI ethics, and how do AI risk mitigation practices differ across roles and other demographic characteristics?

\begin{table}[htbp]
\centering
\caption{Comparison with Prior Empirical Studies on AI Ethics Practice}
\label{tab:comparison_prior_studies}
\small
\begin{tabularx}{\textwidth}{p{1.8cm} p{0.5cm} p{1.0cm} X p{1.5cm} p{1.0cm} p{0.7cm} p{1.0cm}}
\toprule
\textbf{Study} & \textbf{N} & \textbf{Regions} & \textbf{Method} & \textbf{Principles} & \textbf{Mixed} \par \textbf{Roles} & \textbf{Gov.} & \textbf{Mitig.} \\
\midrule
Vakkuri \newline et al.~\cite{vakkuri2020aiethics,vakkuri2022how}
  & 211 & 1 & Survey+CS & Sub. & $\cdot$ & $\cdot$ & $\cdot$ \\
Sanderson \newline et al.~\cite{sanderson2023ai}
  & 21 & 1 & Interviews & 8 & Y & P & P \\
Griffin \newline et al.~\cite{griffin2024ethical}
  & 40 & Intl. & Interviews & Gen. & $\cdot$ & $\cdot$ & $\cdot$ \\
Khan \newline et al.~\cite{khan2023aiethics}
  & 99 & 20 & Survey & 5 & Y & P & $\cdot$ \\
Pant \newline et al.~\cite{pant2024ethicsai}
  & 100 & 8+ & Survey+SLR & Gen. & $\cdot$ & $\cdot$ & $\cdot$ \\
Olson \newline et al.~\cite{olson2025speaks}
  & 217 & Mult. & Survey & Gen. & Y & $\cdot$ & $\cdot$ \\
Reuel \newline et al.~\cite{reuel2025responsibleai}
  & 1000 & 19 & Survey & RAI & Y & Y & P \\
\textbf{Our} \newline \textbf{study}
  & \textbf{414} & \textbf{43} & \textbf{Survey} & \textbf{9} & Y & Y & Y \\
\bottomrule
\end{tabularx}
\begin{tablenotes}
\small
\item CS = case study. SLR = systematic literature review. Mixed Roles = multiple  SDLC role types. Principles = ethics principles covered (Sub.\ = subset, Gen.\ = generic). Gov.\ = governance initiatives. Mitig.\ = role-specific risk mitigation. Y = covered. P = partially covered. $\cdot$ = not covered.
\end{tablenotes}
\end{table}

Our analysis suggests that participants tend to associate AI with process automation and performance enhancement (\textbf{RQ1}). 
Even those not involved in AI development use AI in their daily routines, often viewing it as beneficial for enhancing productivity and accuracy. Data protection and security concerns remain prominent, particularly among those who have not yet adopted AI in their development practices.
 
Our findings reveal disparities in familiarity with AI ethics principles across roles and other demographics (\textbf{RQ2}). Participants in oversight roles (e.g., product managers or requirements analysts), researchers, those in government or mid-sized companies, and female participants generally reported higher familiarity with most AI ethics principles, though our Intra-Class Correlation (ICC) revealed greater variability within roles than between them. These principles are also applied differently in practice: AI managers and requirements analysts are more proactive in embedding ethics guidelines into development processes and documentation, often emphasizing user rights, whereas academics and researchers primarily incorporate technical aspects such as transparency, explainability, and fairness. According to our LASSO and mixed-effects analyses, role is a major predictor of how frequently ethics principles are considered in practice, and qualitative responses suggest organizational size and sector matter as well, with female participants and those from government or mid-sized companies placing greater emphasis on ethical considerations. Finally, our LASSO analysis showed that familiarity with AI governance initiatives, particularly the EU AI Act, is the factor most strongly associated with higher familiarity with AI ethics principles.


Familiarity with AI governance initiatives likewise varies across roles and demographics (\textbf{RQ3}). AI managers were the most familiar with governance initiatives, whereas those in quality assurance and information security roles were the least, and North American participants generally reported greater familiarity than their European counterparts, except for the EU AI Act, which was most familiar to participants in the EU, EEA, and UK. Familiarity with initiatives also tracked closely with familiarity with ethics principles and with hands-on use of AI tools. Perceptions of these initiatives are largely positive: a majority (66.5\%) view the impact of AI regulation favorably, though some practitioners, particularly in smaller companies, worry that compliance could slow innovation. Our LASSO analysis indicates that a positive regulatory outlook is most strongly associated with how frequently practitioners consider ethics in their work ($\beta = 0.333$), suggesting that hands-on experience with ethical practice reduces regulatory anxiety.

Risk mitigation practices vary notably by role and demographic context (\textbf{RQ4}). Individuals in administrative roles, such as product managers, often focus on establishing clear ethical guidelines, providing AI ethics training to their employees, and communicating AI ethics principles across their organizations. Along with requirements analysts, they frequently lead (ethics-related) risk and impact assessments of AI systems. 
Those in development roles employ complementary strategies, such as evaluating model outputs or assessing model performance across diverse populations, to mitigate biases in AI systems. 
Information security and privacy practitioners' tend to focus mainly on traditional security measures, such as encryption and access control, rather than other aspects of AI ethics. Researchers rely on institutional guidelines, policies, or personal judgment when developing AI tools and conducting AI-relevant research. Yet, they report challenges in writing ethics statements due to the difficulty of addressing multiple ethical considerations and the lack of precise guidance.

Overall, these findings underscore the importance of structured ethics guidelines and robust cross-role communication to ensure that diverse perspectives inform ethical decision-making and support the effective integration of ethics into AI development. We provide detailed research and educational directions based on these findings.

\section{Related Work and Background}
\label{sec:related-works}

\subsection{AI Ethics Principles}

In recent years, various governments and agencies created regulations or guidelines to incorporate ethics principles in AI development \cite{NSTC2017,EUDigital,Biden-Harris-1,Biden-Harris-2}, including the EU’s AI Act \cite{AIAct2024}. 
Research also proposed various AI ethics principles and guidelines \cite{algorithmwatch2024global,worsdorfer2023aibill,attard2022ethics,fjeld2020principled,hagendorff2020ethics,jobin2019global,leslie2021artificial}.   

Attard-Frost et al. \cite{attard2023ethics} reviewed 47 AI guidelines and introduced the FAST principles: Fairness, Accountability, Sustainability, and Transparency. They found that fairness and accountability were mentioned in 20 of the 47 documents, sustainability in 11, and transparency in 14. Analyzing 36 influential AI ethics principles documents produced by various stakeholders, such as governments, corporations, professional bodies, etc reveals eight recurring themes \cite{fjeld2020principled}: privacy (featured in 97\% of documents), accountability (97\%), safety and security (81\%), transparency and explainability (94\%), fairness and non-discrimination (100\%), human oversight (69\%), professional responsibility (78\%), and the promotion of human values (69\%). Leslie et al. \cite{leslie2021artificial} categorize core AI ethics principles into key values like human dignity, individual freedom and autonomy, the prevention of harm, non-discrimination, equality and diversity, transparency and explainability in AI systems, data protection and privacy rights, accountability and responsibility, and the upholding of democracy and the rule of law.

Wörsdörfer \cite{worsdorfer2023aibill} combines various AI ethics principles and guidelines and proposes a list of ordoliberal-inspired AI ethics principles that could form the basis for ``AI guardrails''~\cite{ayyamperumal2024current,dong2024building} and ``constitutional AI''~\cite{abiri2024public,huang2024collective}. These principles include respect for human rights, data protection and privacy, harm prevention and beneficence, non-discrimination and freedom from privileges, fairness and justice, transparency and explainability, accountability and responsibility, democracy and the rule of law, and environmental and social sustainability. A human-centered approach to AI prioritizes respect for fundamental rights and dignity, aiming to preserve human agency, control, and responsibility. Central to this philosophy is the principle of beneficence, which mandates that AI systems do no harm by avoiding discrimination, manipulation, and negative profiling while respecting personal privacy. This involves actively tackling biased algorithmic decision-making to ensure freedom from bias and protecting the most vulnerable and marginalized groups, thereby addressing exclusion and inequality. Adopting ``explainable AI'' promotes algorithmic transparency and accountability, enabling the opening of black-box algorithms and enhancing trust. Integrating ``ethics by design'' ensures that ethical considerations such as fairness and justice are woven into the fabric of AI development. This necessitates embedding AI within participatory democracy and the rule of law, emphasizing the importance of stakeholder dialogue and public debates to navigate the complexities posed by AI technologies. 

Recent work has also examined the operationalization of AI ethics principles. Zieli\'{n}ski et al.~\cite{zielinski2026ramp} present RAMP, an LLM-based system that extracts policy statements from governance documents (e.g., EU AI Act, NIST AI RMF), decomposes them into atomic obligations, and proposes system-specific, testable compliance rules. Demonstrating how governance requirements can be translated into actionable engineering artifacts while maintaining human oversight. On the user-facing side, H\"{o}ltervennhoff et al.~\cite{holtervennhoff2026doom} conducted a large-scale study (5 focus groups, $N=1{,}354$ survey participants) on AI-generated image labels as a transparency mechanism, finding that while users consider labeling helpful, it can lead to overreliance and unintended side effects---underscoring the complexity of operationalizing transparency in practice.

In this paper, we adopt Wörsdörfer's \cite{worsdorfer2023aibill} AI ethics principles due to their comprehensive scope, which encompasses those found in other frameworks, as well as their practical applicability (see Appendix \ref{Appendix:Defs} - Table \ref{tab:AIEthics-Definition} for more detail).

\subsection{Ethics in AI Development}


Several studies examined developers' understanding and challenges, gaps in the practices, and organizational structures in implementing ethics principles in AI systems through user studies with developers and practitioners~\cite{sanderson2023ai,vakkuri2022how,vakkuri2019implementing,khan2023aiethics,reuel2025responsibleai,vakkuri2020aiethics,pant2024ethics,morley2023operationalising,rakova2021where,griffin2024ethical,holstein2019improving,vakkuri2019ethically,deng2022understanding,vakkuri2019ethicallyaligned,agbese2023implementing,pant2024ethicsai}. 

Vakkuri et al.~\cite{vakkuri2020aiethics,vakkuri2022how} investigated developers' awareness and implementation of AI ethics principles. They found that developers often treat AI features like other software functionalities and that existing AI ethics guidelines have a limited impact on their practices. They further observed that many companies struggle to integrate AI ethics into existing workflows, particularly in balancing ethics principles with efficiency and business goals, leading to the deprioritization of moral considerations. Pant et al. \cite{pant2024ethicsai,pant2024ethics} examined AI developers’ awareness of AI ethics principles through a user study with 100 participants and analysis of 38 academic papers. They identified that awareness varied demographically, as those with more experience, a Ph.D., or identified as female reported higher awareness, and it was mainly influenced by workplace policies, personal interest, and media coverage. The study revealed that cultural and personal biases shape AI outcomes, while unclear role boundaries create uncertainty about responsibility, often shifting accountability to users or upper management. Participants were most familiar with data protection and privacy but had a limited understanding of human-centered principles such as fairness, accountability, transparency, and explainability, which aligns with the findings in \cite{sanderson2023ai} and \cite{morley2023operationalising}. Human-centered values were viewed as the hardest to implement or uphold due to inconsistent perceptions, limited knowledge and resources, a lack of practical tools and guidelines, insufficient organizational support, inherent human biases, and trade-offs between utility and explainable AI systems; similar to \cite{rakova2021where}. The study emphasized the importance of diverse teams, explicit role definitions, and structured approaches to ethical decision-making. Griffin et al.~\cite{griffin2024ethical}, and Agbese et al.~\cite{agbese2023implementing} complemented these findings through interviews with AI developers, showing that while most are aware of the existing issues, they lack formal training and support to navigate them consistently and often rely on informal advice due to insufficient organizational guidelines. 

Other studies have revealed specific challenges associated with implementing AI ethics principles. 
Khan et al.~\cite{khan2023aiethics} conducted a user study with developers and law practitioners, revealing that the biggest challenges in implementing AI ethics principles are a lack of ethical knowledge and unclear regulations. 
Hartikainen et al. \cite{hartikainen2022human} found that developers are often responsible for the early design of AI systems, where they follow flexible practices rather than strict design processes, and test with clients rather than end-users. Stahl et al.~\cite{stahl2022organisational} emphasized the need for organizational awareness and stakeholder communication to implement AI ethics principles. Kawakami et al.'s~\cite{kawakami2024studyingup} study on how power dynamics influence decision-making in AI design highlights the importance of training leadership to effectively communicate and engage diverse stakeholders in ethical decision-making. A study by Olson et al. \cite{olson2025speaks} shows that marginalized demographics report ethical concerns more frequently and are more aware and sensitive to ethical implementations in software development. Reuel et al.~\cite{reuel2025responsibleai} propose a formal maturity model to measure the organizational and operational maturity of Responsible AI (RAI) and highlight that companies may appear to be more prepared for RAI in terms of organizational structure, internal processes, and frameworks than their practices reveal. 
For many organizations, RAI was seen as a means of improving revenue, brand performance, or technical performance.



\subsection{Positioning of This Study}

Table~\ref{tab:comparison_prior_studies} compares our study with key prior empirical studies on AI ethics practices across eight dimensions. Several gaps emerge from this comparison. First, prior studies have typically focused on a single stakeholder group (e.g., developers) or combined at most two (e.g., developers and lawyers); none have systematically compared familiarity, practices, and risk mitigation strategies across the full range of SDLC roles. Second, no prior work has jointly examined familiarity with AI ethics principles \emph{and} governance initiatives (e.g., EU AI Act, NIST AI RMF), despite growing evidence that regulatory awareness shapes ethical practices. Third, past research relied mainly on descriptive statistics or qualitative analysis; predictive modeling (e.g., LASSO, mixed-effects) to identify which factors most strongly drive familiarity and practice has not yet been applied. Fourth, the dynamic between perceptions of regulation, role-specific risk mitigation strategies, and actual ethics consideration frequency has not been examined in an integrated study. Our work addresses these gaps by combining a seven-role breakdown with mixed-methods analysis---including LASSO regression, linear mixed-effects models, Kruskal-Wallis and Chi-square tests, and thematic qualitative coding---across nine ethics principles, eight governance initiatives, and role-specific risk mitigation strategies, drawing on 414 participants from 43 countries.

\section{Study Design and Analysis}
\label{sec:survey-methods}

This section describes our survey design methodology and analysis processes. 

\subsection{Survey Design}
\label{sec:survey-design}

\subsubsection{Overview}
\label{sec:survey-overview}

We conducted two large-scale studies, using Qualtrics: one on the Prolific  platform~\cite{palan2018prolific,Tahaei-CHI2022} and another across various forums, including AI-related subreddits, Quora, X, LinkedIn, HuggingFace, and Kaggle, with AI practitioners in 43 countries. Our studies were completed between August and September 2024. 
We used Prolific as it maintains a pool of active participants who are regularly screened and vetted by the platform, offers a higher pay rate, allows a selection of a more specific pool of participants with basic programming skills and AI familiarity, and is recommended by prior work~\cite{Tahaei-CHI2022,kaur2022recruit}. 

\subsubsection{Recruitment Process and Filtering}
\label{sec:recruitment-filtering}

The sample size was determined in accordance with Prolific’s guidelines and informed by prior similar research~\cite{prybylo2024evaluating}, which recommends a minimum of 300 representative participants. We initially pre-screened the Prolific participants based on the following requirements: (a) being at least 18 years old, (b) being fluent in English, (c) using AI tools at least once a week, (d) possessing knowledge of common programming languages, (e) having more than two years of work experience, and (f) working in software-related industries. We used their industry, rather than their professional role, as a filter since Prolific does not allow a role-based selection. We paid an average of \$12.58 per hour to those who completed the survey - above the minimum recommended hourly rate by Prolific (\$8 per hour). Prybylo et al.~\cite{prybylo2024evaluating} discuss that although pre-screening via programming questions, such as those recommended by Danilova et al.~\cite{danilova2021you}, has its advantages~\cite{Tahaei-CHI2022,kaur2022recruit,danilova2021you,serafini2023recruitment}, it poses several risks, such as overusing the same questions in multiple surveys, their limited scope to only developers and not other roles (e.g., testers or privacy experts), and the accessibility of AI tools like ChatGPT~\cite{brown2020language}, which could answer those questions with high accuracy~\cite{danilova2021you}. We discuss how we mitigated the pre-screening challenges below and in Section~\ref{sec:ethical-considerations}.
After the initial pre-screening, we recruited 450 participants across the EU, the US, and other countries (i.e., ``world pool'') on Prolific. Of those who started the survey, 120 did not finish it and were thus excluded from our analysis. 

We carefully selected the forums related to AI development, e.g.,  Quora, HuggingFace, and Kaggle, as well as subreddits, like (\texttt{r/LanguageTechnology}, \texttt{r/PromptEngineering}, \texttt{r/MachineLearningJobs}, and \texttt{r/ComputerVision}).
We also posted the survey on our LinkedIn and academic X profiles, which mainly consist of computer science researchers and developers. We recruited participants who were (a) at least 18 years old and (b) had at least two years of AI development or research experience. Participants could enter a raffle to win one of the four \$25 Amazon gift cards (as described in the consent form). We received 356 responses, and excluded those who did not provide consent (as required by our IRB policy) (2) or did not complete the survey (154). Overall, we received 200 complete responses from the forums. 

\begin{table}[h]
\centering
\caption{Breakdown of Participants' Roles}
\begin{tabular}{p{7cm}ccc}
 \toprule
\textbf{Role} & \textbf{Total} & \textbf{Group A} & \textbf{Group B} \\ 
\midrule
AD: AI/Software Developer, Architect, Data Scientist & 195 (47.1\%) & 53 & 142 \\ 
AM: AI (Product/Project) Manager, Administrative Role (e.g., CEO, CTO)             & 76 (18.4\%)  & 14 & 62  \\ 
RA: Requirements Analyst            & 54 (13.0\%) & 15 & 39  \\ 
AR: AI Researcher                                   & 26 (6.3\%)  & 4 & 22   \\ 
QA: Quality Assurance Engineer, IT Maintenance           & 29 (7.0\%)  & 9  & 20  \\ 
ISec: Information Security/Privacy Expert         & 19 (4.6\%)  & 6  & 13  \\ 
AE: AI Ethicist, Legal Team                       & 5 (1.2\%)   & 2  & 3   \\ 
Other: Other, please specify                         & 10 (2.4\%)  & 4  & 6   \\ 
\midrule
\textbf{Total}                                     & \textbf{414} & \textbf{107} & \textbf{307} \\ 
\bottomrule
\end{tabular}%
\label{table:participant_roles}
\end{table}
 
We then conducted another filtering process to remove unreliable responses from both sources. We first removed those who completed the survey in under three minutes (twelve from online forums and Prolific), had duplicate answers from other participants (25 from online forums), and those whose written responses were not in English (nine from online forums). We also attempted to remove AI-generated answers from both sources. We identified such responses based on prior research on analyzing differences in linguistic patterns between human and LLM-generated text~\cite{tang2024science}. Munoz et al.~\cite{munoz2024contrasting} found that LLMs tend to have a similar length distribution for similar tasks with a lower lexical diversity than human responses. Their findings are complemented by Guo et al.~\cite{guo2023how}, who found that LLMs tend to have longer answers with a less diverse vocabulary in Q\&A tasks than humans. We flagged and removed participants with abnormally long answers and those with multiple answers of similar length, vocabulary, grammatical structure, and syntactical patterns that appeared to be AI-generated (i.e.,  9 participants from Prolific and 15 from online forums). Ultimately, the final participant pool consisted of 276 individuals from Prolific and 138 from online forums. 

\subsubsection{Survey Questions}
\label{sec:survey_questions}

Our survey features a mix of demographic, perception, expertise, and practice questions, all designed in alignment with our RQs (see Section~\ref{sec:introduction}). The survey design builds upon the methodological framework established by Prybylo et al.,~\cite{prybylo2024evaluating}, extending their role and demographic-based analysis of privacy perceptions and practices to the broader domain of AI ethics. Following this precedent, we used Likert-scale items to quantify perceived impact, familiarity, and the frequency of ethical practices, enabling robust statistical comparisons across diverse demographics. We also leveraged the findings and the gaps identified in prior research, such as the implementation of AI ethics principles~\cite{morley2023operationalising,rakova2021where,vakkuri2022how,griffin2024ethical}, familiarity with AI  regulations and their impact~\cite{pant2024ethics}, the organization of AI development teams and culture \cite{morley2023operationalising,rakova2021where,vakkuri2022how,griffin2024ethical,agbese2023implementing}, and team diversity and stakeholder involvement in decision making~\cite{hartikainen2022human,stahl2022organisational,sanderson2023ai}to formulate the questions. While we did not have a formal external pilot study, we conducted an informal pilot with members of our research lab, who completed the full survey and provided detailed feedback. 
Based on their feedback, we iteratively refined the questions to resolve ambiguous wording, improve question ordering, and reduce completion time. Our survey covers nine AI ethics principles, and we initially asked participants to write open-ended responses about the risks associated with each and their mitigation strategies. Even after applying role-specific branching to shorten the instrument (following Prybylo et al.~\cite{prybylo2024evaluating}), the pilot revealed that answering open-ended risk-related questions for all nine principles made the survey too long. In response, for the main survey we (i) randomized the principles and presented each participant with only a subset of the open-ended risk-based questions, and (ii) supplemented these with multiple-choice questions, retaining coverage across all principles while reducing completion time.
Lastly, as one of the authors teaches computer and business ethics courses, the survey questions were informed by his experience and interactions with both computer science and business students in his classes. 

All participants were required to provide informed consent in accordance with our university's IRB policy before proceeding to the main survey. After obtaining consent, we asked all participants the same set of twelve questions, partly related to demographics and their professional experience with AI development. We then divided the remaining questions into two groups: one for those who did not have professional experience developing AI systems (\textbf{Group A - No AI Development Experience}) and another for those with experience (\textbf{Group B - With AI Development Experience}), based on the responses to questions \textit{P11} and \textit{P12}. We further divided the Group B participants into seven subgroups, based on the roles defined in Table \ref{table:participant_roles}, except for the ``Others'' category, and 
asked role-specific questions to ensure the survey was not too long. Previous research~\cite{prybylo2024evaluating} shows that participants may not finish a survey if it is longer than 30 minutes. 
For example, we asked the ISec roles about addressing ethical concerns regarding data privacy and security (B.13.5) and the AI researchers about their challenges in writing ethics statements (B.16.5). Ten participants who chose ``Other, please specify'' could not be attributed to any specific role and were thus assigned to Group A. 
Our breakdown loosely follows the software development lifecycle (SDLC) phases. However, we separated the Information Security/Privacy (ISec) and AI Researcher and Ethicist roles to evaluate their perspectives on AI ethics. 
The complete list of questions (except questions 1-3, which are the required Prolific identification questions and our consent form) can be found here.\footnote{Survey questions: https://tinyurl.com/232zaruh}

\subsection{Study Analysis Process}
\label{sec:survey-results}

Our survey results are organized around our RQs, focusing on how various roles perceive, comprehend, and implement AI ethics throughout the SDLC, across various demographics. 
Our analysis employs a mixed-method approach, combining both quantitative and qualitative methodologies. 

\subsubsection{Qualitative Analysis} We evaluated the descriptive and open-ended questions through open coding procedures and iterative processes. For questions related to \emph{The perceived risks and mitigation strategies associated with each AI ethics principle}, the first and second authors conducted a thematic analysis to find recurring patterns and themes. They independently assigned categories for the first 100 responses, then met to discuss their results, resolve the discrepancies, and create a guideline (see Appendix~\ref{Appendix:Defs}). This consensus-based approach to inter-coder agreement follows established qualitative methodology~\cite{cascio2019team, mcdonald2019reliability}. Rather than computing statistical inter-rater reliability metrics such as Cohen's kappa---which assume predefined categories and may not align with inductive thematic analysis where codes emerge iteratively from the data~\cite{oconnor2020intercoder}, we ensured reliability through independent coding followed by structured discussion, discrepancy resolution, and codebook creation. This negotiated agreement approach is widely used in qualitative software engineering research~\cite{mcdonald2019reliability,gonzalez2023reliability}.
For participants' definitions of AI within their industries, however, the first author used the NIST Taxonomy of Human-AI Use Cases~\cite{nist2024taxonomy} to categorize their responses, and the last author reviewed the overall results. For the remaining open-ended questions, the first author conducted a thematic analysis of each question, and the last author (who is an ethics researcher) examined these results to resolve inaccuracies. 

\subsubsection{Quantitative Analysis} 
We conducted a non-parametric \textit{Kruskal-Wallis test}~\cite{kruskal1952use} to analyze responses to Likert-scale questions and a \textit{Chi-Square test} \cite{greenwood1996guide} to assess the association between AI ethics principles or AI governance initiatives familiarity and various demographic factors. We hypothesize, based on our RQs, that factors such as company size, professional roles, location, and gender may influence individuals’ practices, perceptions, and familiarity with AI ethics principles and governance initiatives, ultimately shaping development practices. To control for Type I errors and reduce the likelihood of false positives, we applied the Bonferroni correction \cite{Bonferroni}. Since the Bonferroni correction is very conservative and may increase Type II errors, we discussed the results with respect to $alpha = 0.05$ and the adjusted value (i.e., $\frac{0.05}{26} = 0.0019$). 

We conducted \textit{least absolute shrinkage and selection operator (LASSO)}~\cite{tibshirani1996lasso} regression to identify which demographics, practices, and perceptions are most associated with: (1) familiarity with AI ethics principles and governance initiatives, (2) frequency of considering AI ethics principles, and (3) perceived impact of AI governance initiatives. 
We used the Likert-scale responses as continuous dependent variables in all models, as recommended by Norman et al. \cite{norman2010likert}. For potential explanatory variables, we used the questions listed in Appendix \ref{Appendix:stats} - Table~\ref{tab:lasso-predictors}.

We preprocessed features according to their measurement type: \textit{ordinal variables} (age groups, company size, experience levels, and frequency scales) were coded numerically to preserve rank order; \textit{categorical variables} (education level, company type, role) were one-hot encoded; \textit{multi-select questions} (AI applications, tools used, mitigation strategies) were converted to binary indicator variables; and 
\textit{binary variables} (e.g., company builds AI) were coded as 0/1. 
This step yielded 118 and 179 features for Groups A and B, respectively. 
Following best practices~\cite{hastie2009elements}, we employed a two-stage procedure: first, we conducted a LASSO analysis with 5-fold cross-validation using the one-standard-error rule to select important predictors; second, we re-estimated coefficients using standard linear regression on only the selected features to obtain unbiased estimates. 
We reported post-LASSO standardized coefficients, which represent the change in the outcome variable (on the original Likert scale) associated with a one-standard-deviation change in each predictor. 
We evaluated the model performance using R² (proportion of variance explained) on a held-out test set (20\% of data).

Finally, we employed \textit{linear mixed-effects models} to analyze our survey data, accounting for the hierarchical structure in which practitioners are nested within professional roles. This approach is appropriate because individuals within the same role may share similar organizational contexts and professional norms~\cite{stroup2012glmm}. 
We specify two models to test our hypotheses about factors influencing familiarity with AI ethics principles and how they are considered in practice \textbf{(RQ2)}.

\textbf{Model 1 (AI Ethics Principle Familiarity)} predicts practitioners' mean familiarity with nine AI ethics principles (scale 1--5) based on years of AI development experience (ordinal 1--4 representing 1--2 years through 10+ years), active AI development involvement (binary), and mean familiarity with eight AI governance initiatives (scale 1--5), with random intercepts for professional roles. 

\textbf{Model 2 (Ethics Consideration Frequency)} predicts how often practitioners consider ethics in their work (scale 1--5) based on mean familiarity with AI ethics principles, mean familiarity with AI governance initiatives, and years of experience, with random intercepts for professional role (as shown in Table \ref{table:participant_roles}).
Prior to interpretation, we performed diagnostics to assess normality assumptions. 

\textit{Outcome Variable Distributions:} We examined distributions of all outcome variables prior to modeling. Familiarity with AI ethics principles shows left skewness (skewness = $-0.790$), as does ethics consideration frequency (skewness = $-0.731$). Familiarity with AI governance initiatives is approximately symmetric (skewness = $-0.152$). Shapiro-Wilk tests rejected the normality assumption for all three outcomes ($p < 0.001$). Model residuals show some deviation from perfect normality: Shapiro-Wilk tests yielded $W = 0.974$ ($p = 0.0001$) for Model 1 and $W = 0.983$ ($p = 0.003$) for Model 2. We consider these models acceptable given the mild nature of departures (W-statistics near 0.98), large sample sizes, and significant effects ($p < 0.01$). 

\textit{Box-Cox Transformation Exploration:} To address non-normality, we used Box-Cox power transformations for each outcome variable. For familiarity with AI ethics principles, the optimal $\lambda = 1.99$ reduced skewness from 0.790 to 0.167, though normality tests remained significant ($W = 0.972$, $p < 0.001$). For frequency of considering AI ethics principles, $\lambda = 1.86$ reduced skewness from 0.731 to 0.195 ($W = 0.873$, $p < 0.001$). For AI governance initiative familiarity, $\lambda = 0.90$ provided a small improvement, with skewness increasing slightly from 0.152 to 0.202 ($W = 0.937$, $p < 0.001$). While Box-Cox transformations reduced skewness for two outcomes, they did not achieve normality; therefore, we did not use these transformed data in our analysis. 

\subsubsection{Methodological Validation}
\label{sec:method-validation}


\subparagraph{\textbf{Survey Instrument Reliability}} To assess the internal consistency of our Likert-scale instruments, we computed Cronbach's alpha for the two main familiarity blocks answered by participants with AI development experience (Group B): the AI ethics principle familiarity scale (9 items, question B.2.1) and the AI governance initiative familiarity scale (8 items, question C.1). Cronbach's alpha is used to measure internal consistency computed over participants' item-level responses, quantifying how strongly the items within a scale co-vary as a single underlying construct, rather than an interrater agreement metric. We mapped each participant's 1--5 Likert familiarity ratings to numeric scores and computed alpha over the participants who provided complete responses on the respective scale (n = 294 for the 9-item principle scale and n = 291 for the 8-item governance scale; participants with any missing item on a scale were excluded listwise for that scale), using the standard Cronbach's alpha formula. The principle familiarity scale yielded $\alpha$ = 0.916 and the governance initiative familiarity scale yielded $\alpha$ = 0.950. Both values exceed the commonly accepted threshold of 0.70, indicating internal consistency.

\subparagraph{\textbf{Likert-as-Continuous Justification}}
Following established practice in empirical software engineering~\cite{prybylo2024evaluating,pant2024ethicsai}, we treat Likert-scale responses as continuous dependent variables in our LASSO regression and mixed-effects models. This approach is well-supported by the methodological literature, which shows that parametric methods are robust to ordinal data under common conditions~\cite{norman2010likert}. Importantly, we complement this treatment with non-parametric tests throughout our analysis: Kruskal-Wallis tests for ordinal comparisons across multiple groups and Chi-square tests for categorical associations, ensuring our conclusions do not rely solely on the continuity assumption.

\subparagraph{\textbf{Prolific vs.\ Forum Sample Comparison}}
To assess potential sampling bias between our two recruitment sources (276 Prolific participants and 138 forum participants), we compared the means for familiarity with AI ethics principles and AI governance initiatives between the two groups using Mann-Whitney U tests. Forum participants reported slightly higher mean principle familiarity than Prolific ones ($M=3.91$ vs.\ $M=3.66$; Mann-Whitney $U=9104.5$, $p=0.018$, $r=0.16$), and substantially higher mean governance initiative familiarity ($M=3.34$ vs.\ $M=2.36$; $U=10137.5$, $p<0.001$, $r=0.44$). These results suggest that forum participants---recruited from AI-focused communities such as HuggingFace, Kaggle, and AI subreddits---may have greater engagement with AI governance topics. We control for this potential source effect in our LASSO and mixed-effects analyses by including demographics and experience as covariates, and we acknowledge this as a limitation.

\subparagraph{\textbf{AI-Generated Response Filtering}}
The 24 responses flagged as potentially AI-generated were independently reviewed by two authors before exclusion to ensure that we did not accidentally remove valid responses. 
We acknowledge that this filtering process, while conservative, may have excluded some genuine responses or missed some AI-generated ones (see Section~\ref{sec:limitations}).

\subsubsection{Ethical Considerations}
\label{sec:ethical-considerations}
This research adheres to our university's ethical guidelines and was approved by our Institutional Review Board (IRB). All participants agreed to a thorough consent form that included information about the investigators, risks, benefits, compensation, and confidentiality. All participants were informed about their voluntary participation, with the emphasis on their right to withdraw at any time. No personally identifiable information was collected, and measures were in place to ensure the anonymity, confidentiality, and security of responses. The contact information of all investigators and the IRB team was also included. No participants contacted the investigators or the IRB about the study or the compensation. 


   



\section{Findings}
\label{sec:findings}

\subsection{Demographics}
After filtering, we accepted 414 responses (see Section~\ref{sec:recruitment-filtering}). 
201 participants worked at companies in North America, 118 in EU+EEA+UK, and 95 in other or unspecified locations. 
$\sim$74\% of them indicated they had experience developing AI products (i.e., Group B) while $\sim$26\% did not have any experience (i.e., Group A). 
Table \ref{table:participant_roles} illustrates that the majority are in AD (e.g., AI/software developer) (47.1\%) and AM (e.g., AI manager) (18.4\%) roles. Appendix \ref{appendix:dem} - Table \ref{Table:Demographics} shows that most participants identify as male (67.6\%), are below the age of 45 ($\sim$88\%), and have completed a Bachelor's or higher degrees ($\sim$89\%), with $\sim$77\% in computer science, information science, software engineering, data science, and electrical, computer, privacy, or security engineering. Among the 13.1\% with a business degree, $\sim$32\% are in AD, $\sim$30\% in AM, and $\sim$24\% in RA roles. 
Of the 10.2\% who did not have degrees within any of the above groups, $\sim$25\% were in health sciences and $\sim$10\% in law and psychology (among others). Most participants ($\sim$60\%) have at least two years of AI development experience.

\begin{tcolorbox}[colback=blue!10,colframe=black!30,boxrule=0.1pt, boxsep=0.1pt]
\textbf{Implications.} The 47\% AD share reflects the developer-heavy composition of the AI field, while the under-representation of AR and AE roles (6.3\% and 1.2\%, respectively) mirrors the field's limited adoption of dedicated ethics personnel~\cite{reuel2025responsibleai}. The predominance of male participants (67.6\%) aligns with industry demographics but underscores the importance of our gender-disaggregated analyses, given findings by Pant et al.~\cite{pant2024ethicsai}, and Olson et al.~\cite{olson2025speaks} that gender correlates with ethics awareness.
\end{tcolorbox}

\subsection{AI Perceptions and Usage}

To address \textbf{RQ1}, we examined participants’ perceptions of AI, its applications, and its impact on their work.

\subsubsection{Defining AI}
\begin{figure}[!htbp]
    \centering
    \includegraphics[width=0.9\textwidth]{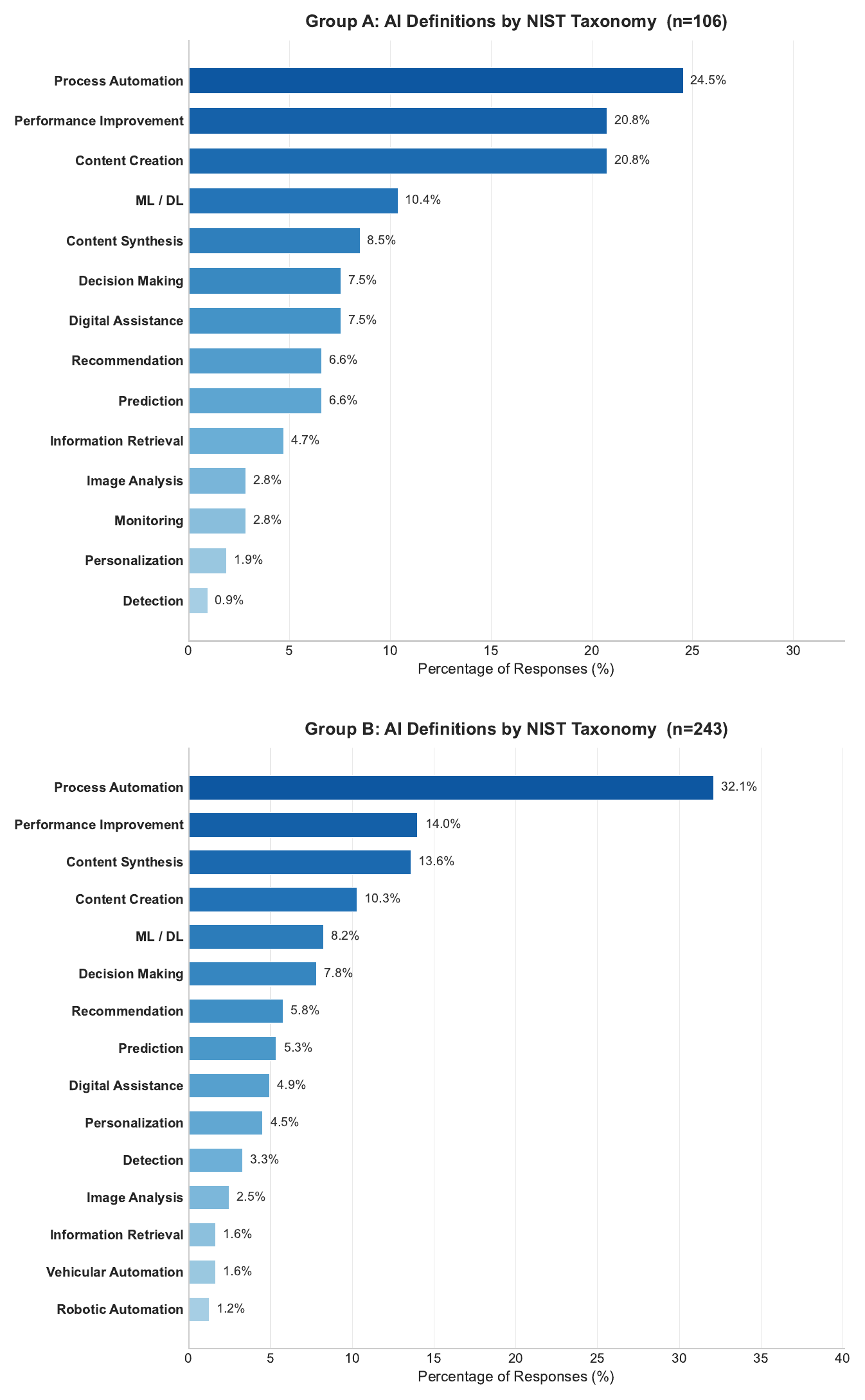} 
    \caption{AI Definitions Based on NIST Human-AI Taxonomy}
    \label{fig:definitions}
\end{figure}

To understand how participants perceive AI, we asked them to define AI within their industry. We used the NIST Taxonomy of Human-AI Use Cases~\cite{nist2024taxonomy} to categorize their responses (see Appendix \ref{Appendix:Defs} - Table \ref{Table:NIST_taxonomy} for more details on the taxonomy and Figure \ref{fig:definitions} for a breakdown of participants' responses). This taxonomy does not capture some technical definitions, such as machine learning. We subsequently created a dedicated category for \textit{Machine/Deep Learning (ML/DL)}. Multiple labels are possible.

Majority of Groups A and B define AI in terms of ``process automation,'' ``performance improvement,'' ``content creation,'' and ``content synthesis,'' with ``ML/DL'' also among the top five definitions. For process automation, one participant explained, \textit{``In our industry, AI is defined as useful assistance systems to make repetitive work easier.''}\footnote{All participant quotes are provided verbatim, including any spelling or grammatical errors.} For performance improvement, another participant described it as, \textit{``A system that acts as a support to eliminate time wastage and allows the team focus on more profit generating ventures.''} Neither group defines AI in terms of ``discovery.'' Interestingly, Group A refers to ``content creation'' roughly twice as often as Group B ($\sim$20.8\% vs $\sim$10.3\%). In contrast, ``content synthesis'' - a more technical term - is mentioned more frequently by Group B, who are generally more AI-savvy than Group A. We observe further disparities in popularity between Groups A and B's definitions. For example, $\sim$3\% of Group A participants define AI as ``monitoring,'' while Group B does not mention this concept. Conversely, $\sim$3\%  of the more AI-savvy Group B define AI as ``robotic or vehicular automation,'' which is not considered by Group A. 

12.26\% and 17.70\% of responses in Groups A and B highlighted a wider range of use cases that could not be mapped to any of the NIST categories. Rather than defining AI, one participant spoke of its implications: \textit{``It is revolutionary and can change everything in this industry.''} We categorized these definitions as ``Others.'' 
Specifically, in Group B, some responses focus on societal or future impacts of AI (e.g., \textit{``corruptive and hugely beneficial in equal measures''}).

We further analyzed Group B by their roles. ``Process automation'' is the most common AI definition among AM roles (41.03\%), but it is considerably less common among ISec (20\%) and AR (18.18\%) roles. This may be because the management team is more heavily involved in process improvement, an area where AI-driven automation can have a positive impact. Participants in QA and AD roles are most likely to describe AI as ``content synthesis,'' which is a more technical term. QAs are also considerably more likely than other roles to mention ``decision making'' (27.78\%). ISec roles are more than three times more likely to say ``content creation'' (33.33\%) than other roles. This may suggest that individuals in these roles are less likely to be involved in AI development activities, and like Group A participants, see AI through the lens of contemporary narratives rather than its technical aspects. Similar to prior work, these findings underscore the need for tools and educational frameworks that focus on privacy and security aspects of AI, as ISec team members play a critical role in addressing emerging threats that developers are often unaware of~\cite{gao2024idk,meisenbacher2024privacy}.

\begin{tcolorbox}[colback=blue!10,colframe=black!30,boxrule=0.1pt, boxsep=0.1pt]
\textbf{Implications.} Our results indicate that those with experience in AI development (i.e., Group B) identify more technical aspects of AI and its abilities in areas such as data analysis and automation 
How AI is viewed and defined depends heavily on the level of involvement in AI development. 
\end{tcolorbox}

\subsubsection{AI Usage}
We asked both groups about the primary applications of AI in their products. As shown in Table \ref{tab:ai_applications}, ``chatbots, personal assistants, or recommender systems'' are considered the most common AI applications across both groups. Group B considers ``robotics and automation,'' ``cybersecurity,'' ``computer vision,'' ``human resources,'' and ``legal'' more than Group A. In comparison, Group A participants are $\sim$13\%, $\sim$13\%, and $\sim$10\% more likely to choose ``translation and text generation,'' ``healthcare,'' and ``chatbots, personal assistants, or recommender systems.'' These results further indicate that those without AI development experience (Group A) are more optimistic about applications tailored for the general public, such as chatbots, while those with AI experience  (Group B) consider more job-related or technical applications.

\begin{table}[ht]
\centering
\caption{Applications of AI in Products}
\label{tab:ai_applications}
\begin{tabular}{lcc}
 \toprule
\textbf{Application} & \textbf{Group A} & \textbf{Group B} \\
\midrule
Chatbots, Personal Assistants, or Recommender Systems
  & 79 (73.8\%) & 193 (63.5\%) \\
Customer Service
  & 58 (54.2\%) & 145 (47.7\%) \\
Code Completion or Code Generation
  & 51 (47.7\%) & 132 (43.4\%) \\
Translation or Text Generation
  & 60 (56.1\%) & 129 (42.4\%) \\
Predictive Analysis
  & 46 (43.0\%) & 123 (40.8\%) \\
Robotics and Automation
  & 36 (33.6\%) & 120 (39.5\%) \\
Cybersecurity
  & 36 (33.6\%) & 113 (37.5\%) \\
Computer Vision
  & 28 (26.2\%) & 101 (33.2\%) \\
Financial Services
  & 38 (35.5\%) & 97 (31.9\%) \\
Healthcare
  & 43 (40.2\%) & 81 (26.6\%) \\
Logistics
  & 29 (27.1\%) & 74 (24.3\%) \\
Entertainment or Communication
  & 29 (27.1\%) & 73 (24.0\%) \\
Personalization and Advertisement
  & 34 (31.8\%) & 72 (23.7\%) \\
Human Resources
  & 25 (23.4\%) & 71 (23.4\%) \\
Legal
  & 18 (16.8\%) & 56 (18.4\%) \\
Retail
  & 17 (15.9\%) & 44 (14.5\%) \\
Other, please explain
  & 3 (2.8\%) & 7 (2.3\%) \\
Prefer not to say
  & --- & 1 (0.33\%) \\
\bottomrule
\end{tabular}
\end{table}

\begin{figure}[!htbp]
    \centering
    \includegraphics[width=0.9\textwidth]{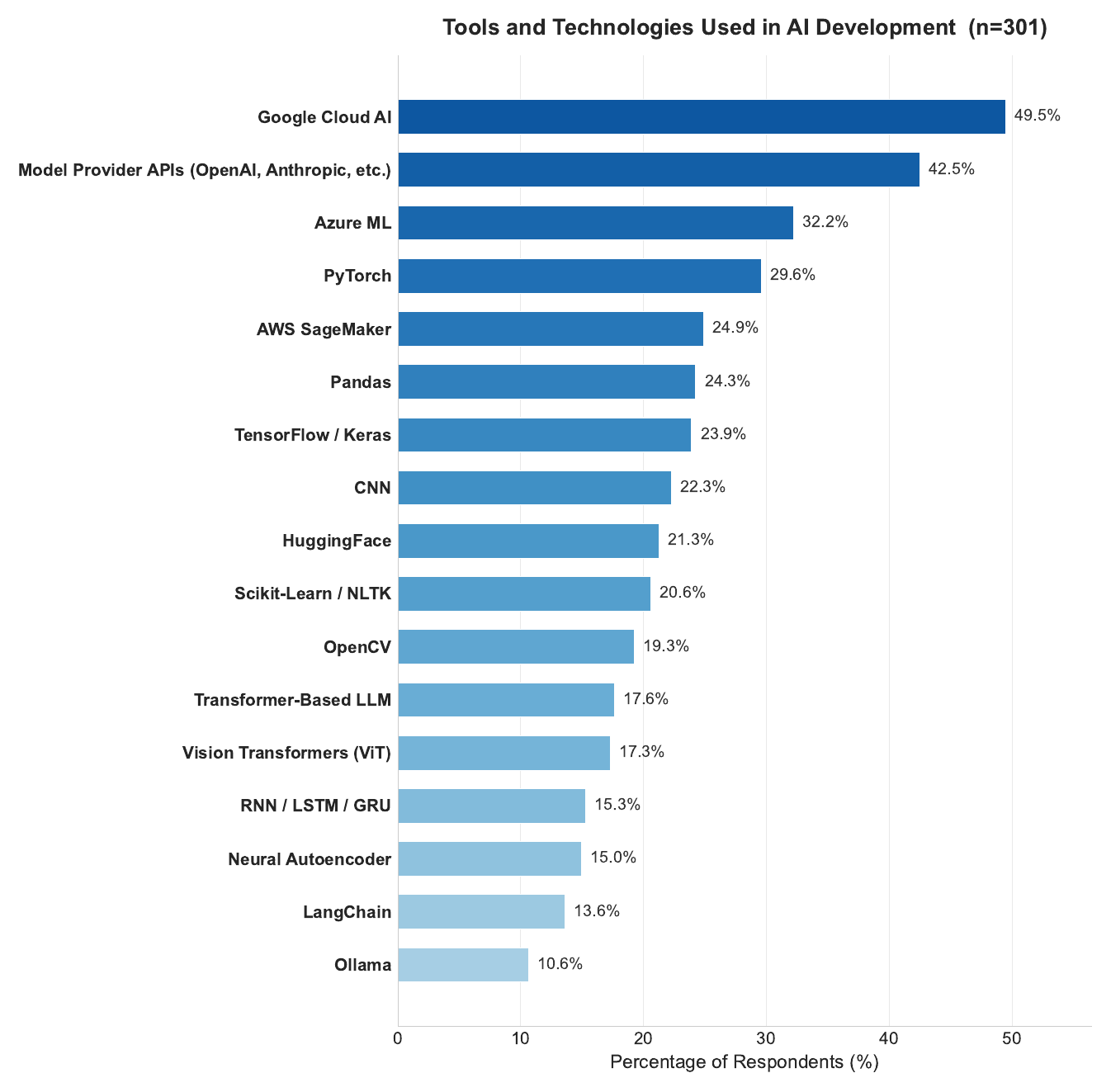} 
    \caption{{Group B - Tools and Technologies Used in Product or Development}}
    \label{fig:b_tools_technologies}
\end{figure}

Group B participants could select multiple AI tools and technologies they use in their products, as shown in Figure \ref{fig:b_tools_technologies}. ``Google Cloud AI'' and ``Model Provider APIs, such as OpenAI, Anthropic, LLama, etc.'' are the most popular AI tools selected by the participants (49.5\% and 42.5\%, respectively). This suggests that practitioners primarily use pre-trained models instead of building or training their own. In addition, ``Azure ML,'' ``Pytorch,'' ``AWS Sagemaker,'' and ``Tensorflow'' are more popular than ``transformer-based LLMs'' and other deep and machine-learning architectures. 

All participants were asked to discuss how often they use AI tools and what they would consider the main impact of AI tools on their work. Tables \ref{tab:freq_ai_usage} and  \ref{tab:ai_effects} show the frequency of AI usage and its impact. Most participants (i.e., 82.24\% of Group A and 96.67\% of Group B) use AI at least 1-3 times per week, while more than 50\% of Group B uses AI tools daily. The difference in usage frequency distributions between groups is statistically significant ($\chi^2(4) = 32.81$, $p < 0.001$, Cram\'{e}r's $V = 0.29$; see Appendix \ref{Appendix:stats} - Table~\ref{tab:new_stat_tests}). 
All participants consider ``increased efficiency'' (71.0\% and 77.5\% of Groups A and B) and ``improved accuracy'' (41.1\% and 55.3\% of Groups A and B) in their work, as the main effect. Interestingly, Group B selected ``cost reduction'' as the third most important impact of AI, with 22.2\% more than Group A ($\chi^2(1) = 14.89$, $p < 0.001$, $V = 0.19$; see Appendix \ref{Appendix:stats} - Table~\ref{tab:new_stat_tests}). In contrast, Group A identified ``innovation and creativity'' as the third most crucial impact. The results show that all participants recognize AI's benefits in enhancing workplace tasks, with experienced users more frequently considering automating cost-intensive tasks. 


\begin{table}[ht]
\centering
\caption{Frequency of AI Tool Usage}
\label{tab:freq_ai_usage}
\begin{tabular}{lcc}
\toprule
\textbf{Frequency} & \textbf{Group A} & \textbf{Group B} \\
\midrule
Daily & 36 (33.64\%) & 167 (54.33\%) \\
4--6 times a week & 30 (28.04\%) & 76 (24.67\%) \\
1--3 times a week & 22 (20.56\%) & 54 (17.67\%) \\
Rarely & 17 (15.89\%) & 7 (2.33\%) \\
Never & 1 (0.93\%) & 2 (0.67\%) \\
Prefer not to say & 1 (0.93\%) & 1 (0.33\%) \\
\bottomrule
\end{tabular}
\end{table}

\begin{table}[!htbp]
\centering
\caption{Effects of AI-Enhanced Tools on Development Process and Workflow}
\label{tab:ai_effects}
\begin{tabular}{p{6cm}p{2.5cm}p{2.5cm}}
\toprule
\textbf{Effect/Impact} & \textbf{Group A} & \textbf{Group B} \\
\midrule
Increased Efficiency             & 76 (71.0\%) & 234 (77.5\%) \\
Improved Accuracy               & 44 (41.1\%) & 167 (55.3\%) \\
Innovation and Creativity       & 43 (40.2\%) & 141 (46.7\%) \\
Enhanced Decision Making        & 42 (39.3\%) & 120 (39.7\%) \\
Improved Research Quality       & 42 (39.3\%) & 140 (46.4\%) \\
Training and Skill Development  & 38 (35.5\%) & 112 (37.1\%) \\
Cost Reduction                  & 33 (30.8\%) & 160 (53.0\%) \\
Streamlined Workflows           & 29 (27.1\%) & 111 (36.8\%) \\
Security and Privacy Concerns   & 17 (15.9\%) & 70 (23.2\%)  \\
No Significant Impact           & 8 (7.5\%)   & 19 (6.3\%)   \\
Other, please explain                  & 2 (1.9\%)   & 5 (1.7\%)    \\
Prefer not to say               & 1 (0.93\%)   & 3 (0.99\%)    \\
\bottomrule
\end{tabular}
\end{table}

\begin{figure}[!htbp]
    \centering
    \includegraphics[width=0.85\textwidth]{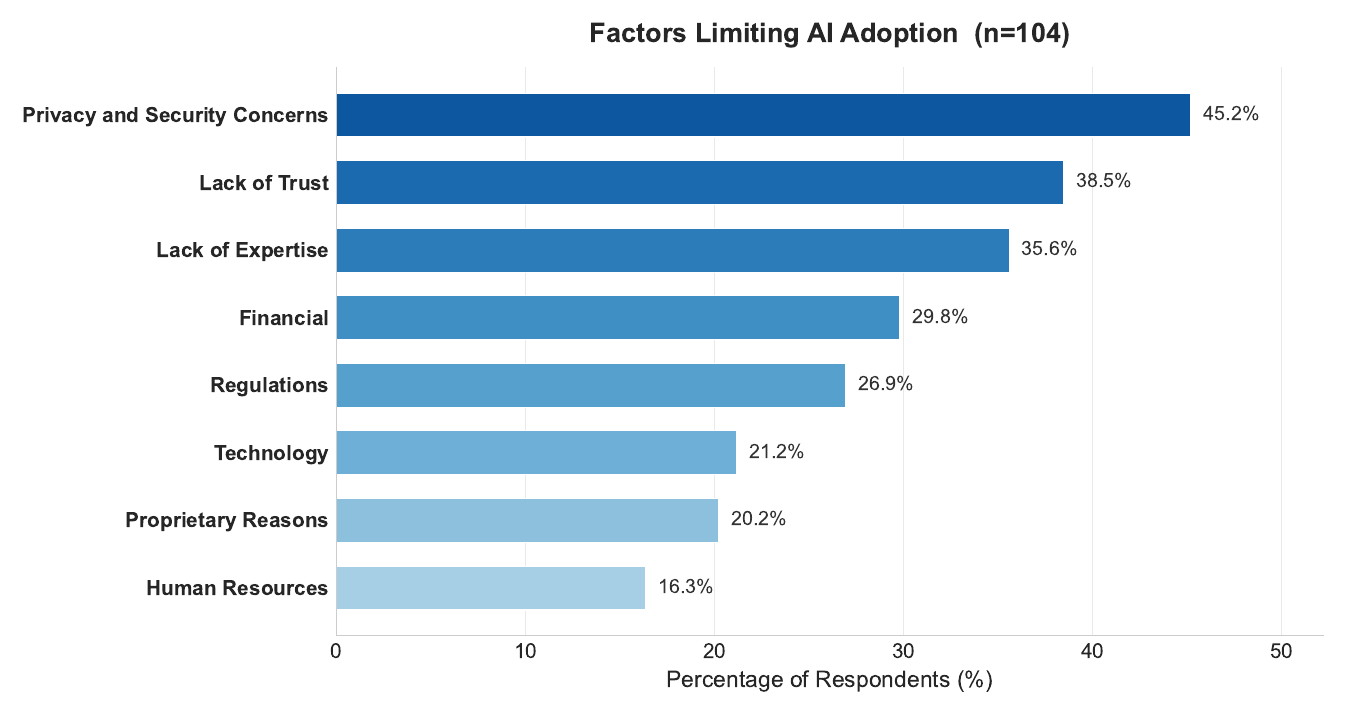} 
    \caption{Factors Limiting AI Adoption in Development Processes}
    \label{fig:reasonsAI}
\end{figure}

Since Group A's companies did not adopt AI in their development processes, we asked them to select factors that limit their companies' involvement with AI. As shown in Figure \ref{fig:reasonsAI}, ``privacy and security concerns'' (45.2\%), ``lack of trust'' (38.5\%), and ``lack of expertise''(35.6\%) are the most significant factors, while ``human resources'' (16.3\%) and ``proprietary reasons'' (20.2\%) appear not as decisive. This highlights the importance of ethics considerations, including privacy, security, and trust, for companies that do not incorporate AI in their products. As a follow-up question, they had to describe why these factors limit their companies' involvement. We received 106 responses, in which we identified seven common themes (see Table \ref{Table:track_a_reasons}). These responses reinforced the previous finding, with $\sim$29\% expressing concerns regarding security and privacy, where many reiterated worries about data sharing and consent (e.g., \textit{``customer data concerns''}) or were uncertain about the security of AI systems (e.g., \textit{``if you have proprietary information and materials, you don't want to feed it into an LLM model that will later regurgitate that information to any user''}). 18.87\% considered ``high costs'' as the main factor, while 12.26\%, 11.32\%, and 10.38\%, mentioned a lack of ``experience,'' ``established AI applications,'' or ``trust with AI systems.'' 

\begin{table}
\centering
\caption{Group A - Reasons for Limiting AI Usage in Development}
\begin{tabular}{p{3.8cm}cp{7cm}}
\toprule
\textbf{Theme} & \textbf{Count} & \textbf{Example} \\
\midrule
Security and privacy concerns & 31 & "Customers have a choice to not consent to ai being used with their sensitive and confidential information" \\
\addlinespace
High costs for reliable AI systems & 20 & "Lack the hardware for building large AI models (GPUs)" \\
\addlinespace
Lack of experience & 13 & "Old school people, not so well informed about possibilities of AI" \\
\addlinespace
Lack of established AI applications & 12 & "in the education industry they prefer proven widely adopted tools ... not bleeding edge technology" \\
\addlinespace
Lack of trust with AI systems & 11 & "Because we deal with real life issues that we can hardly trust AI with" \\
\addlinespace
Not ready for integration & 6 & "Large government bureaucracy can't jump on the bandwagon" \\
\addlinespace
Regulatory limitations & 5 & "The insurance industry has many regulations and compliance concerns" \\
\addlinespace
Other & 18 &  \\
\addlinespace
\bottomrule
\end{tabular}
\label{Table:track_a_reasons}
\end{table}

\begin{tcolorbox}[colback=blue!10,colframe=black!30,boxrule=0.1pt, boxsep=0.1pt]
\textbf{Implications.} Most participants, regardless of their AI development experience, use AI very frequently, especially to increase efficiency in their workflow. However, growing concerns about security and privacy create an opportunity for companies to formalize the integration of AI into their software development activities through policies, access controls, and approved toolchains that ensure sensitive and proprietary data remains protected.
\end{tcolorbox}

Table \ref{tab:rq1_summary} summarizes the key findings for RQ1.

\begin{table}[ht]
\centering
\caption{Summary of Key Findings for RQ1: AI Perceptions and Usage}
\label{tab:rq1_summary}
\small
\begin{tabularx}{\textwidth}{X p{2.5cm} p{1.5cm}}
\toprule
\textbf{Finding} & \textbf{Result} & \textbf{Ref.} \\
\midrule
\textbf{Defining AI} & & \\
- Both groups primarily define AI in terms of process automation, performance improvement, content creation, and content synthesis & 70--75\% (both) & Fig.~\ref{fig:definitions} \\
- Gr.~B uses more technical terminology (e.g., ``content synthesis''), while Gr.~A refers to ``content creation'' roughly twice as often & 21\% vs 10\% & Fig.~\ref{fig:definitions} \\
- AI definitions vary by role: AM roles favor ``process automation'' (41\%); ISec roles favor ``content creation'' (33\%); QA roles emphasize ``decision making'' (28\%) & Role-dependent & Sec.~4.2.1 \\
\midrule
\textbf{AI Usage and Impact} & & \\
- Daily AI tool use is significantly more common among experienced practitioners (Gr.~B) than those without AI development experience (Gr.~A) & 54\% vs 34\%$^{***}$ & Table~\ref{tab:freq_ai_usage} \\
- Increased efficiency is the top perceived impact of AI tools across both groups & 71--78\% & Table~\ref{tab:ai_effects} \\
- Experienced practitioners (Gr.~B) value cost reduction significantly more than Gr.~A & 53\% vs 31\%$^{***}$ & Table~\ref{tab:ai_effects} \\
- Chatbots, personal assistants, and recommender systems are the most commonly identified AI applications across both groups & 74\% / 64\% & Table~\ref{tab:ai_applications} \\
- Google Cloud AI and Model Provider APIs (e.g., OpenAI, Anthropic) are the most-used tools among Gr.~B & 50\%; 43\% & Fig.~\ref{fig:b_tools_technologies} \\
\midrule
\textbf{Barriers to AI Adoption (Gr.~A)} & & \\
- Privacy and security concerns are the most significant factor limiting AI adoption among companies not yet using AI & 45\% & Fig.~\ref{fig:reasonsAI} \\
- Lack of trust in AI systems and lack of expertise are the second and third most cited barriers & 38\%; 36\% & Fig.~\ref{fig:reasonsAI} \\
- Qualitative responses reinforce security/privacy concerns (29\%) and cite high costs (19\%) as key themes & 106 responses & Table~\ref{Table:track_a_reasons} \\
\bottomrule
\end{tabularx}
\begin{tablenotes}
\small
\item $^{***}p < 0.001$. Full test statistics reported inline in Sec.~4.2.
\end{tablenotes}
\end{table}


\begin{tcolorbox}[colback=gray!10,colframe=black!30,boxrule=0.1pt, boxsep=0.1pt]
\textbf{RQ1 Summary - AI Perception \& Usage:} All participants, regardless of their experience, mainly define AI as process automation or performance enhancement, report daily use of AI tools, and perceive a positive impact on efficiency and accuracy. Those with AI development experience tend to engage with more technical aspects of AI in their work, whereas those without such experience typically utilize AI tools, like chatbots or recommender systems, often for innovation and creativity. Concerns regarding security and privacy remain the primary reasons some companies have not yet adopted AI tools in their development.
\end{tcolorbox}

\subsection{AI Ethics Familiarity, Practices and Perceptions}

We assessed participants’ familiarity with AI ethics principles~\cite{worsdorfer2023aibill} and governance initiatives and how their practices and perceptions regarding these principles and initiatives vary across demographics (i.e., \textbf{RQ2}). 

\subsubsection{Familiarity with AI Ethics Principles}

\begin{figure}[!htbp]
         \centering
         \includegraphics[width=0.88\textwidth]{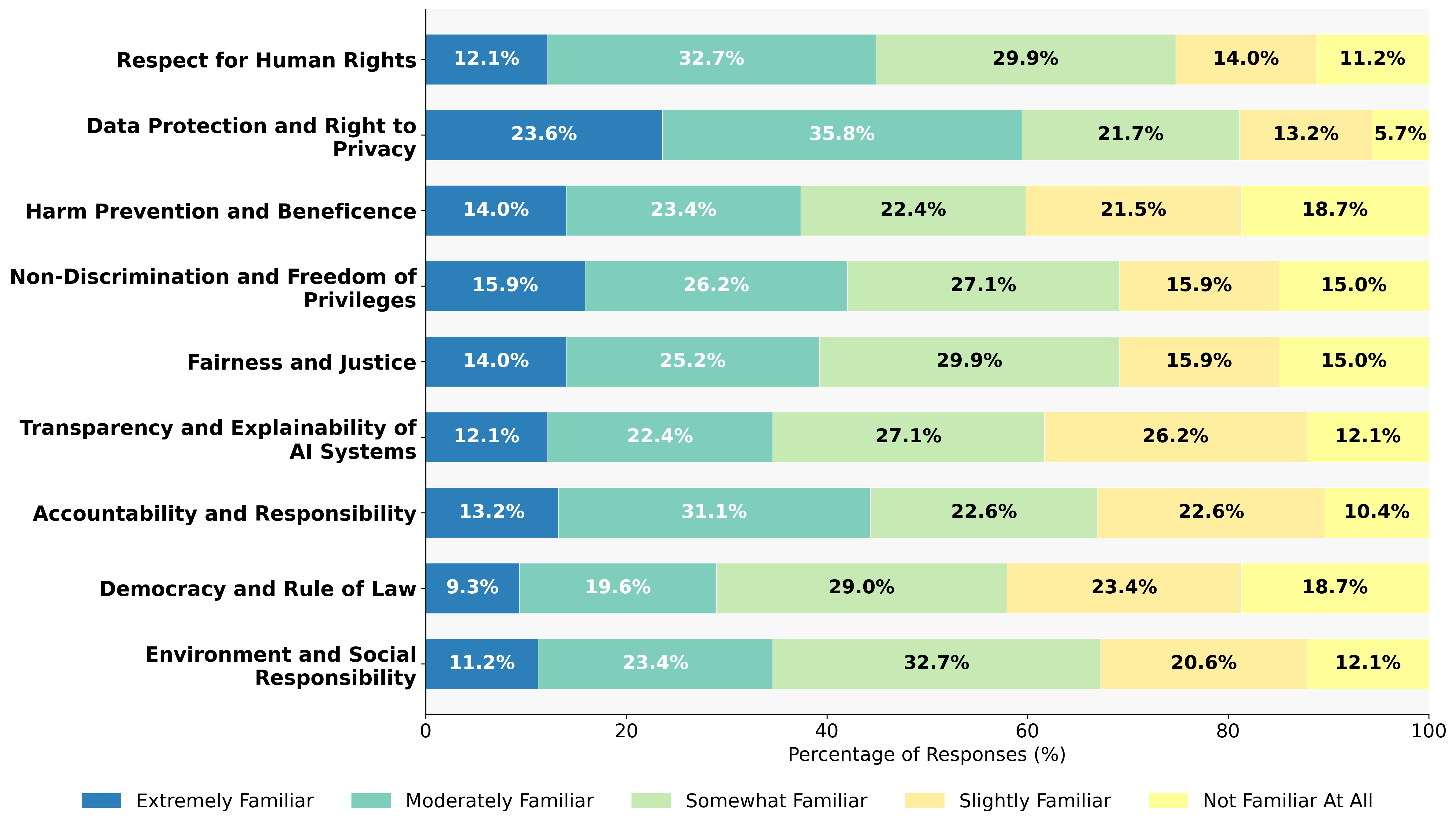}
         \caption{Group A - Degree of Familiarity with AI Ethics Principles}
         \label{fig:A:Principle_Familiarity}
\end{figure}

\begin{figure}[!htbp]
         \centering
         \includegraphics[width=0.88\textwidth]{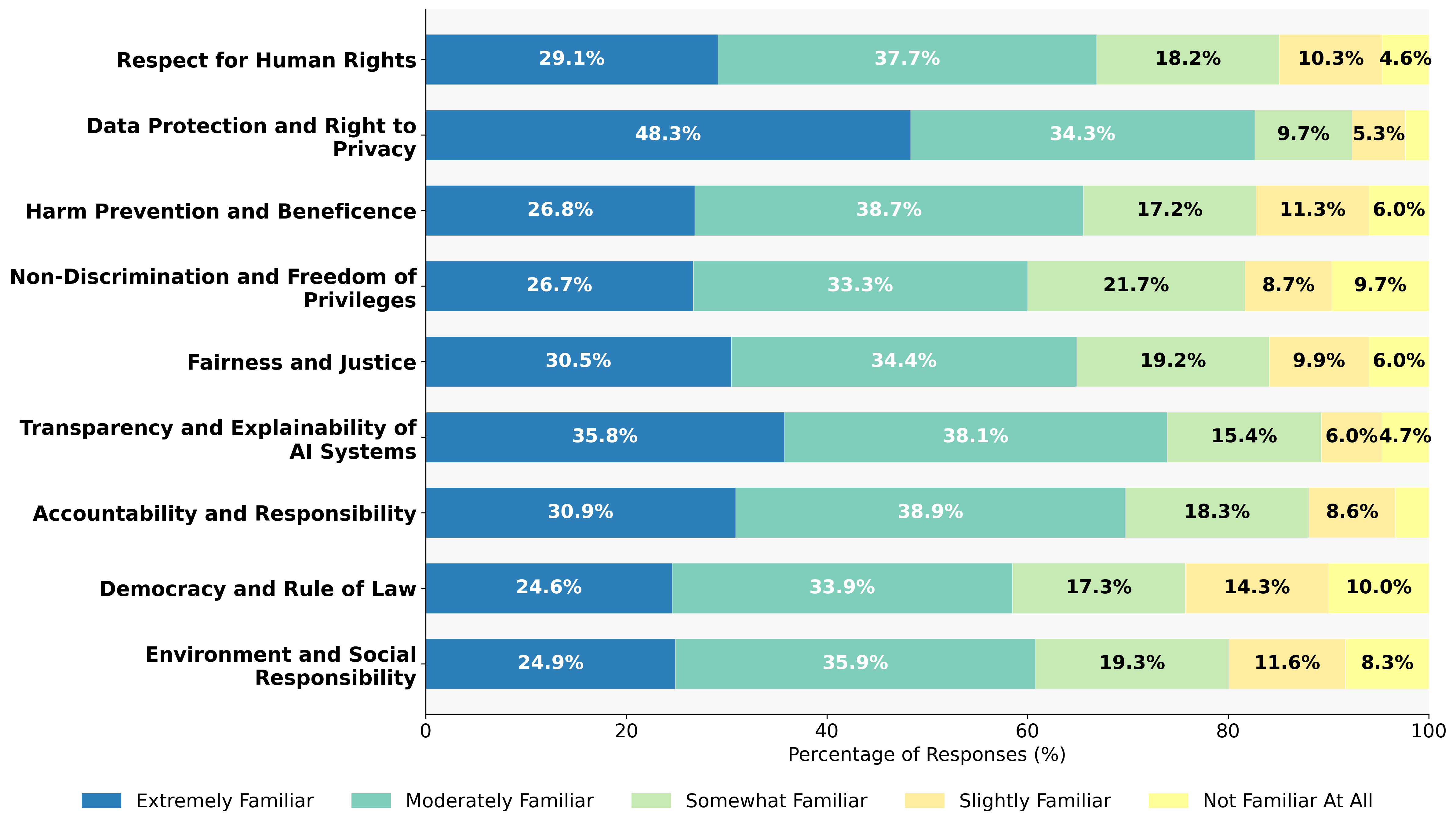}
         \caption{Group B - - Degree of Familiarity with AI Ethics Principles}
         \label{fig:B:Principle_Familiarity}
\end{figure}

We asked participants to rate their familiarity with various AI ethics principles (see Figures \ref{fig:A:Principle_Familiarity} and \ref{fig:B:Principle_Familiarity}). We combine the responses of \emph{somewhat, moderately,} and \emph{extremely familiar}. The results show that both groups are most familiar with ``data protection and the right to privacy'' (81.2\% and 92.3\%) and least familiar with ``democracy and rule of law'' (58.0\% and 75.7\%). Overall, Group B reports significantly higher mean principle familiarity than Group A ($U = 9684.5$, $p < 0.001$, $r = 0.31$; see Appendix \ref{Appendix:stats} - Table~\ref{tab:new_stat_tests}), likely because the concept of privacy is more tangible for those actively involved in development. Those with AI development experience are more familiar with ``transparency and explainability of AI systems'' (89.3\%) and ``accountability and responsibility'' (88.0\%). Those in Group A are more familiar with ``respect for human rights'' (74.8\%), while for all other principles, their familiarity is below 70\%. 

\begin{figure}[!htbp]
    \centering
    \includegraphics[width=1.0\textwidth]{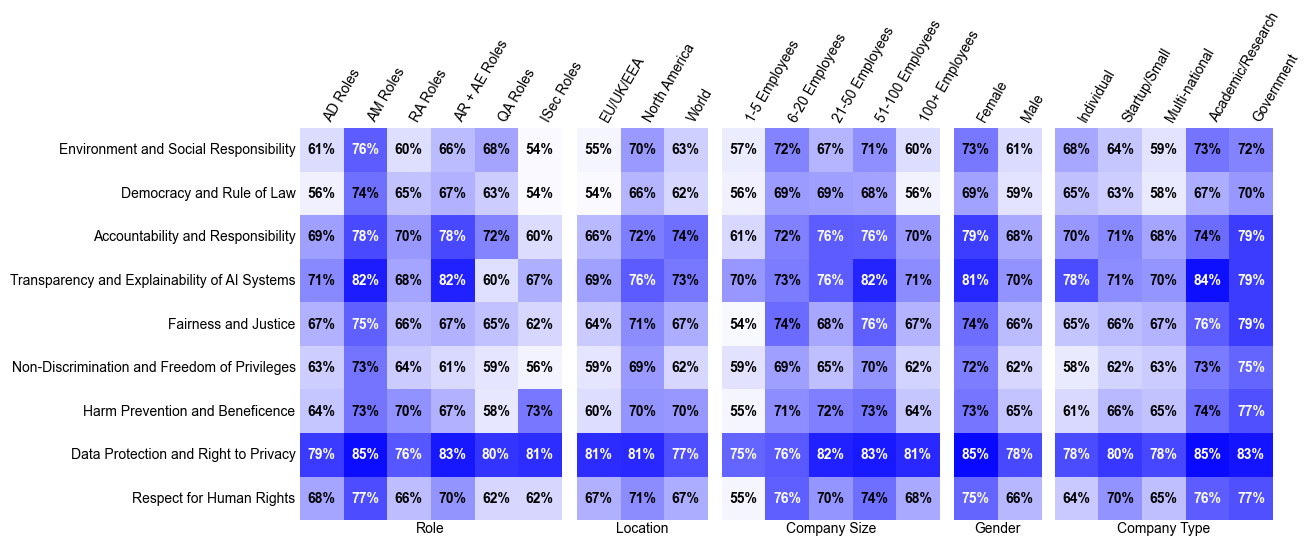}
    \caption{Group B - Familiarity With AI Ethics Principles by Demographics}
\label{fig:D:Principles_Familiarity_B_Demographics}
\end{figure}

For Group B, we further investigated how various demographics differ across the principles. As shown in Figure \ref{fig:D:Principles_Familiarity_B_Demographics}, those in AM roles are the most familiar with all AI ethics principles, followed closely by AI researchers and ethicists. Interestingly, those in ISec roles are least familiar with most principles except ``data protection and the right to privacy'' and ``harm prevention and beneficiary.'' RA, AD, and QA team members are comparable to each other; however, AD roles are slightly more familiar with principles that directly impact datasets and model training (i.e., ``transparency and explainability of AI systems'' and ``fairness and justice'') than AR and QA roles. 

An analysis of the company types indicates that participants from government ($M = 4.07$) or academic/research ($M = 4.04$) organizations are the most familiar with all principles, while those from multinational corporations ($M = 3.63$) are the least familiar ($p = 0.046$; see H1c in Table~\ref{table:demographics-familiarity}). This result aligns with company size, where those working in companies with 1-5 or 100+ are least familiar with all principles. This may indicate that AI practitioners in midsize companies are more likely to be involved in the human-focused design aspects of AI systems than those in larger corporations, where such tasks might be delegated to policy teams, or smaller companies that may lack the resources to implement these features. 

On average, EU+EEA+UK participants ($M = 3.55$) are the least familiar with all principles except for ``data protection and the right to privacy,'' where their familiarity matches that of their North American counterparts ($M = 3.87$); this location effect is statistically significant ($p = 0.010$; see H1b in Table~\ref{table:demographics-familiarity}). This is likely explained by the GDPR, which has enforced strong data protection norms across the EU since 2018 and is the most probable driver of familiarity with this specific principle. The EU AI Act may serve as a complementary factor. Lastly, female participants are more familiar with all principles than male participants ($p = 0.0003$; see H1e in Table~\ref{table:demographics-familiarity}).  

As shown in Table \ref{table:demographics-familiarity}, we found no evidence of a correlation between demographics and average familiarity among Group A. Without applying the Bonferroni correction, we observe correlations between location, company type, and gender, and Group B's average familiarity ($p-value=0.0102, 0.0463$ and $0.003$), while with the Bonferroni correction, only the correlation with gender remains statistically significant. 

Our analysis of ethics principles familiarity complements prior work, emphasizing the importance of diversity in demographics and expertise within AI development teams~\cite{pant2024ethics,vakkuri2022how,olson2025speaks}. Others have also highlighted the need for effective communication among development teams~\cite{stahl2022organisational}, and the critical role of management in engaging their teams and stakeholders from various demographics in ethical decision-making~\cite{kawakami2024studyingup}. 

We conducted a LASSO analysis to identify the predictive factors within demographics, practices, and perceptions that are most strongly associated with \textit{familiarity with AI ethics principles}. For Group A, the LASSO model selected no features at the optimal regularization parameter, resulting in a null model. The model yielded a negative test R² (-0.06), indicating performance worse than a baseline model that predicts the mean. For Group B, the model selected 9 of 141 features (test $R^2 = 0.56$). LASSO model indicates that increases in familiarity with AI ethics principles are most strongly associated with greater familiarity with the \textit{EU AI Act} ($\beta = 0.154^*$). Additional predictors include more positive \textit{perceptions of regulatory impact} ($\beta = 0.152^{**}$), consideration of \textit{environmental and social responsibility} in practice ($\beta = 0.140^*$), and the perceived effect of \textit{cost reduction} for AI adoption ($\beta = 0.124^*$). 
The full coefficient table is provided in Appendix~\ref{Appendix:stats} - Table~\ref{tab:lasso_trackb_principles}.

\begin{tcolorbox}[colback=blue!10,colframe=black!30,boxrule=0.1pt, boxsep=0.1pt]
\textbf{Implications.} The consistently higher familiarity among AM roles suggests that managers' exposure to organizational governance and stakeholder communication naturally builds ethics awareness. This finding supports investing in management-led ethics initiatives rather than relying solely on developer training. The lower familiarity among ISec and QA roles is concerning because these roles are critical gatekeepers for trust, safety, and security in deployed AI systems~\cite{ala-luopa2024trusting, prybylo2024evaluating}. Targeted training for these roles that focus on principles beyond data protection could address this gap. The gender effect ($p = 0.0003$) reinforces Olson et al.'s~\cite{olson2025speaks} findings that diverse teams bring broader ethical perspectives, providing empirical support for diversity initiatives in AI development.
\end{tcolorbox}

\begin{table}[]
\centering
\caption{P-Values for Hypotheses H1a to H1e}
\begin{tabular}{lccccc}
\toprule
\textbf{} & \textbf{Role} & \textbf{Location} & \textbf{Company Type} & \textbf{Company Size} & \textbf{Gender} \\
\midrule
\textbf{Group A} & {0.3903} & {0.0523} & {0.6912} & {0.6070} & {0.1433}\\
\textbf{Group B} & {0.1040} & \textbf{*0.0102} & \textbf{*0.0463} & {0.1412} & \textbf{**0.0003} \\
\bottomrule
\end{tabular}
\label{table:demographics-familiarity}
\begin{tablenotes}
\centering
\small
\item {Note:} * $p < 0.05$; ** $p < 0.0019$ (Bonferroni correction)
\end{tablenotes}
\end{table}

\subsubsection{Practice and Experience with AI Ethics Principles}

\begin{figure}[!htbp]
        \centering
        \includegraphics[width=0.9\textwidth]{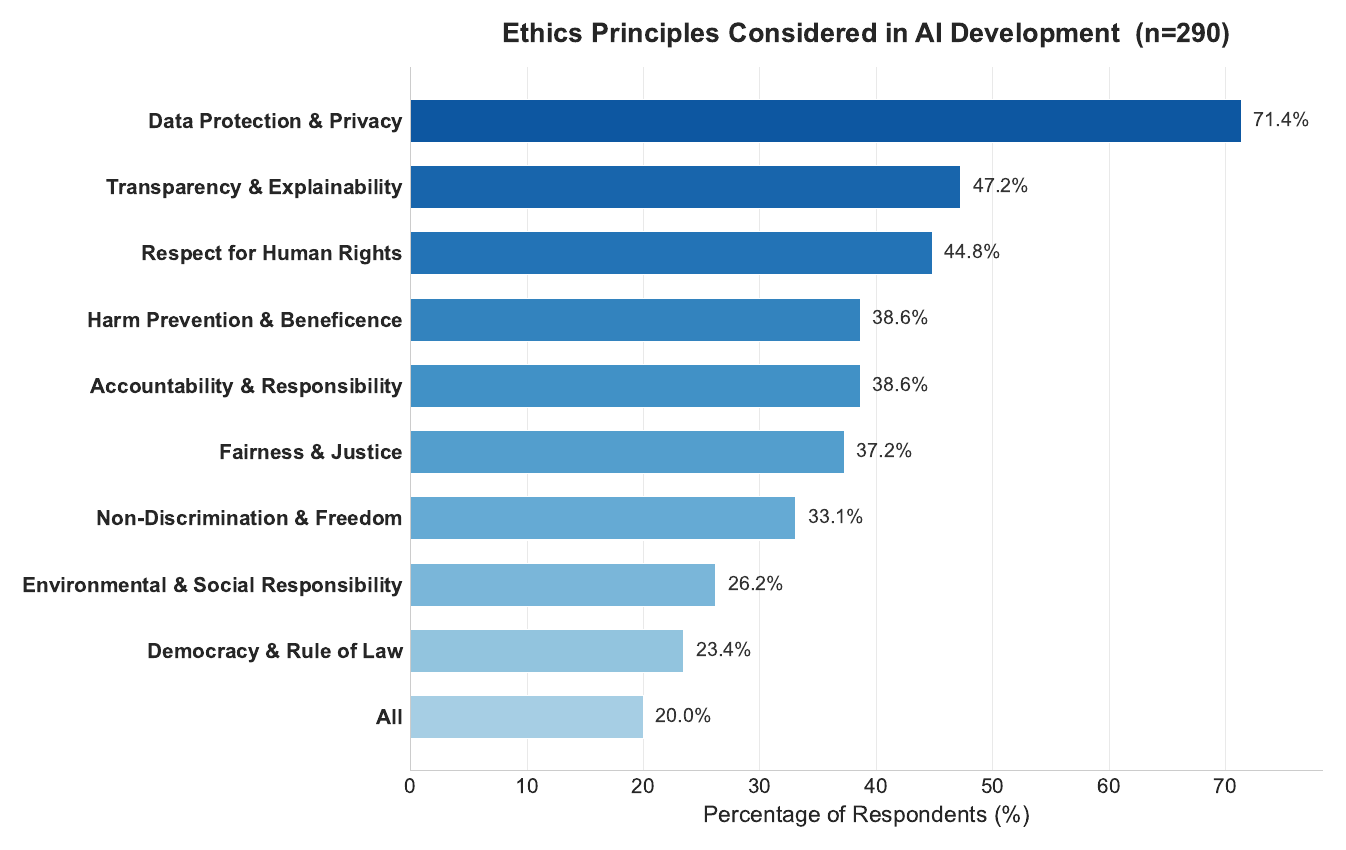} 
        \caption{{Group B - Ethics Principles in AI Development Activities}}
        \label{fig:considered_principles}
    \hfill
\end{figure}

We further evaluated whether Group B considers these principles in their AI development. Figure \ref{fig:considered_principles} shows the results. As expected, ``data protection and the right to privacy'' is considered the most important. This is probably because the concept of digital privacy has existed for much longer than other computer ethics aspects. ``Transparency and explainability of AI systems'' is regarded as the second most important principle, while ``democracy and rule of law'' is considered the least important. The results match the degree of familiarity, as shown in Figure \ref{fig:B:Principle_Familiarity}. 
We tested whether there is a correlation between the average familiarity and the use of principles in AI development (H2a). With  $p-value=0.0000$, we found a correlation. 


In a more role-specific analysis, we observed that those in AM, AD, AR+AE, and RA roles consider a variety of principles in their AI development activities. For example, those in AM roles consider the principles of ``respect for human rights,'' ``non-discrimination and freedom of privileges,'' ``harm prevention and beneficence,'' and ``fairness and justice'' whereas those in AR+AE ensure that the ``accountability and responsibility'' is applied in the workflow. In contrast, those in QA or ISec roles do not consider these principles as much, except for ``data protection and the right to privacy.'' This highlights areas for improvement, as QA roles are crucial for maintaining trust in AI systems~\cite{ala-luopa2024trusting} and collaborating with ISec teams is essential to ensure the security and privacy of AI models~\cite{prybylo2024evaluating}. 

Participants working in \emph{academic/research} institutions prioritize more technical principles, like ``data protection and the right to privacy,'' ``fairness and justice,'' and ``transparency and explainability of AI systems,'' when developing or researching AI models or datasets. In contrast, those employed in \emph{government} agencies are likelier to emphasize abstract principles, including ``respect for human rights,'' ``non-discrimination and freedom of privileges,'' and ``harm prevention and beneficence.'' 

Those in EU+EEA+UK consider ``data protection and the right to privacy'' and ``democracy and rule of law'' more than those working in North America and other locations. This is expected since privacy is considered a fundamental (human) right and is regulated by the GDPR \cite{GDPR}. Female participants are more likely than their male counterparts to consider AI ethics principles in AI development activities.

We asked Group B how often they consider AI ethics principles in their work. Most participants consider them ``Always'' (23.92\%) or ``Often'' (44.52\%), while only 9.97\% and 2.33\%, respectively, consider them ``Rarely'' or ``Never.'' Our results show that those in QA (92.3\%), ISec (91.7\%), AM (91.5\%), and AR+AE (90\%) roles are more likely to consider a subset of AI ethics principles in their work at least occasionally, whereas those in AD (85.4\%) and RA (75\%) roles report slightly lower engagement. However, this role-based difference in consideration frequency is not statistically significant at $\alpha = 0.05$ ($H(6) = 12.34$, $p = 0.055$, $\epsilon^2 = 0.04$; see Appendix \ref{Appendix:stats} - Table~\ref{tab:new_stat_tests}). Those with a BSc.\ or higher degree are $\sim$10\% more likely to consider AI ethics principles in their work than those without, though this difference is not statistically significant ($U = 3876.0$, $p = 0.815$, $r = 0.01$; see Appendix \ref{Appendix:stats} - Table~\ref{tab:new_stat_tests}). Female participants include AI ethics principles in their work $\sim$13\% more than their male counterparts ($U = 6264.0$, $p < 0.001$, $r = 0.25$) and consider a greater number of distinct ethics principles ($U = 7013.0$, $p = 0.011$, $r = 0.15$).

We conducted a LASSO analysis to examine the predictive factors most strongly associated with how frequently AI developers consider ethics principles in their work (ranging from never to always). The results show that more frequent ethical consideration is most strongly associated with higher overall \textit{familiarity with AI ethics principles} (+0.281), more positive \textit{perceptions of regulatory impact} (+0.226), and familiarity with the \textit{US Executive Order on Safe AI} (+0.210) (see Appendix \ref{Appendix:stats} - Table~\ref{table:lasso_ethics_frequency}). Tables \ref{tab:rq2_summary} and \ref{tab:rq2_summary-2} show the summary of the result.

\begin{tcolorbox}[colback=blue!10,colframe=black!30,boxrule=0.1pt, boxsep=0.1pt]
\textbf{Implications.} The finding that 68\% of practitioners consider ethics ``Always'' or ``Often'' is encouraging, but the non-significant role differences ($p = 0.055$) suggest that within-role variability is high. Some individuals in every role engage deeply while others do not, regardless of job title. This implies that interventions should target individual awareness (e.g., through AI policy education) rather than assuming role-based training will suffice. The non-significant education effect ($p = 0.815$) further suggests that formal education alone does not drive ethical practice; rather, as our LASSO results show, familiarity with specific governance initiatives is the stronger predictor.
\end{tcolorbox}

\begin{table}[!htbp]
\centering
\caption{Summary of Key Findings for RQ2: Familiarity with Ethics Principles}
\label{tab:rq2_summary}
\small
\begin{tabularx}{\textwidth}{X p{2.5cm} p{1.5cm}}
\toprule
\textbf{Finding} & \textbf{Result} & \textbf{Ref.} \\
\midrule
\textbf{Familiarity with AI Ethics Principles} & & \\
- Data protection and the right to privacy is the most familiar principle across both groups, likely because digital privacy has been a concern longer than other AI ethics aspects & 81\% / 92\% & Fig.~\ref{fig:B:Principle_Familiarity} \\
- Democracy and rule of law is the least familiar principle, potentially because it is more abstract and less directly applicable in daily development work & 58\% / 76\% & Fig.~\ref{fig:B:Principle_Familiarity} \\
- Gr.~B (with AI development experience) reports significantly higher overall familiarity than Gr.~A, with a medium effect size & $p<0.001$, $r=0.31$ & Sec.~4.3.1 \\
\midrule
\textbf{Familiarity by Professional Role} & & \\
- AM roles report the highest familiarity across all 9 principles, followed closely by AI researchers and ethicists & Highest on 9/9 & Fig.~\ref{fig:D:Principles_Familiarity_B_Demographics} \\
- ISec and QA roles are the least familiar with most principles, except data protection and harm prevention & Lowest on 7/9 & Fig.~\ref{fig:D:Principles_Familiarity_B_Demographics} \\
- AD roles are slightly more familiar than RA/QA with principles directly impacting model training (transparency, fairness) & & Fig.~\ref{fig:D:Principles_Familiarity_B_Demographics} \\
\midrule
\textbf{Demographic Correlations with Familiarity} & & \\
- Gender is significantly correlated with principle familiarity (Bonferroni-significant); female participants report higher familiarity across all principles & $p=0.0003^{**}$ & Table~\ref{table:demographics-familiarity} \\
- Location is significantly correlated; EU+EEA+UK participants are least familiar overall ($M{=}3.55$) except for data protection, likely driven by GDPR awareness & $p=0.010^{*}$ & Table~\ref{table:demographics-familiarity} \\
- Government and academic/research organizations report the highest familiarity; multinationals and startups report the lowest & $p=0.046^{*}$ & Table~\ref{table:demographics-familiarity} \\
\midrule
\bottomrule
\end{tabularx}
\begin{tablenotes}
\small
\item $^{*}p < 0.05$; $^{**}p < 0.0019$ (Bonferroni).
\end{tablenotes}
\end{table}

\begin{table}[htbp]
\centering
\caption{Summary of Key Findings for RQ2: Ethics Principles in Practice}
\label{tab:rq2_summary-2}
\small
\begin{tabularx}{\textwidth}{X p{2.5cm} p{1.5cm}}
\toprule
\textbf{Finding} & \textbf{Result} & \textbf{Ref.} \\
\midrule
\textbf{Ethics Consideration Frequency} & & \\
- Most Gr.~B practitioners consider AI ethics principles at least occasionally, with the majority reporting ``Always'' or ``Often'' & 68\% Always/Often & Sec.~4.3.2 \\
- QA (92\%), ISec (92\%), and AM (92\%) roles are most likely to consider ethics at least occasionally, while RA roles report lower engagement (75\%) & Role-dependent & Sec.~4.3.2 \\
- Role-based differences in consideration frequency are marginally non-significant, suggesting within-role variability exceeds between-role differences & $p=0.055$ & Sec.~4.3.2 \\
\midrule
\textbf{Gender Differences in Ethics Practice} & & \\
- Female practitioners consider ethics in their work significantly more frequently than male counterparts, with a small-to-medium effect & $p<0.001$, $r=0.25$ & Sec.~4.3.2 \\
- Female practitioners consider a greater number of distinct ethics principles in development activities & 5.5 vs 4.6, $p=0.011$ & Sec.~4.3.2 \\
\midrule
\textbf{Education Level} & & \\
- Having a BSc.\ or higher degree does not significantly predict how often practitioners consider AI ethics, despite a $\sim$10\% descriptive difference & $p=0.815$ & Sec.~4.3.2 \\
\midrule
\textbf{Location and Organizational Context} & & \\
- Practitioners in the EU+EEA+UK prioritize data protection and democracy \& rule of law, reflecting the influence of GDPR and the EU AI Act & GDPR-driven & Sec.~4.3.2 \\
- North American practitioners are more comfortable with technical solutions and tend to look to external support for implementing AI ethics & & Sec.~4.3.2 \\
- Those in academic/research institutions and government agencies report more robust ethical frameworks compared to startups or small companies & & Sec.~4.3.2 \\
\bottomrule
\end{tabularx}
\begin{tablenotes}
\small
\item $^{*}p < 0.05$; $^{**}p < 0.0019$ (Bonferroni).
\end{tablenotes}
\end{table}

\begin{tcolorbox}[colback=gray!10,colframe=black!30,boxrule=0.1pt, boxsep=0.1pt] 
\textbf{RQ2 Summary - AI Ethics Principles.} Familiarity with AI ethics principles proved to be more dependent on individuals than job titles: it varied more within roles than between them (ICC $\approx$ 3\%), meaning deeply engaged and largely unaware practitioners coexist in every role. Whether that knowledge is put into practice depends far more on role (ICC $\approx$ 16\%), indicating that ethical practice is governed by team and organizational norms rather than personal knowledge alone. This contrast highlights concerning trends: AM roles were the most familiar and the most likely to embed principles into their work, while QA and ISec practitioners---the roles responsible for software safety and security---were the least familiar, except on ``data protection and the right to privacy.'' Familiarity and its application further varied by location, company type, and gender, with female participants reporting both higher familiarity and more consistent integration across all principles. The strongest predictor of principle familiarity was neither experience nor formal education but familiarity with AI governance initiatives, especially the EU AI Act. This suggests that exposure to governance initiatives is valuable for building ethical literacy.
\end{tcolorbox}%


\subsection{AI Governance Initiatives Familiarity, and Perceptions}

We assessed participants’ familiarity with AI  governance initiatives, and how their perceptions regarding these initiatives vary across demographics (i.e., \textbf{RQ3}). 

\subsubsection{Familiarity with AI Governance Initiatives} 
\label{sec:familiarity-governance-ini}

\begin{figure}[!htbp]
         \centering
         \includegraphics[width=0.88\textwidth]{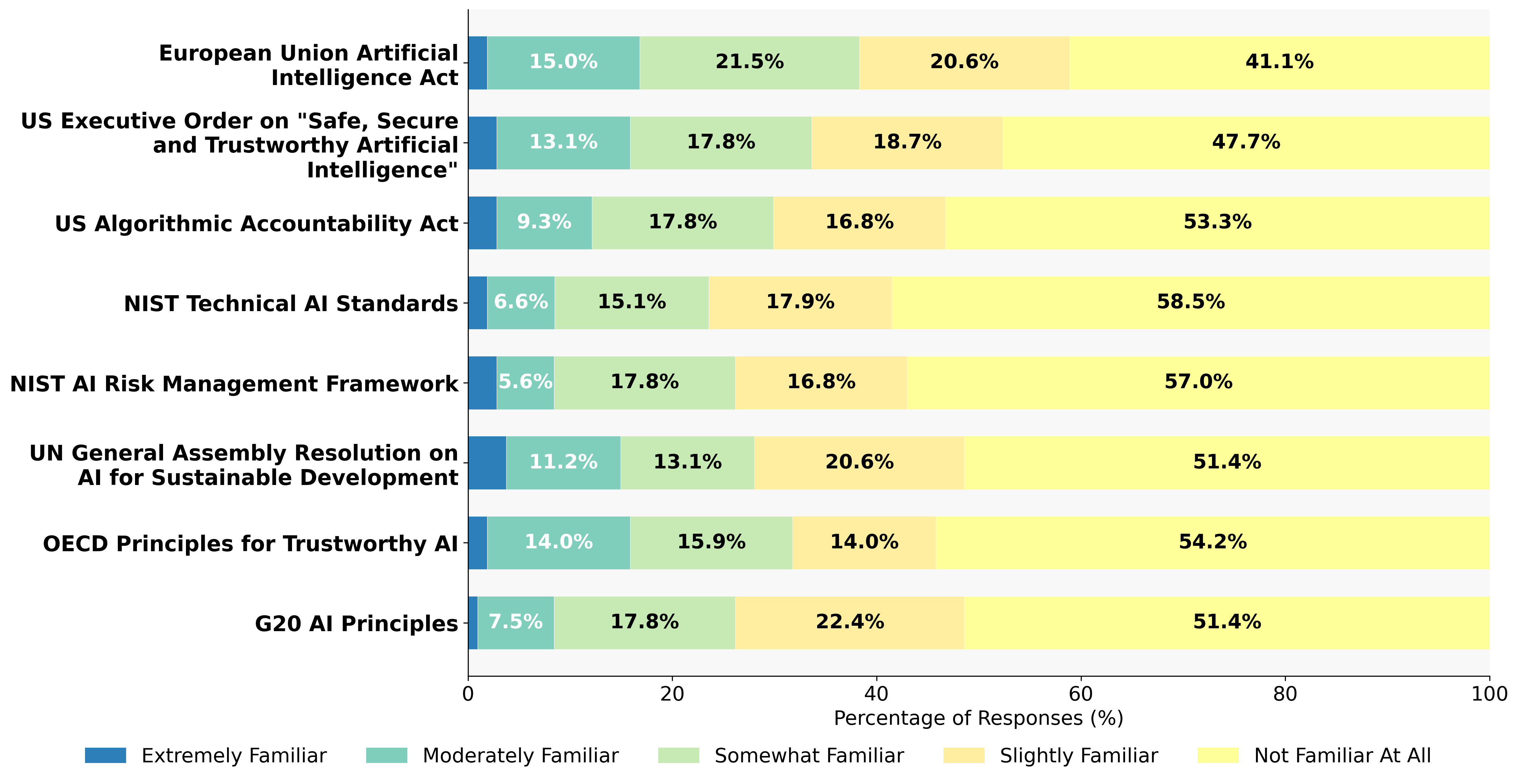}
         \caption{Group A - Degree of Familiarity with Governance Initiatives}
         \label{fig:Regulation_Familiarity_A}
\end{figure}

\begin{figure}[!htbp]
         \centering
         \includegraphics[width=0.88\textwidth]{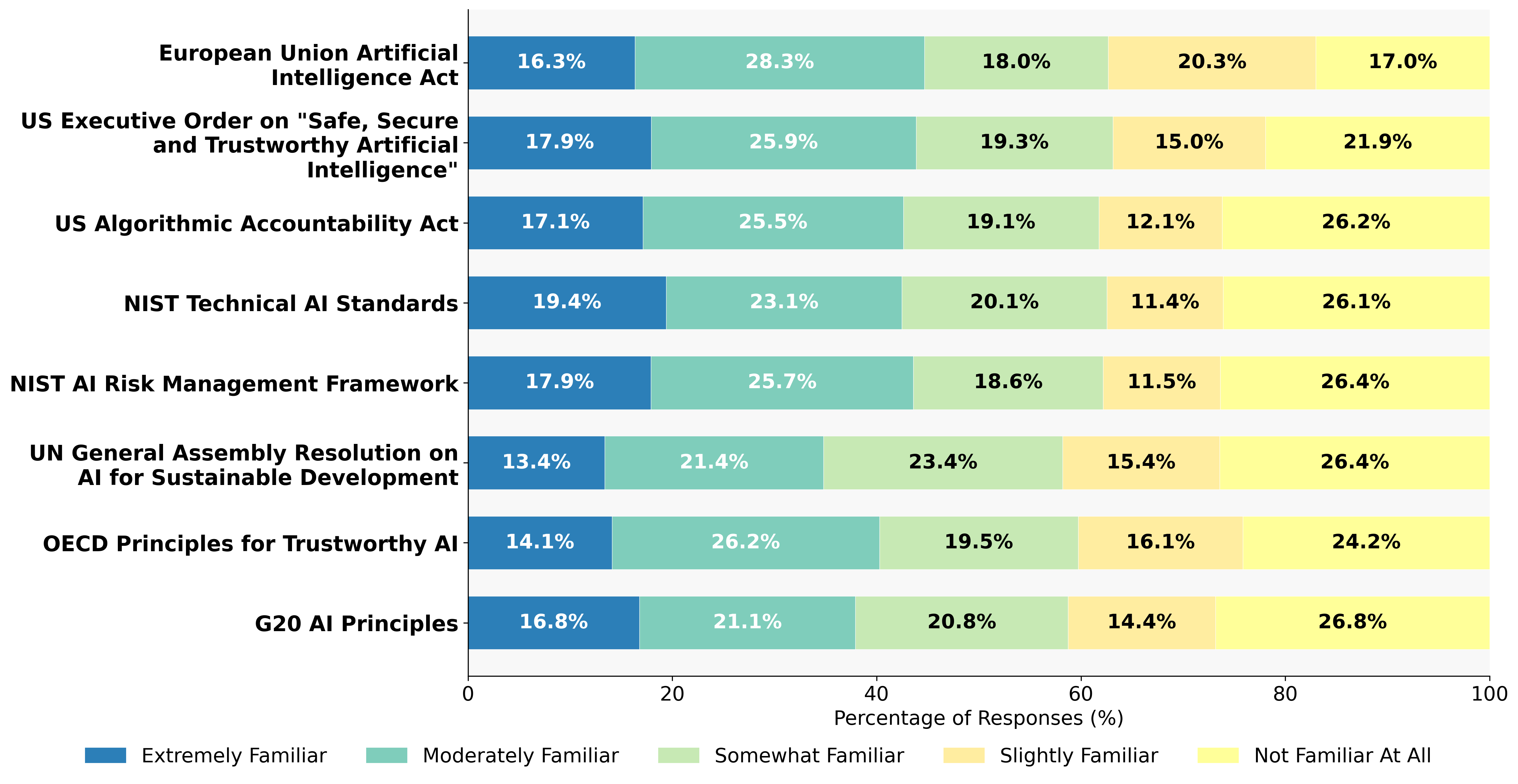}
         \caption{Group B - Degree of Familiarity}
         \label{fig:B:Regulation_Familiarity}
\end{figure}

We asked all participants to rate their familiarity with common AI ethics initiatives (see Figures \ref{fig:Regulation_Familiarity_A} and \ref{fig:B:Regulation_Familiarity}). Between $\sim$58\% and $\sim$62\% of Group B are at least ``somewhat familiar'' with all initiatives. Group A scored $>25\%$  lower than their Group B counterparts. The initiatives with the largest disparities between the two groups are the ``NIST Technical AI Standards'' and the ``NIST AI Risk Management Framework'' (with a difference of 39.7\% and 35.9\%, respectively). Our results show that participants are generally less familiar with the initiatives than with the principles.

\begin{table}[t]
\centering
\caption{P-Values for Hypotheses H3a to H3e}
\begin{tabular}{lccccc}
\toprule
\textbf{} & \textbf{Role} & \textbf{Location} & \textbf{Company Type} & \textbf{Company Size} & \textbf{Gender} \\
\midrule
\textbf{Group A} & {0.1818} & {0.1770} & {0.2924} & \textbf{*0.0179} & {0.0787}\\
\textbf{Group B} & \textbf{**0.0008} & \textbf{**0.0003} & {0.3207}  & \textbf{**0.0000} & \textbf{**0.0001} \\
\bottomrule
\end{tabular}
\label{table:demographics-familiarity-ai-initiatives}
\begin{tablenotes}
\centering
\small
\item {Note:} * $p < 0.05$; ** $p < 0.0019$ (Bonferroni correction)
\end{tablenotes}
\end{table}

\begin{figure}[!htbp]
    \centering
    \includegraphics[width=1.0\textwidth]{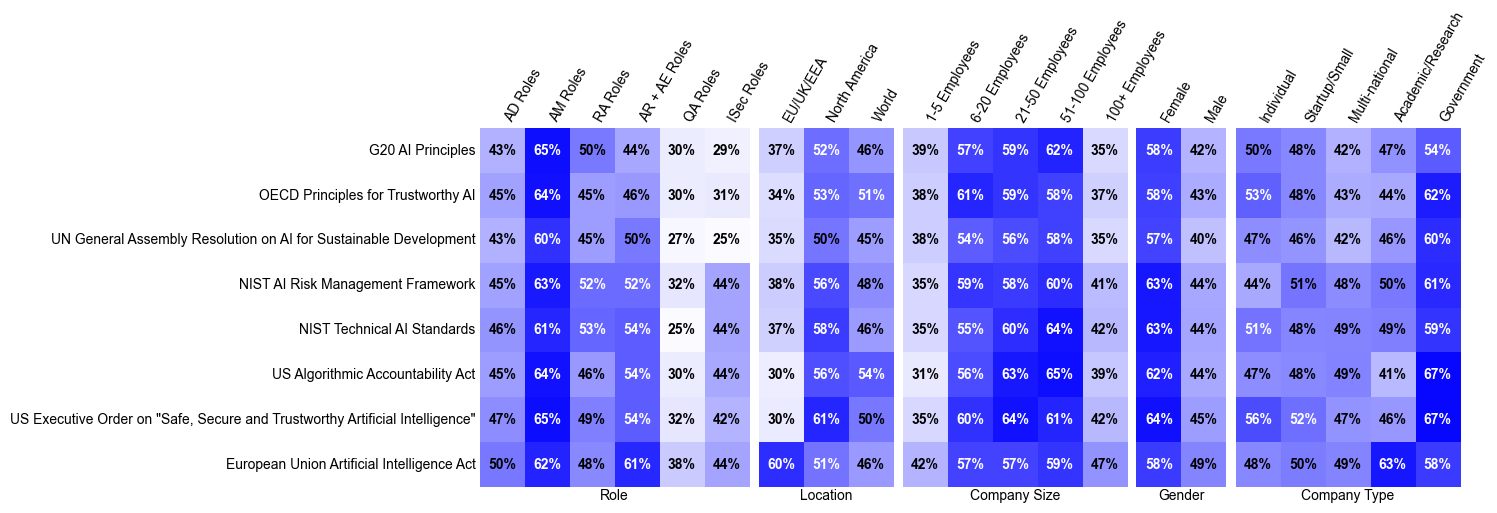} 
    \caption{Group B - Familiarity with AI Governance Initiatives by Demographics} \label{fig:Initiatives_Familiarity_B_Demographics}
\end{figure}

As shown in  Figure \ref{fig:Initiatives_Familiarity_B_Demographics},  AM roles in Group B are most familiar with all initiatives, while those in QA and ISec roles are the least familiar. Those in North America are more familiar with all initiatives than other participants, except for the EU's AI Act, which participants from the EU, EEA, and UK are most familiar with. A similar trend appears across company types: government employees are the most familiar with all AI ethics initiatives except the EU's AI Act, for which research and academic employees are the most familiar. Similarly to the principles, participants in companies with 1-5 and 100+ employees are the least familiar with all initiatives. We evaluated the correlation between familiarity and role (\textbf{H3a}), location (\textbf{H3b}), company type (\textbf{H3c}), company size \textbf{H3d}), and gender (\textbf{H3e}). As shown in Table \ref{table:demographics-familiarity-ai-initiatives}, we did not find any correlation for Group A, after Bonferroni correction. Without the correction, there is a correlation between location and the average familiarity for Group A. However, for Group B, we found significant correlations between familiarity with AI governance initiatives and participants' roles, location, company size, and gender ($p-values = 0.0008, 0.0003, 0.000,$ and $0.0001$, respectively), both with and without the Bonferroni correction. No significant correlation was observed with company type. 

We conducted a LASSO analysis to identify the predictive factors within demographics, practice, and perceptions most strongly associated with familiarity with AI governance initiatives. 
For Group B, our model indicates that greater familiarity with AI governance initiatives is most strongly associated with the use of specific AI tools, i.e., \textit{Google Cloud AI} (+0.251) and \textit{LangChain} (+0.240) (see Appendix \ref{Appendix:stats} - Table~\ref{table:lasso_regulation_familiarity_b}). Participants who used \textit{Neural Auto Encoders} (+0.192) and those who developed \textit{contingency plans} for ethical risk mitigation (+0.181) also exhibited higher familiarity with AI governance initiatives. Familiarity with the \textit{harm prevention and beneficence} principle showed a marginally significant positive association (+0.221). In contrast, participants with only a \textit{high school degree} showed lower levels of familiarity (-0.213). 

\begin{tcolorbox}[colback=blue!10,colframe=black!30,boxrule=0.1pt, boxsep=0.1pt]
\textbf{Implications.} The substantially lower familiarity with governance initiatives compared to ethics principles (58--62\% vs 75--92\%) reveals a concrete intervention opportunity: practitioners know \emph{what} ethical AI means but are less aware of \emph{how} regulations operationalize it. Since our LASSO and mixed-effects models show that governance initiative familiarity is the strongest predictor of both ethics principle familiarity and ethics consideration frequency, organizations could achieve impact by investing in regulatory literacy training.
\end{tcolorbox}

\begin{table}[ht]
\centering
\caption{Impact of Future AI Regulations - All Participants}
\label{tab:regulation_perception}
\begin{tabular}{ccccc} 
\toprule
\textbf{Highly +ive} & \textbf{Somewhat +ive} &  \textbf{Neutral} & \textbf{Somewhat -ive} & \textbf{Highly -ive} \\ 
\midrule
30.90\%   & 35.60\%   & 23.10\%   & 6.40\% &  4.00\%       \\
\bottomrule
\end{tabular}
\end{table}

\subsubsection{Perceptions of AI Governance Initiatives}

We evaluated whether participants perceived AI governance initiatives as negatively affecting their work. 48\%, 25\%, and 15\% of participants responded \emph{no}, \emph{yes}, and \emph{unsure}, respectively. The remaining participants chose not to respond. We then examined how they perceive the impact of these initiatives on shaping the future of technology in their field (see Table~\ref{tab:regulation_perception}). The participants could respond on a Likert Scale with an option to describe the impact in more detail. The majority (66.5\%) sees at least a somewhat positive impact of the initiatives on the future of technology. However, many participants were uncertain or neutral (23.1\%). About 10\% consider a negative impact.
One participant noted: \textit{``We already have that in Europe. I'm all hears for data and privacy protection laws, but sometimes these laws are maded by people outside of the industry and this has had a negative impact on the European Tech Market before.''}
About half of those who responded \emph{yes} are mostly concerned that regulations may limit innovation. One participant remarked: \textit{``I believe that regulations for mitigating AI ethics risks might initially pose challenges, such as increased compliance costs and slower innovation. However, they can ultimately benefit the industry by building trust, ensuring fair practices, and fostering a more responsible and sustainable approach to AI development.''} Others are concerned about operational challenges and costs associated with regulations. According to one participant: \textit{``If the company fails to comply with the regulations, it may face legal action, fines or other legal consequences. This not only causes direct financial losses, but may also damage the company's reputation.''} Among those who do not believe regulation negatively impacts their work, a common theme is the belief that regulation can enhance trust across the AI ecosystem. In the words of one participant: \textit{``As said, we need to ensure the data protection before using AI for our daily business. Having a higher authority to regulate that safety will increase our and our client's trust in the systems.''} Among those who are unsure, we observe a balance between recognizing the positive aspects of regulations and concerns about limiting innovation and competitiveness. One respondent commented: \textit{''Regulations for mitigating AI ethics risks might increase compliance costs and slow innovation, potentially creating competitive disadvantages for smaller companies. However its also important to maintain certain standards.''} 

We conducted a LASSO analysis to identify predictive factors most strongly associated with participants' perceived impact of AI regulations, ranging from highly negative to highly positive. 
For Group B, our model shows that more positive perceptions of regulatory impact are most strongly associated with the \textit{frequency of considering ethics in development} (+0.333) and with familiarity with the \textit{environment and social responsibility} principle (+0.284) (see Table~\ref{table:lasso_regulation_impact_b}). 

\begin{table}[ht]
\centering
\caption{LASSO Results: Predictors of Regulatory Impact Perception (Gr. B)}
\begin{tabular}{lcc}
\hline
\textbf{Predictor} & \textbf{Coefficient} & \textbf{Std. Error} \\
\hline
\multicolumn{3}{l}{\textit{AI Practice}} \\
\quad Frequency of considering ethics in development & $0.333^{**}$ & $(0.109)$ \\
\\[-6pt]
\multicolumn{3}{l}{\textit{Familiarity with AI Ethics Principles}} \\
\quad Environment and Social Responsibility & $0.284^{**}$ & $(0.105)$ \\
\hline
Overall Mean & \multicolumn{2}{c}{3.68} \\
Observations & \multicolumn{2}{c}{137} \\
R² (test) & \multicolumn{2}{c}{0.10} \\
Features Selected & \multicolumn{2}{c}{2 / 150} \\
\hline
\multicolumn{3}{p{12cm}}{\footnotesize \textit{Note:} $^*p < 0.05$; $^{**}p < 0.01$; $^{***}p < 0.001$. Dependent Var.: Perceived impact of AI regulations (1--5 scale, highly -ive to highly +ive). Coefficients are standardized (per 1-SD change in predictor).} \\
\end{tabular}
\label{table:lasso_regulation_impact_b}
\end{table}



\begin{tcolorbox}[colback=blue!10,colframe=black!30,boxrule=0.1pt, boxsep=0.1pt]
\textbf{Implications.} The tension between positive regulatory outlook (66.5\%) and concerns about innovation slowdown reflects a challenge for software engineering practices, how to implement compliance without creating development bottlenecks. Our finding that practitioners who consider ethics more frequently also perceive regulations more positively ($\beta = 0.333$) suggests that hands-on experience with ethical practice reduces regulatory anxiety.
\end{tcolorbox}

Table \ref{tab:rq3_summary} shows the summary of the result.

\begin{table}[!htbp]
\centering
\caption{Summary of Key Findings for RQ3: Familiarity with Governance Initiatives}
\label{tab:rq3_summary}
\small
\begin{tabularx}{\textwidth}{X p{2.5cm} p{1.5cm}}
\toprule
\textbf{Finding} & \textbf{Result} & \textbf{Ref.} 
\\
\midrule
\textbf{Familiarity with Governance Initiatives} & & \\
- Participants are generally less familiar with governance initiatives than with ethics principles, with Gr.~A scoring $>$25\% lower than Gr.~B & 58--62\% vs 75--92\% & Fig.~\ref{fig:B:Regulation_Familiarity} \\
- For Gr.~B, role, location, company size, and gender are all significantly correlated with governance initiative familiarity (all Bonferroni-significant) & All $p<0.001^{**}$ & Table~\ref{table:demographics-familiarity-ai-initiatives} \\
- The majority of participants perceive a positive impact of future AI regulations, though 23\% are uncertain and 10\% perceive a negative impact & 66.5\% positive & Table~\ref{tab:regulation_perception} \\
\midrule
\bottomrule
\end{tabularx}
\end{table}

\begin{tcolorbox}[colback=gray!10,colframe=black!30,boxrule=0.1pt, boxsep=0.1pt]
\textbf{RQ3 Summary - AI Governance Initiatives:} Overall, the average familiarity with the AI governance initiatives is lower than the familiarity with the principles among various roles and demographics. However, most AI practitioners have a positive outlook on how regulations will impact the future of AI development and their companies. The field is expected to benefit from increased user trust fostered by regulations and their associated safety guidelines. However, many express concern about the effects of these regulations, particularly regarding the challenges of implementing guidelines into AI technology, which may result in slowing innovation and reducing competitiveness. 
\end{tcolorbox}%

\subsection{Predictors of AI Ethics Principle Familiarity and Practice}

Here, we present the results of linear mixed-effects models to identify statistically significant predictors of familiarity, practice, and perceptions related to AI ethics principles and initiatives (i.e., \textbf{RQ2} and \textbf{RQ3}).

\begin{table}[ht]
\centering
\caption{Fixed Effects in Linear Mixed-Effects Models}
\begin{tabular}{p{6.5cm}cccc}
\toprule
\textbf{Variable} & \textbf{Effect} & \textbf{SE} & \textbf{$z$} & \textbf{$p$-value} \\
\midrule
\multicolumn{5}{l}{\textit{Model 1: Principle Familiarity ($n=267$, $k=6$ roles)}} \\
Intercept & 2.465 & 0.164 & 15.03 & $<0.001$ \\
Experience & 0.074 & 0.045 & 1.64 & 0.101 \\
Develops AI & $-0.056$ & 0.114 & $-0.49$ & 0.623 \\
Governance Initiative Familiarity & 0.419 & 0.036 & 11.65 & $<0.001$ \\
\\[-6pt]
\midrule
\multicolumn{5}{l}{\textit{Model 2: Ethics Consideration Frequency ($n=267$, $k=6$ roles)}} \\
Intercept & 2.009 & 0.309 & 6.50 & $<0.001$ \\
Experience & 0.019 & 0.057 & 0.34 & 0.735 \\
Principle Familiarity & 0.296 & 0.078 & 3.80 & $<0.001$ \\
Governance Initiative Familiarity & 0.238 & 0.056 & 4.23 & $<0.001$ \\
\bottomrule
\end{tabular}
\label{table:mixed_model_results}
\end{table}

\subparagraph{\textbf{Predictors of AI Ethics Principle Familiarity}} 
We evaluate which factors predict familiarity with AI ethics principles (i.e., \textbf{Model 1}). Model 1 reveals that familiarity with AI governance initiatives is the only statistically significant predictor ($\beta = 0.419$, $p < 0.001$), whereas the factor, years of experience in AI development, is not ($\beta = 0.074$, $p = 0.101$). Each 1-point increase in governance initiative familiarity predicts a 0.419-point increase in principle familiarity, indicating that knowledge of initiatives and principles is strongly linked. 

\subparagraph{\textbf{Predictors for Considering AI Ethics Principles}} We examine which factors predict the frequency of considering AI ethics principles (i.e., \textbf{Model 2}). Model 2 shows that both \textit{principle familiarity} ($\beta = 0.296$, $SE = 0.078$, $z = 3.80$, $p < 0.001$) and \textit{governance initiative familiarity} ($\beta = 0.238$, $SE = 0.056$, $z = 4.23$, $p < 0.001$) are significant predictors, while experience is not ($\beta = 0.019$, $SE = 0.057$, $p = 0.735$). The likelihood ratio tests confirmed that both types of familiarity independently improve model fit: adding governance initiative familiarity to a model of only principles ($\chi^2(1) = 12.45$, $p < 0.001$) and principle familiarity to a model of only familiarity with AI governance initiatives ($\chi^2(1) = 10.69$, $p = 0.001$) both significantly improved fit. The model that includes both familiarity types ($AIC = 701.34$) fit substantially better than models with only principles ($AIC = 711.80$) or only initiatives ($AIC = 710.03$). 

\subparagraph{\textbf{Hypotheses Testing}} We conduct hypothesis tests to analyze the impact of various factors on familiarity and consideration of AI ethics principles. 

\textit{H4a: Does AI development experience predict familiarity with AI ethics principles?}
The experience effect is $\beta = 0.182$, $SE = 0.054$, $z = 3.37$, corresponding to \textbf{$p = 0.001$}. We \textbf{reject $H4a\_0$} (no effect) at $\alpha = 0.05$. Each additional experience level predicts a 0.182-point increase in familiarity with principles. A practitioner with 10+ years scores 0.55 points higher than one with 1--2 years---a small but statistically significant effect.

\textit{H4b: Does familiarity with AI governance initiatives predict frequency of considering AI ethics principles in work?} The AI governance initiative familiarity effect is $\beta = 0.348$, $SE = 0.047$, $z = 7.37$, corresponding to \textbf{$p < 0.001$}. We \textbf{strongly reject $H4b\_0$} (no effect) at $\alpha = 0.05$. Each 1-point increase in initiative familiarity predicts a 0.348-point increase in ethics consideration frequency---a moderate to large effect. 

\textit{H4c: Does professional role predict familiarity and consideration of AI ethics?}
The Intraclass Correlation Coefficient (ICC) quantifies variance due to role clustering. We got $ICC = 0.029$ (2.9\% between roles) and $ICC = 0.155$ (15.5\% between roles) for AI ethics principle familiarity and AI ethics frequency, respectively.
We \textbf{partially reject $H4c\_0$}, suggesting that role has a negligible impact on knowledge but a moderate impact on practice frequency.

\subparagraph{\textbf{Model Selection and Interaction Effects}} 
To validate our findings regarding the significance of fixed effects, we compared nested models using Likelihood Ratio Tests (LRT), with significance level $\alpha = 0.05$. 

Appendix \ref{Appendix:stats} - Table~\ref{table:model_comparison_interactions} shows the result of the model comparison, including interaction models. For familiarity with AI ethics principles, adding experience improved fit over the null model (LRT $\chi^2(1) = 12.04$, $p = 0.001$), but adding AI development involvement did not (LRT $\chi^2(1) = 1.20$, $p = 0.271$). We tested two interactions: 1. experience $\times$ location, and 2. experience $\times$ \texttt{develops\_ai}. Neither significantly improved model fit (LRT $p = 1.000$ and $p = 0.309$), and both increased AIC ($\Delta$AIC = $-10.45$ and $-2.40$), indicating worse fit. Thus, the experience effect on familiarity with AI ethics principles is consistent across locations and in AI development involvement.  

For the frequency of use of AI ethics principles, adding familiarity with governance initiatives significantly improved fit over the null model (LRT $\chi^2(1) = 52.03$, $p < 0.001$), but adding experience did not (LRT $\chi^2(1) = 1.24$, $p = 0.265$). We tested two interactions: (1) \texttt{governance\_initiative\_familiarity} $\times$ location, and (2) \texttt{governance\_initiative\_familiarity} $\times$ experience. Neither interaction significantly improved fit (LRT $p = 1.000$ and $p = 1.000$), and both increased AIC ($\Delta$AIC = $-13.62$ and $-7.02$), indicating worse fit. Thus, the initiative familiarity effect on ethics consideration is consistent across locations and experience levels.


\subparagraph{\textbf{Effect Sizes}} 
We calculated marginal $R^2$ and conditional $R^2$ to quantify the variance in the outcome that each model explains~\cite{nakagawa2013general}. Marginal $R^2$ represents variance explained by fixed effects alone (i.e., the predictors of interest), while conditional $R^2$ includes variance explained by both fixed and random effects (i.e., accounting for clustering within roles). The difference indicates the extent to which additional variance is attributable to role-based grouping. For Model 1, fixed effects explained 6.3\% of variance (marginal $R^2 = 0.063$), increasing to 9.4\% when accounting for role clustering (conditional $R^2 = 0.094$). This shows that practitioners within the same role tend to exhibit similar levels of familiarity with the principles. 
For Model 2, familiarity with initiatives explained 17.4\% of variance (marginal $R^2 = 0.174$), increasing to 27.8\% with role clustering (conditional $R^2 = 0.278$)--a moderate effect. 
The larger gap between the marginal and conditional $R^2$  suggests that professional roles play a substantial role in explaining variation in how frequently AI ethics principles are considered.

\subparagraph{\textbf{Interpretation}} 
These findings reveal asymmetry: experience predicts knowledge ($\beta = 0.182$, $p = 0.001$) but accounts for only a small portion of variance (6.3\%), whereas familiarity with governance initiative strongly predicts ethical practice ($\beta = 0.348$, $p < 0.001$) with moderate explained variance (17.4\%). This indicates that simply having more years of experience does not necessarily lead to more frequent consideration of ethics; rather, specific knowledge of AI governance initiatives drives ethical practice. Role-based clustering is minimal for knowledge (ICC = 2.9\%) but moderate for practice (ICC = 15.5\%), suggesting that organizational and role-specific norms influence behavior more than individual knowledge acquisition. These effects remain consistent across locations and experience levels.

\begin{tcolorbox}[colback=blue!10,colframe=black!30,boxrule=0.1pt, boxsep=0.1pt]
\textbf{Implications.}
Linear mixed-effects modeling shows that AI development experience predicts familiarity with ethics principles, while  AI governance initiatives familiarity is the strongest predictor of how frequently practitioners consider ethics. Professional role has minimal impact on knowledge but a meaningful influence on practice. These relationships hold across locations and experience levels, which suggests that education, and role-specific interventions may promote ethical AI development more effectively than experience-based training alone. 
\end{tcolorbox}

\subsection{Risk Mitigation Strategies} 

In \textbf{RQ4}, our goal is to understand the actual practices employed by AI practitioners to mitigate AI ethics risks. We tailored questions on AI ethics risk mitigation strategies to each role within Group B, as detailed in Table \ref{table:participant_roles}. 

\begin{figure}[!htbp]
    \centering
    \includegraphics[width=0.9\textwidth]{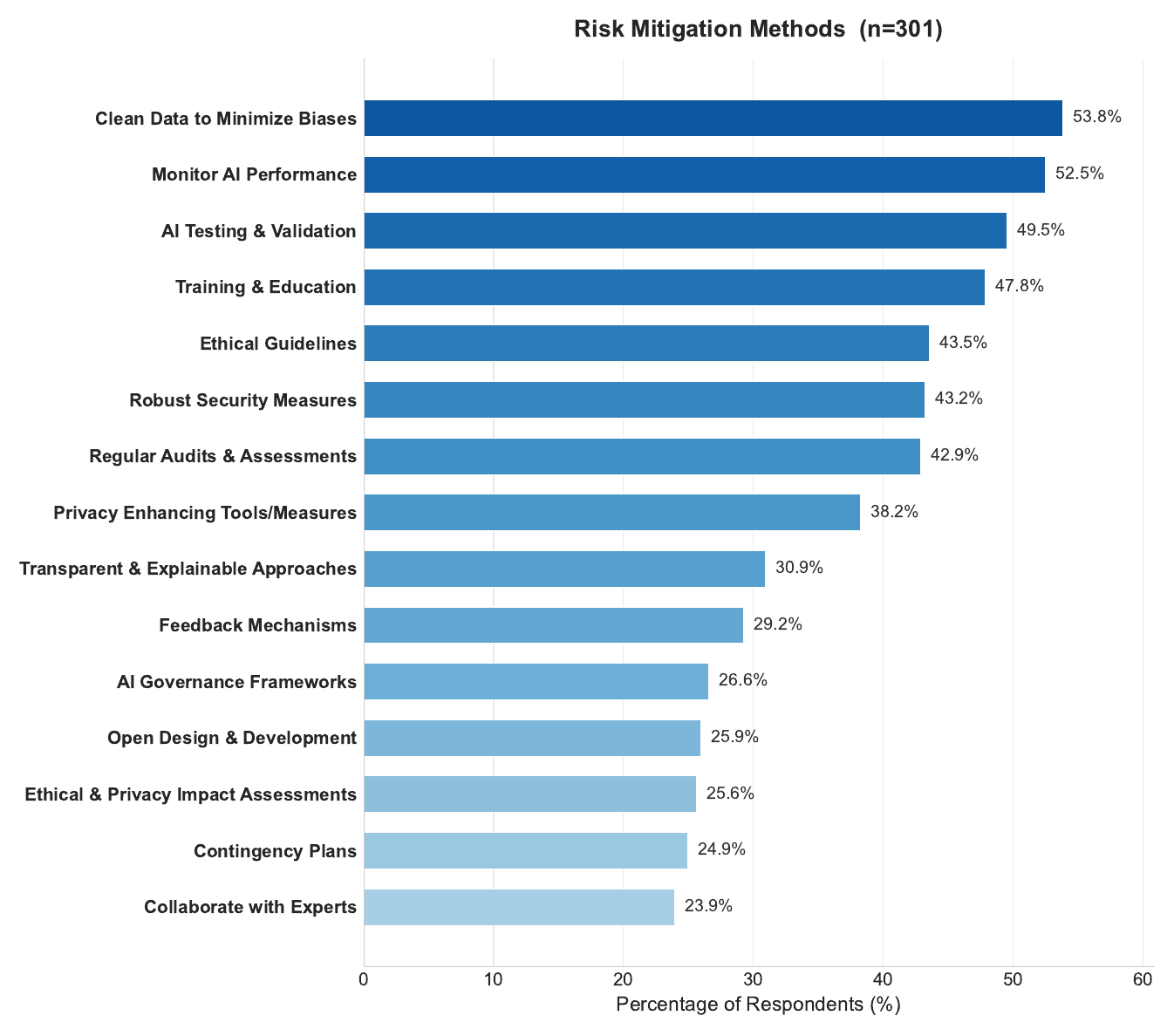}
    \caption{Group B - Risk Mitigation Methods Used}
    \label{fig:D:mitigation_methods}
\end{figure}

Overall, all participants in Group B identified essential risk mitigation strategies to address ethical concerns, as shown in Figure \ref{fig:D:mitigation_methods}. Technical solutions like ``clean data for biases,'' ``monitor AI system performance,'' and ''use AI testing and validation'' are the most commonly selected, while policy and legal-based approaches, such as ``collaborate with experts,'' ``create contingency plans,'' and ``conduct ethical and privacy impact assessments,'' are some of the least used strategies. 

\subsubsection{AI Manager and Administrative (AM) Roles} Table~\ref{tab:ai_ethics_effectiveness} shows how AM roles
rate the effectiveness of ethics principle integration in their products.

\begin{table}[ht]
\centering
\caption{Effectiveness of AI Ethics Principles Integration}
\label{tab:ai_ethics_effectiveness}
\begin{tabular}{p{1.8cm}p{1.8cm}p{1.8cm}p{1.8cm}p{1.8cm}p{1.5cm}} 
\toprule
\textbf{Extremely Effective} & \textbf{Moderately Effective} & \textbf{Somewhat Effective} & \textbf{Moderately Ineffective}& \textbf{Somewhat Ineffective} & \textbf{Prefer not to say}\\ 
\midrule
27.87\%  & 29.51\%  & 27.87\%   & 8.20\% &  4.92\%   & 1.64\%      \\
\bottomrule
\end{tabular}
\end{table}


Next, we analyzed whether, among AI management roles, higher familiarity with AI ethics principles is associated with a higher perceived effectiveness of principle integration (\textbf{H5a}) (see Table \ref{Table:H4}). 
The Chi-Square test indicates an association between familiarity with AI ethics principles and the perceived effectiveness of their integration into development; however, this association is not statistically significant after applying the Bonferroni correction ($alpha = 0.0019$).

\begin{table}[h!]
\centering
\caption{P-Value and Chi-Square Statistic for Hypotheses H5a to H5e}
\begin{tabular}{cccccc}
\toprule
& \textbf{H5a} & \textbf{H5b}  & \textbf{H5c} & \textbf{H5d} & \textbf{H5e} \\ \midrule
\textbf{P-Value} & \textbf{*0.0080} & \textbf{*0.0126} & \textbf{**0.0001} & \textbf{*0.0050} & 0.8270  \\ 
\textbf{Chi-Square Statistic} & 13.78 & 12.75 &  45.8273 & 23.6054 & 5.0833  \\ \bottomrule
\end{tabular}%
\label{Table:H4}
\begin{tablenotes}
\centering
\small
\item {Note:} * $p < 0.05$; ** $p < 0.0019$ (Bonferroni correction)
\end{tablenotes}
\end{table}

We followed up by asking them to identify the actions their company is taking to promote AI ethics. Table \ref{tab:am_ai_ethics_actions} shows the results. Only $\sim$50\% of the AM roles conduct regular audits of the AI systems for ethics compliance, which is concerning given that one of the mitigation strategies in OWASP Top 10 for LLMs and AI models \cite{owasp2025llm} is to \emph{``implement robust logging and monitoring systems to detect unusual patterns.''} 

Breaking down these responses by location, we observe that participants from North America are most likely to establish an AI ethics committee (57.14\%), while those from the EU+EEA+UK are more likely to implement ethics guidelines (63.64\%). In contrast, participants outside of these two regions take other actions, e.g., ``conduct regular audit AI systems for ethical compliance'' (86.7\%), ``provide AI ethics training for employees'' (73.3\%), and/or ``engage with external experts on AI ethics'' (73.3\%). 

They then selected who is responsible for ensuring AI ethics principles are followed. The results are shown in Table \ref{tab:ai_ethics_responsibility}  (note that participants were allowed to select multiple options): 63.93\% pointed to the AI development team and upper management, while only 42.62\% pointed to the internal legal and compliance team or the AI ethics committee.  
Previous research has highlighted the critical role of dedicated teams in ensuring software/AI compliance~\cite{prybylo2024evaluating}. Other studies have emphasized the challenges of enforcing accountability for AI ethics and regulatory standards and the importance of collaborating with policymakers to implement them~\cite{orr2020attributions}.

\begin{table}[ht]
\centering
\caption{AM Roles: Actions Taken at Company to Promote AI Ethics}
\label{tab:am_ai_ethics_actions}
\begin{tabular}{lcc} 
\toprule
\textbf{Action Taken}                                              & \textbf{Counts} & \textbf{Percentage} \\ \midrule
Provide AI ethics training for employees                          & 36                & 60.00\%             \\
Implement ethics guidelines for AI development and deployment & 36                & 60.00\%             \\
Engage with external experts on AI ethics                         & 33                & 55.00\%             \\
Establish an AI ethics committee                                 & 33                & 55.00\%             \\
Conduct regular audits of AI systems for ethics compliance                 & 31                & 51.67\%             \\
Prefer not to say                                                  & 3                 & 5.00\%              \\ \bottomrule
\end{tabular}
\end{table}

\begin{table}[ht]
\centering
\caption{AM Roles: Primary Responsibility for Following AI Ethics Principles}
\label{tab:ai_ethics_responsibility}
\begin{tabular}{p{8cm}cc} 
\toprule
\textbf{Responsible Party}             & \textbf{Counts} & \textbf{Percentage} \\ \midrule
AI Development Team                    & 39                & 63.93\%             \\
Upper Management                       & 39                & 63.93\%             \\
AI Ethics Committee                    & 26                & 42.62\%             \\
Internal Legal and Compliance Team     & 26                & 42.62\%             \\
Prefer not to say                      & 1                 & 1.64\%              \\ \bottomrule
\end{tabular}
\end{table}

Lastly, we asked them to provide examples of how they ensure their team adheres to AI ethics principles in development and describe how they promote and enforce them. Four mentioned \emph{``developing ethical guidelines,''} while one participant mentioned they have been \emph{``developing AI-specific red-teams.''} 
Although the responses about promoting and enforcing the principles varied, we observed that 30.3\%, 21.21\%, and 21.21\% of participants mentioned implementing ethical guidelines, communicating AI guidelines with employees, and educating employees on AI ethics, respectively. Of those who mentioned communicating with employees, participants highlighted, among others: \textit{``understanding and teaching staff that AI is there to aid the system, not BE the system''} and \textit{``use of stringent measures on failure to adopt the principles.''} We also gained insights into some of the challenges faced during this process. For instance, one participant noted: \textit{``our role mainly revolves around building what the client requires, and we cannot control the end usage.''} Another added, \textit{``we are a small company still with less than 5 workers so to keep the information up to date is easy right now.''}

\begin{table}[ht]
\centering
\caption{Frequency of Ethical Consideration in AI Requirements Documentation}
\label{tab:ai_ethics_frequency-req}
\begin{tabular}{p{2.cm}p{1.5cm}p{1.5cm}p{2cm}p{1.5cm}p{1.5cm}} 
\toprule
\textbf{} & \textbf{Always} & \textbf{Often} & \textbf{Sometimes} & \textbf{Rarely} & \textbf{Never} \\ 
\midrule
\textbf{Percentage}  & 20.6\% &  35.3\% &  20.6\%  &  17.6\% & 5.9\% \\
\bottomrule
\end{tabular}
\end{table}

\begin{tcolorbox}[colback=blue!10,colframe=black!30,boxrule=0.1pt, boxsep=0.1pt] \textbf{Implications.} AM roles set ethical standards for teams to follow, establishing guidelines, educating staff, and communicating principles across the organization. Despite this, only half conduct regular ethics audits of deployed systems, even though such audits are a recommended safeguard (e.g., in the OWASP Top~10 for LLMs). This gap between policy-setting and monitoring means stated guidelines may go unenforced in practice. Pairing AM roles' guideline-setting with standards of routine audit and compliance monitoring may help close the loop between an organization's declared ethical standards and its operational reality. \end{tcolorbox}

\subsubsection{Requirements Analyst (RA) Roles} 
Those in RA roles described how often they include AI ethics principles in their requirements documentation.
Interestingly, fewer than 25\% of them rarely or never consider ethics in their requirements documentation (see Table \ref{tab:ai_ethics_frequency-req}). Similarly, $\sim$30\% believe that AI ethics requirements hinder, to some degree, the project's outcome (see Table \ref{tab:ar_ai_ethics_impact}). Regarding whether their company conducts (ethics-related) risk and impact assessments of their products, $\sim$53\% of responses were positive, with one participant noting that \emph{``another unit is responsible.''} The rest responded ``no'' or ``unsure.'' 

We evaluated whether there is a correlation between their average reported familiarity with AI ethics principles and their frequency of including them in requirements documentation \textbf{(H5b)} (see Table \ref{Table:H4}). 
With $p-value = 0.0126$, we failed to reject the null hypothesis with the Bonferroni correction ($alpha = 0.0019$). The Chi-Square statistic suggests a possible association between RAs’ familiarity and their inclusion in requirements documentation, though this result should be interpreted with caution.

We examined which AI ethics principles RAs prioritize when defining AI requirements and found notable variation by geographic region. In North America and the EU+EEA+UK, 66.67\% and 88.89\% of them identified ``data protection and the right to privacy'' as their top priority. In contrast, those from other regions ranked it third, behind ``democracy and rule of law'' (62.50\%) and ``harm prevention and beneficence'' (50\%). ``Transparency and explainability of AI systems'' ranked second among RAs in both North America (53.33\%) and the EU+EEA+UK (55.56\%), but dropped to fifth place among RAs from other regions (25\%). These differences in prioritization may reflect the influence of existing regulatory frameworks on developers’ ethical considerations. RAs in North America and the EU+EEA+UK tended to emphasize principles that align closely with established governance frameworks, including the GDPR \cite{GDPR2018}, CCPA \cite{CCPA}, the EU AI Act \cite{AIAct2024}. By contrast, RAs from other regions emphasized principles, such as ``democracy and rule of law'' and ``harm prevention and beneficence,'' which align more with universal ethical standards than specific regional regulations. 




\begin{table}[h]
\centering
\caption{RA Roles: Impacts of Ethics Requirements on AI Projects}
\label{tab:ar_ai_ethics_impact}
\begin{tabular}{p{7cm}cc} 
\toprule
\textbf{Action Taken}  & \textbf{Counts} & \textbf{Percentage} \\ 
\midrule
Significantly enhancing the project outcomes     & 3   & 8.8\%  \\
Somewhat enhancing the project outcomes 
& 17       & 50.00\%             \\
Somewhat hindering the project outcomes          & 9  & 26.4\%             \\
Significantly hindering the project outcomes       & 1 & 3.00\%             \\
No impact at all on the project outcomes         & 3 & 8.8\%             \\
Prefer not to say                                 & 1  & 3.00\%             \\ 
\bottomrule
\end{tabular}
\end{table}

\begin{tcolorbox}[colback=blue!10,colframe=black!30,boxrule=0.1pt, boxsep=0.1pt] \textbf{Implications.} RA roles are uniquely positioned to embed ethical considerations from the early phases of SDLC, rather than treating them as afterthoughts. By conducting ethics impact assessments and incorporating ethics requirements into documentation, organizations can operationalize ethics-by-design. This necessitates investing in training RAs on AI ethics regulations, ensuring that ethical AI is built into software projects from the requirements phase forward.
\end{tcolorbox}



\subsubsection{AI Developer, Architect, and Data Scientist (AD) Roles} 

\begin{figure}[!htbp]
        \centering
        \includegraphics[width=0.85\textwidth]{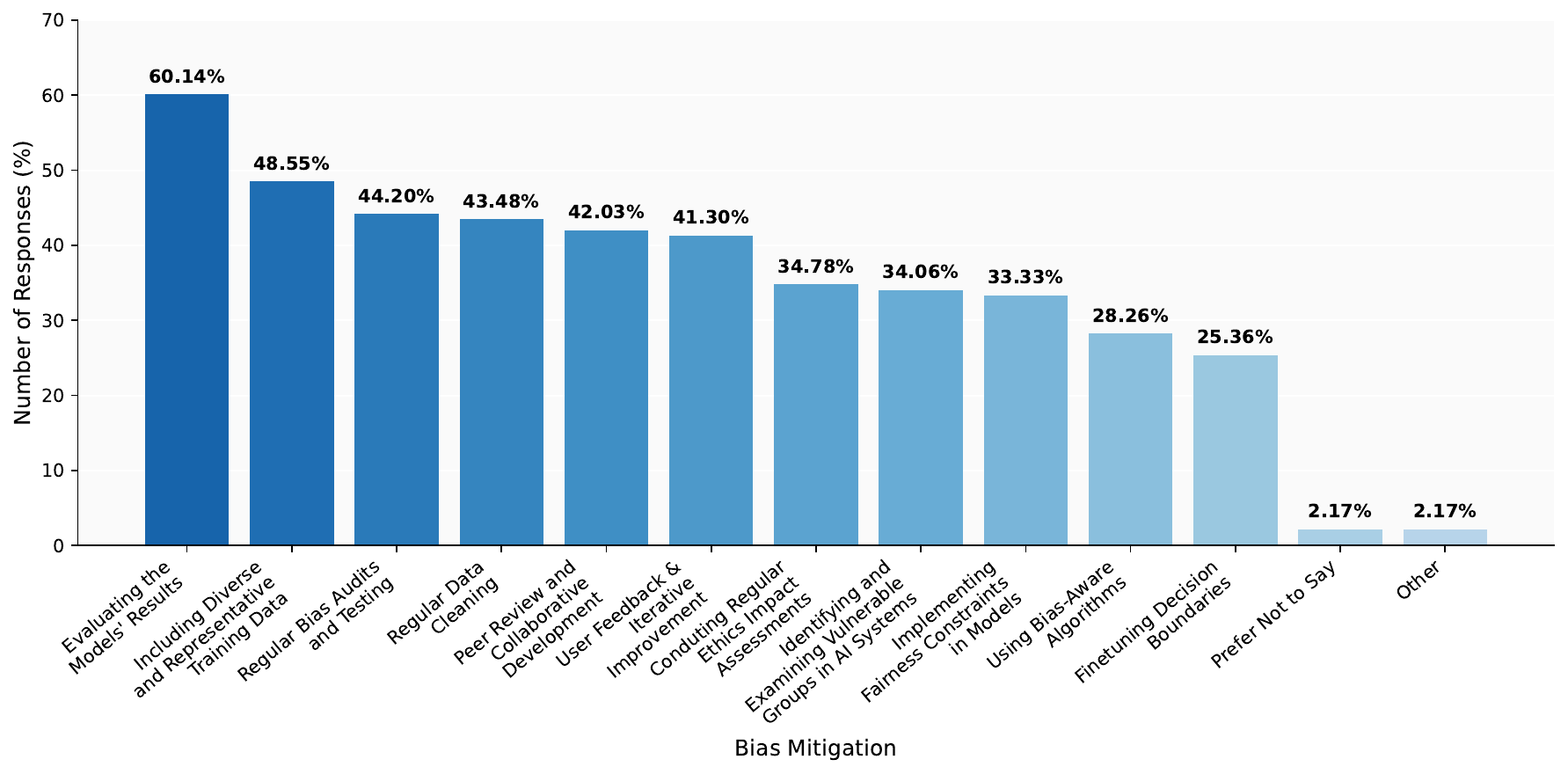} 
        \caption{{AD Roles: Methods for Bias Mitigation}}
        \label{fig:AD:mitigation_strategies}
\end{figure}

Participants in AD roles were asked to select methods for mitigating algorithmic biases. Figure \ref{fig:AD:mitigation_strategies} presents the results, showing that the top strategies focus primarily on technical solutions related to the models and training data. The most common approach to mitigate biases is to ``evaluate the models' results'' (60.14\%), followed by ``including diverse and representative training data'' (48.55\%). Policy-related solutions, such as ``peer-review and collaborative development'' or ``conducting regular ethics impact assessments,'' are less common. Finally, options related to algorithmic or parameter fine-tuning (e.g., ``fine-tuning decision boundaries'')or ``using bias-aware algorithms'' are amongst the least commonly employed.

In the location-based analysis, ADs from the EU+EEA+UK were more likely to select externally focused methods that involve diverse stakeholders, such as ``user feedback and iterative improvement'' (55.26\%) and ``identifying and examining vulnerable groups in AI systems'' (43.06\%). In contrast, ADs from North America tended to favor technical solutions, e.g., ``regular data cleaning'' (47.22\%) and ``regular bias audits and testing'' (43.06\%). ADs from other regions most frequently chose ``ensuring diverse and representative training data'' (58.62\%) and ``regular bias audits and testing.'' 

The majority of those in AD roles (76.2\%) consider transparency and explainability of AI systems at least ``moderately important,'' while only about 8.7\% consider it either ``not at all important'' or ``not very important'' (see Table \ref{tab:transparency_importance}). This aligns with our findings on familiarity with AI ethics principles, where AD roles, on average, are very familiar with this principle (see Figure \ref{fig:D:Principles_Familiarity_B_Demographics} for more details).  Similarly, a large portion of participants (64.1\%) are at least ``moderately confident'' in their ethical decisions during AI development. A small number (12.2\%) appears to be not very confident. 

\begin{table}[ht]
\centering
\caption{AD Roles: Importance of Transparency and Explainability of AI Systems}
\label{tab:transparency_importance}
\begin{tabular}{p{2cm}p{2cm}p{2cm}p{2cm}p{2.5cm}} 
\toprule
\textbf{Extremely Important} & \textbf{Moderately Important} & \textbf{Somewhat Important} & \textbf{Not Very Important}& \textbf{Not At All Important} \\ 
\midrule
52.5\%   & 23.7\%  & 15.1\%    & 5.1\% &  3.6\%       \\
\bottomrule
\end{tabular}
\end{table}

We then evaluated whether there is a correlation between  AD roles’ average familiarity with AI ethics principles and two factors: the importance they assign to AI systems being transparent and explainable (\textbf{H5c}), and their confidence in making ethical decisions during AI development (\textbf{H5d}) (see Table \ref{Table:H4}). With a $p-value < 0.0001$ and $\chi^2 = 45.8273$ for \textbf{H5c}, we can reject the null hypothesis and conclude that there is a correlation between average familiarity with  AI ethics principles and the importance of transparent and explainable AI systems. With a $p-value = 0.0050$ for \textbf{H5d}, we cannot reject the null hypothesis after applying the Bonferroni correction.  However, this value is below the conventional alpha threshold of 0.05, which may suggest the presence of a correlation, though not statistically significant.

\begin{figure}[!htbp]
        \centering
        \includegraphics[width=0.9\textwidth]{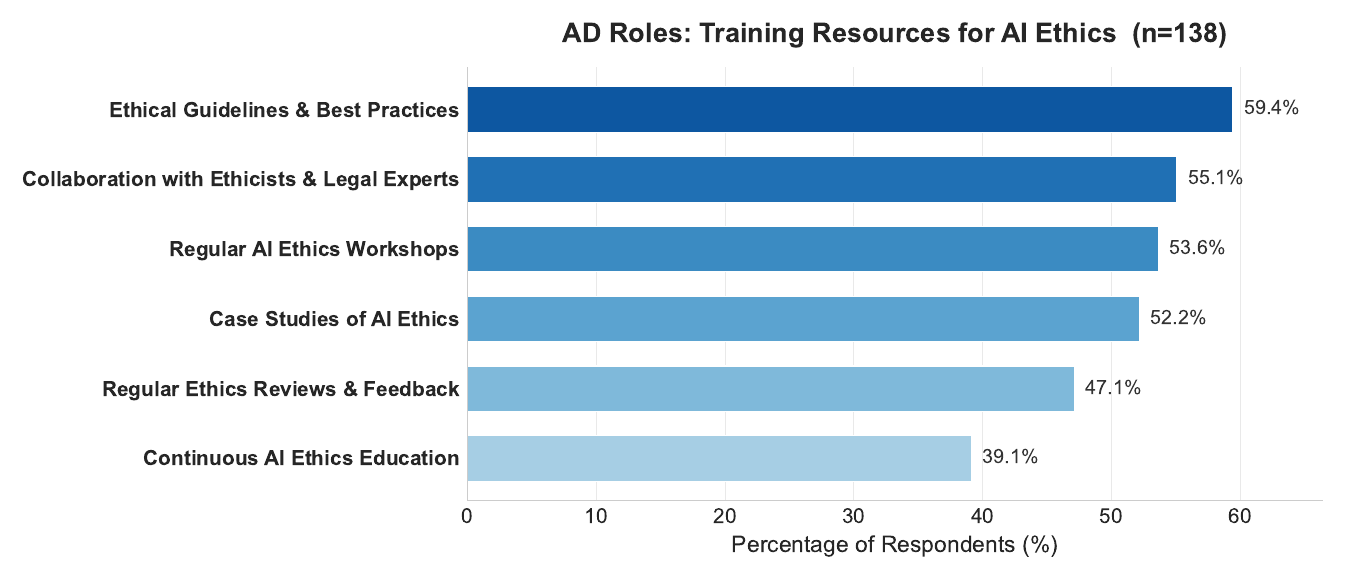} 
        \caption{AD Roles: Training Resources to Help with AI Ethics}
        \label{fig:AD:Training-Resources}
\end{figure}

We analyzed the types of training or resources AD roles may use to help them better integrate AI ethics. Figure \ref{fig:AD:Training-Resources} shows the result.
``Access to ethical guidelines and best practices'' and ``collaboration with AI ethicists and legal experts'' ranked the highest  (59.85\% and 55.47\%, respectively). Interestingly, ``support for continuous education related to AI ethics'' appears to be the least popular resource (39.42\%). 

To gain insights into the individual practices of AD roles, we asked them to describe how they integrate AI ethics principles into their coding practices and algorithm designs. Among 115 answers, we found several themes. 22 ADs described methods completed during the data processing phases. We observed that data processing is essential for mitigating biases in AI systems today, as one AD mentioned: \emph{``I ensure that the data used to train AI models is diverse and representative to prevent bias. I also use techniques like re-sampling, data augmentation, or bias-correction algorithms to minimize any inherent biases in the data.''} Another participant replied: \emph{``When collecting data, ensure the legitimacy and compliance of the data source and avoid violating user privacy.''} Yet another common approach is establishing and adhering to AI guidelines. These may coincide with the education and training for developers, as one noted: \emph{``We have take mandatory courses on AI ethics. There are also somewhat frequent workshops on AI ethics. Also try to align the system along existing frameworks on AI ethics.''} These guidelines may take inspiration from the regulatory framework as one AD responded: \emph{``maintaining defined protocols and regulations.''} 

Similar to RA roles, we asked those in AD roles if their company conducts (ethics-related) risk and impact assessments. Of the 107 responses, 64.5\% indicated ``yes,'' while 24.3\% and 6.5\% indicated ``no'' or were ``unsure,'' respectively. A small number preferred not to respond. The positive responses are slightly higher than those from RA roles, suggesting that both roles are actively involved in risk assessments and ensuring ethical guidelines are considered during the requirements and development phases. There is a variety in how they understood and described these activities, with one saying: \emph{``The security team in my company conduct regular security and compliance assessments of all of our software that are shipped to production.''} One participant noted, \emph{``Peer reviews,''} while another added: \emph{``Yes, we do, by looking at previous case studies and possible pitfalls to make sure we stay within parameters.''} 

\begin{table}[ht]
\centering
\caption{AD Roles: Confidence in Making Ethical Decisions During AI Development}
\label{tab:ethical_confidence}
\begin{tabular}{p{1.8cm}p{1.8cm}p{1.8cm}p{1.8cm}p{1.8cm}p{1.5cm}} 
\toprule
\textbf{Extremely Confident} & \textbf{Moderately Confident} & \textbf{Somewhat Confident} & \textbf{Not Very Confident}& \textbf{Not Confident At All} & \textbf{Prefer not to say}\\ 
\midrule
25.2\%  & 38.9\%  & 23.0\%   & 7.9\% &  4.3\%   & 0.7\%      \\
\bottomrule
\end{tabular}
\end{table}

\begin{tcolorbox}[colback=blue!10,colframe=black!30,boxrule=0.1pt, boxsep=0.1pt] \textbf{Implications.} AD roles carry out most of the technical risk mitigation, yet only a quarter reported being ``extremely confident'' in their ethical decisions. Their most-requested supports were concrete ethical guidelines and direct collaboration with ethicists and legal experts. Embedding the ethics review process and access to AI ethicists into the development workflow can enable AD roles to be more confident in ethical decision making today. \end{tcolorbox}

\subsubsection{Quality Assurance Engineer and IT Maintenance (QA) Roles} 

\begin{figure}[!htbp]
        \centering
        \includegraphics[width=0.9\textwidth]{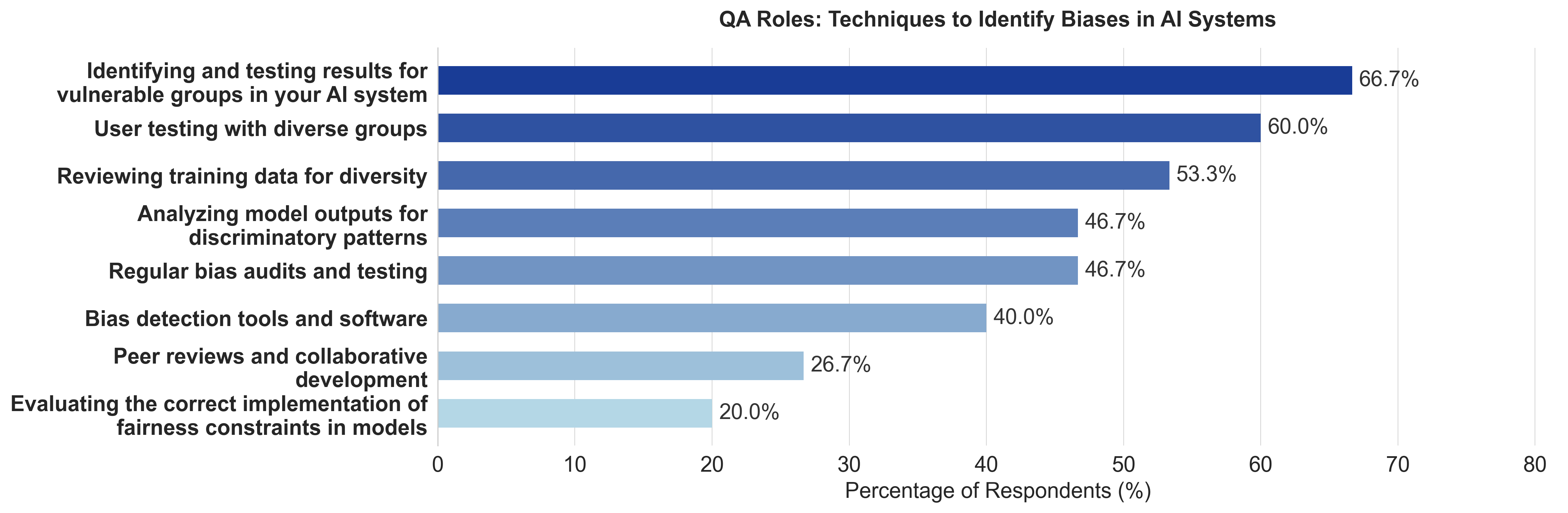} 
        \caption{QA Roles: Techniques to Identify Biases in AI Systems}
        \label{fig:QARoles:Techniques}
\end{figure}

We aimed to understand which techniques QA roles use to identify algorithmic biases. As shown in Figure \ref{fig:QARoles:Techniques}, QA roles prioritize testing for vulnerable groups (67\%) and user testing with diverse groups (60\%)—strategies that are notably less common among AD roles (see Fig.~\ref{fig:AD:mitigation_strategies}).. 


We then inquired about AI ethics training and its importance in their roles. Out of the 20 participants in Group B, only 15 responded, with the majority (73.33\%) indicating that they find it at least ``somewhat important'' (see Table \ref{tab:QA:ai_ethics_training}). 
 
We hypothesized that a positive relationship exists between the average familiarity with AI ethics principles of those in QA roles and the importance they attribute to AI ethics in their role \textbf{(H5e)}.  With $p-values = 0.8270$, we cannot reject the null hypothesis and see no statistical relationship between QA's familiarity with principles and their perceived importance of AI ethics training. 


\begin{table}[ht]
\centering
\caption{QA Roles: Importance of Training on AI Ethics for QA or Tester Role}
\label{tab:QA:ai_ethics_training}
\begin{tabular}{p{2.6cm}p{2.6cm}p{2.6cm}p{2.8cm}} 
\toprule
\textbf{Extremely Important} & \textbf{Moderately Important} & \textbf{Somewhat Important} & \textbf{Not Very Important} \\ 
\midrule
40.00\%   & 33.33\%  & 13.33\%    & 13.33\%       \\
\bottomrule
\end{tabular}
\end{table}



\begin{figure}[!htbp]
        \centering
        \includegraphics[width=0.85\textwidth]{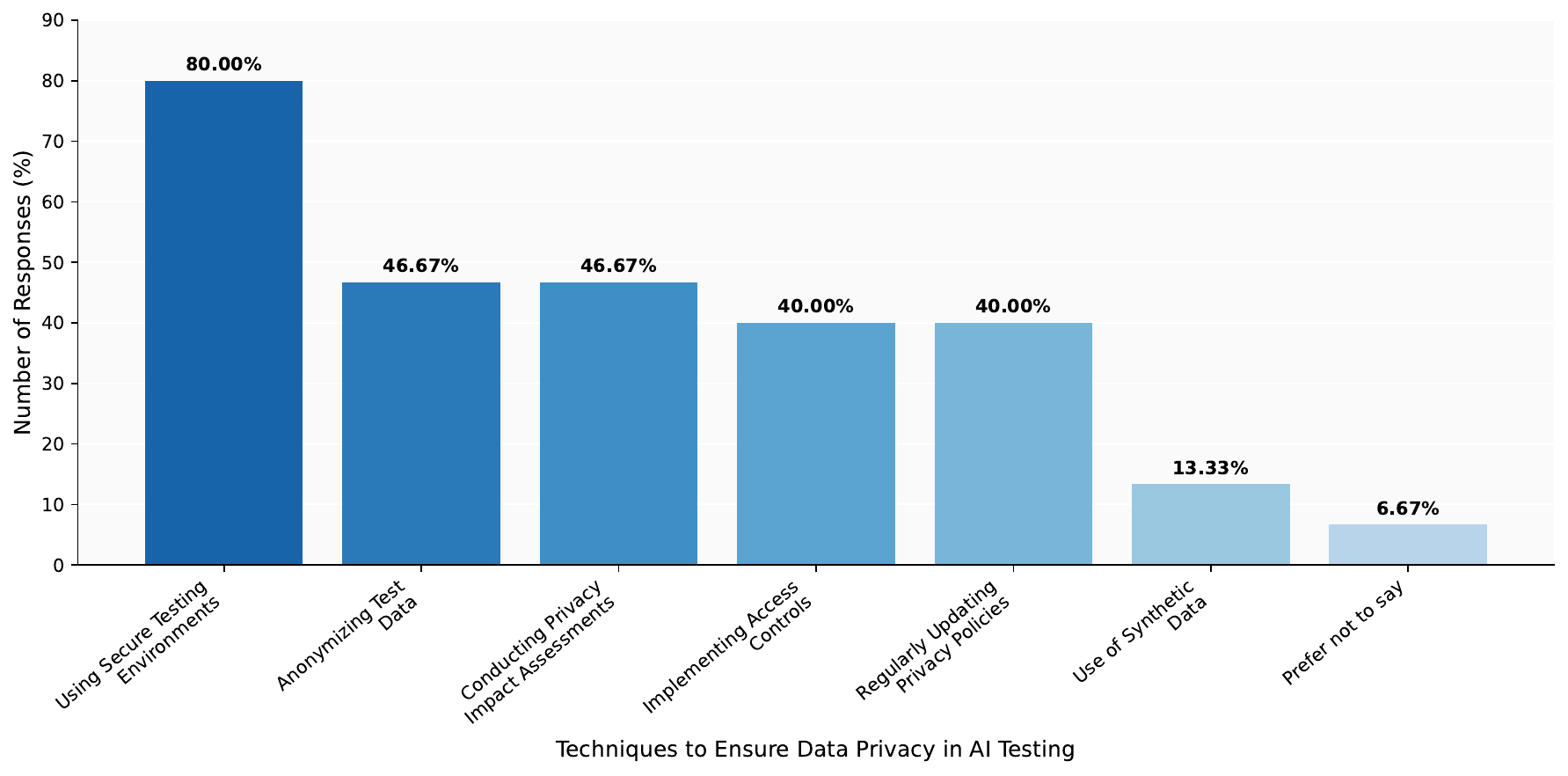} 
        \caption{QA Roles: How Data Privacy is Maintained in AI Testing}
        \label{fig:QA:data_privacy}
\end{figure}

Thirteen participants in QA roles described various steps they take to ensure that AI systems comply with ethical guidelines and prevent biased outputs. One participant responded: \emph{``We user test with a large, diverse sample. Also, we are constantly updating and enhancing diversity training.''} This approach aligns with the importance of diverse voices in development teams for minimizing biases, emphasized in previous research, such as \cite{kawakami2024studyingup,olson2025speaks}. Other participants provided insights into their team's internal processes, with one responding: \emph{``We have already got data which can be used internally and there is a team which regulates this part of testing. hence, we take care of this.''} Another QA responded \emph{``I always follow data privacy policies at regular interval. My first duty is to follow and use data security policies.''} 

Furthermore, we asked them about their approaches to maintaining data privacy during AI testing. Most of them (80\%) mentioned ``using secure testing environment,'' while other privacy-related techniques, such as ``test data anonymization'' and ``conducting privacy impact assessment (PIA),'' were also considered. PIA is generally required to be conducted prior to planning data collection, rather than during the QA and testing phases. However, consistent with findings in \cite{prybylo2024evaluating}, we observed that PIA is often performed at other stages of development. Lastly, ``using synthetic data'' was not a commonly adopted method (see Figure \ref{fig:QA:data_privacy}).

Similar to RA and AD roles, we asked QAs if their company conducts (ethics-related) risk and impact assessments. Of the thirteen, eight said ``yes.'' However, their perspectives varied, with one saying, \emph{``We need to comply with EU laws. It is needed to be checked if our app doesn't suggest wrong products or suggests something wrong to the user.''} Another participant added: \emph{``This was done as part of the initial part, now things are streamlined,''} while another said: \emph{``No, my company does not conduct risk and impact assessments or human rights due diligence. This is because we currently lack the resources and processes needed to carry out these evaluations effectively.''}

\begin{tcolorbox}[colback=blue!10,colframe=black!30,boxrule=0.1pt, boxsep=0.1pt] \textbf{Implications.} QA roles play an important part in catching ethical failures before deployment, but they tend to conduct privacy impact assessments during testing rather than at design, and are rarely included in formal ethics impact assessments. Shifting impact assessments earlier and extending QA's responsibilities from data privacy to ethics testing (e.g., fairness across user subgroups) could improve their ability to function at this role. \end{tcolorbox}

\subsubsection{Information Security and Privacy Expert (ISec) Roles}
ISec members had varied perspectives on the impact of AI technologies on data privacy. 46.14\% mentioned that AI ``somewhat'' or ``significantly compromise data privacy,'' while  38.26\% noted they ``somewhat'' or ``significantly improve data privacy.'' The remaining 15.38\% said there is ``no impact on data privacy at all'' (See Table \ref{tab:ISec:AI_privacy_impact}).

\begin{table}[ht]
\centering
\caption{ISec Roles: Impact of AI on Data Privacy within the Company}
\label{tab:ISec:AI_privacy_impact}
\begin{tabular}{p{7.5cm}cc}
\toprule
\textbf{Response Option} & \textbf{Count} & \textbf{Percentage} \\ \midrule
Somewhat compromise data privacy      & 5              & 38.46\%           \\
Somewhat improve data privacy        & 4              & 30.77\%           \\
No impact on data privacy at all      & 2              & 15.38\%           \\
Significantly improve data privacy    & 1              & 7.69\%            \\
Significantly compromise data privacy & 1              & 7.69\%            \\ \bottomrule
\end{tabular}
\end{table}

\begin{figure}[!htbp]
        \centering
        \includegraphics[width=0.85\textwidth]{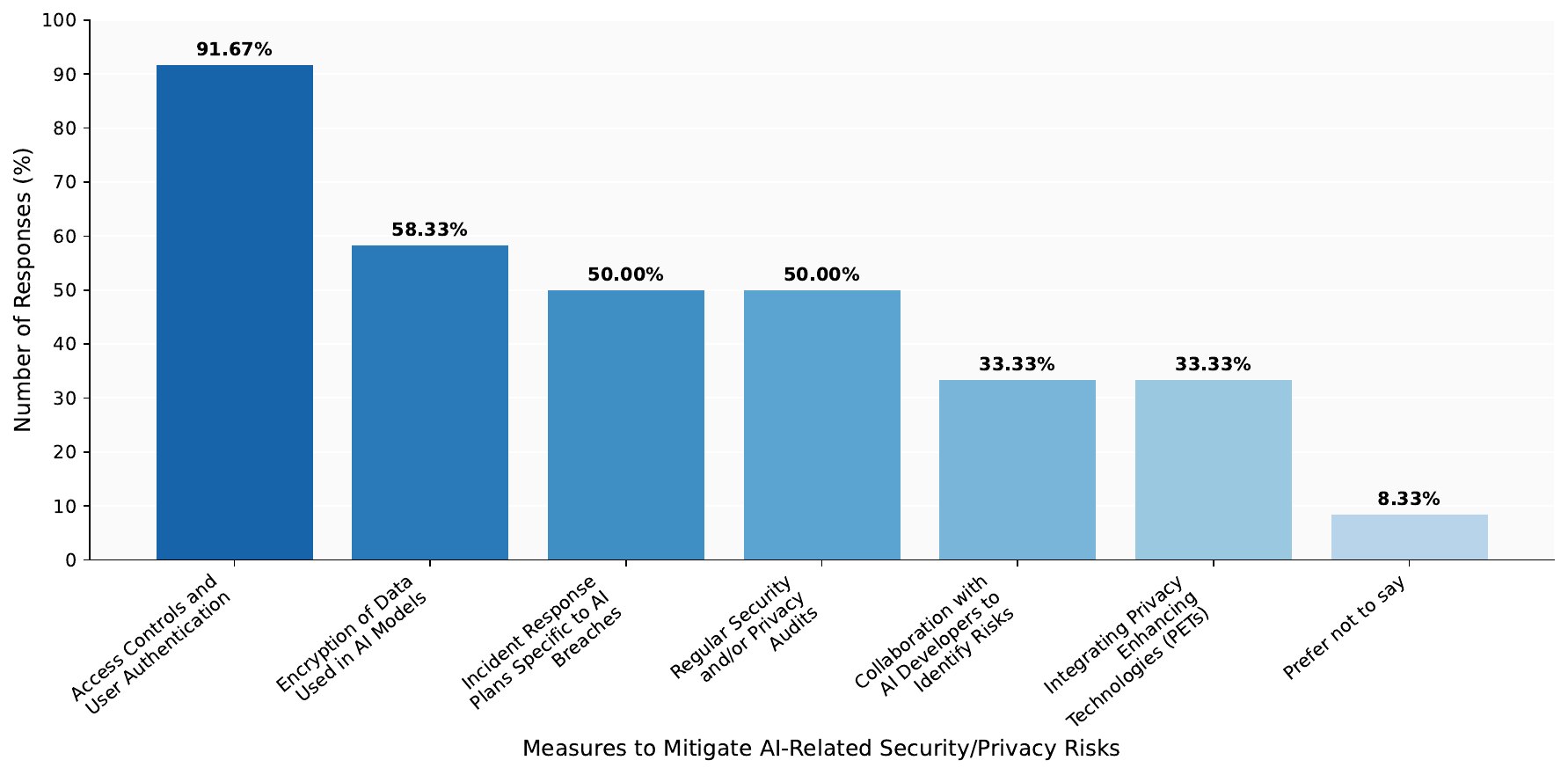} 
        \caption{ISec Roles: Measures to Mitigate AI-Related Security and Privacy Concerns}
\label{fig:ISec:measures_mitigating_concerns}
\end{figure}

When asked what measures their company implements to mitigate AI-related security or privacy risks (see Figure \ref{fig:ISec:measures_mitigating_concerns}), traditional security measures, such as ``access controls and user authentication'' and ``encryption of data used in AI models,'' are the top two approaches (91.67\% and 58.33\%, respectively). This result aligns with prior research suggesting that security concepts are generally more tangible than privacy concepts \cite{hadar2018privacy,prybylo2024evaluating}. Half of them  consider implementing ``incident response plans specific to AI security/privacy breaches,'' or conducting ``regular security and/or privacy audits of AI systems.'' Only 33.33\%  said they ``integrate privacy-enhanced technologies'' or ``collaborate with AI developers to identify security/privacy risks.'' This suggests a potential need for ISec members to collaborate more closely with developers while implementing incident response plans and conducting AI audits. 

More than 80\% of ISec members said their companies ensure compliance with data privacy regulations at least ``moderately well,'' while the rest preferred not to respond. About 70\% said they are at least ``somewhat involved''  in integrating privacy and security principles into the AI development process. However,  15.38\% and 7.69\% reported being ``rarely involved'' or ``neutral,'' respectively. Although these numbers are not high, it may still be concerning if security or privacy experts are not actively involved in related security and privacy practices during AI development.

ISec members described how they address potential ethical concerns about data privacy and security in AI systems. From twelve responses, we observed a variety of approaches, with five participants mentioning the use of various security protocols. For instance, one participant responded: \emph{``To address ethical concerns related to data privacy and security in AI systems, I implement strong data protection measures, ensure compliance with privacy regulations, use encryption and access controls, and regularly audit and update security practices. Additionally, I advocate for transparency and user consent regarding data usage.''} However, three of these five responses limited their scope to only traditional security methods (e.g., \emph{``Use encryption, access controls, and other security measures to protect data from unauthorized access''}). Another group member noted: \emph{``I try to use common sense, but It is handled by other areas.''}

They also explained how they or their organizations ensure the protection of personal data and uphold the privacy rights of affected communities. Six out of the eleven responses mentioned traditional security or privacy techniques, e.g., \emph{``By introducing encryption to every system''} or \emph{``Collect only the necessary data for the intended purpose and avoid excessive data collection.''} Three of them also mention ``following policies or guidelines'', with one ISec member responding: \emph{``Yes, we have AI policy in place that is strictly adhere by all employees. We also have configured DLP tools that help detect misuse of PII data.''} Another participant simply said: \emph{``we follow main GDPR recommandations.''} These responses highlight the importance of following standard data security and privacy practices in AI development and the need for enhancing awareness around emerging ethical concerns associated with AI systems.


\begin{table}[ht]
\centering
\caption{ISec Roles: Involvement in Integrating Privacy and Security Principles}
\label{ab:ISec:involvement}
\begin{tabular}{p{2.5cm}p{3cm}p{2.5cm}p{1.5cm}p{1.cm}} 
\toprule
\textbf{Very Involved} & \textbf{Somewhat Involved} & \textbf{Rarely Involved} & \textbf{Neutral} & \textbf{PnS} \\ 
\midrule
30.77\%   & 38.46\%    & 15.38\%    & 7.69\%  & 7.69\%        \\
\bottomrule
\end{tabular}
\end{table}


\begin{tcolorbox}[colback=blue!10,colframe=black!30,boxrule=0.1pt, boxsep=0.1pt] \textbf{Implications.} ISec professionals' tendency to equate ethics with privacy and security measures signals a notable gap in broader AI ethics literacy. Unfamiliarity with most AI ethics principles except privacy and data protection is concerning in roles that are critical to organizational risk management. Companies could prioritize education initiatives, particularly around established frameworks such as OWASP Top 10 for LLMs and NIST guidelines for ethical AI, to broaden ISec teams' ethical reasoning beyond a security-centric lens.
\end{tcolorbox}


\subsubsection{AI Researcher (AR) and AI Ethicist (AE) Roles} 
AI researchers develop models, tools, and datasets for research or public use. Hence, our goal was to understand their practices in mitigating AI risks and applying AI ethics principles. 

\begin{figure}[!htbp]
        \centering
        \includegraphics[width=0.8\textwidth]{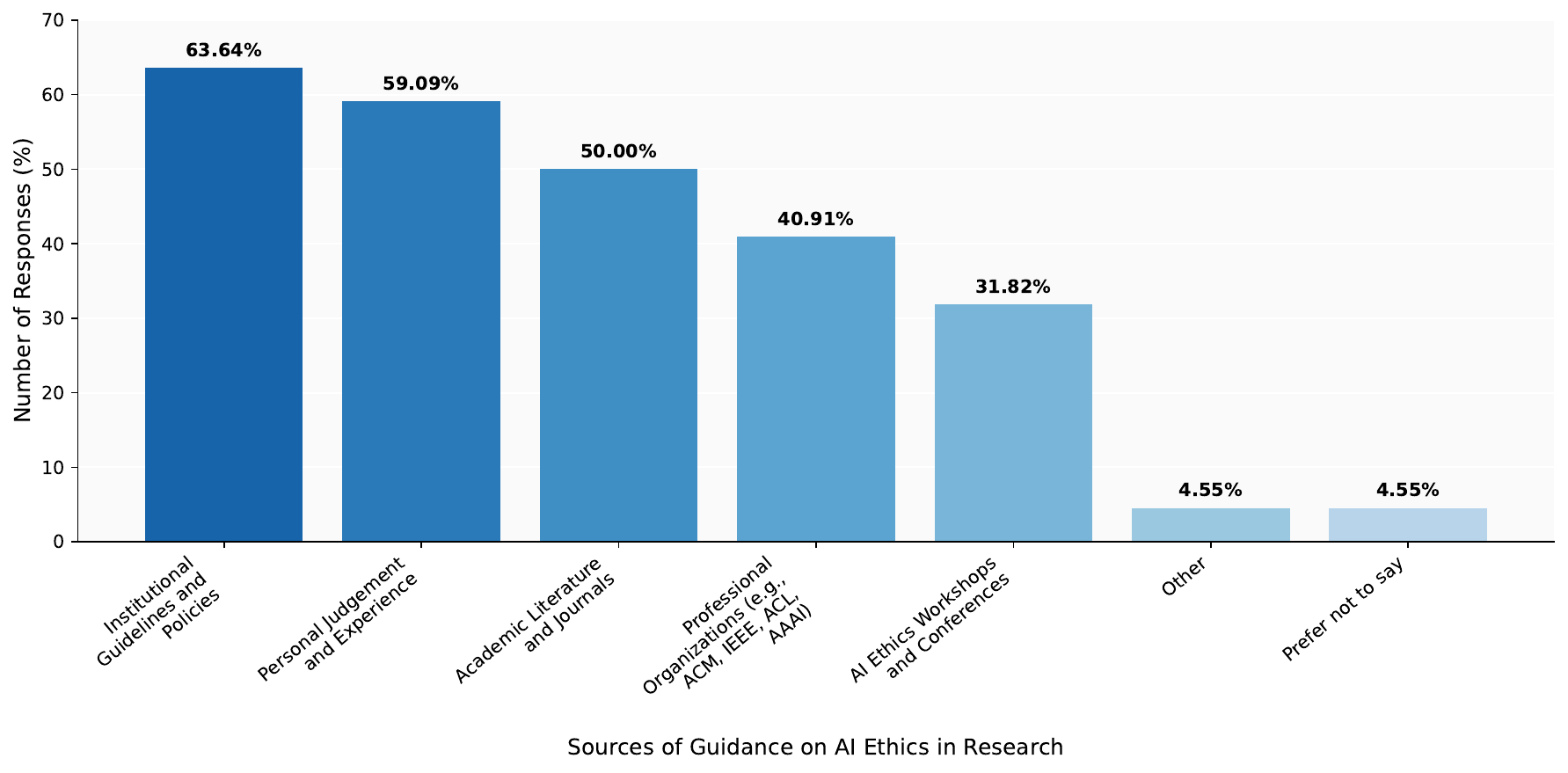} 
        \caption{AR Roles: Sources of Guidance on AI Ethics in Research}
        \label{fig:AR:guidance}
\end{figure}

When asked how often they incorporate AI ethics principles into their research methodology, $\sim$85\% mentioned   ``always,'' ``often,'' or ``sometimes,'' while only $\sim$15\%  said ``rarely'' or ``never.'' In terms of seeking guidance on AI ethics in their research, the top three choices are ``institutional guidelines and policies'' (63.64\%), ``personal judgment and experience'' (59.09\%), and ``academic literature and journals'' (50.0\%).  
One mentioned \emph{``government regulations''} as a source of guidance (see Figure \ref{fig:AR:guidance}). 

When examining regional differences among ARs, we found disparities that may reflect how researchers from regions with varying levels of access and opportunity approach their work. Participants from North America were more likely to rely on ``academic literature and journals'' (77.78\%) and ``AI ethics workshops and conferences'' (55.56\%), compared to their counterparts in the EU+EEA+UK (40\% and 20\%, respectively) and other regions (25\% and 12.5\%, respectively). In contrast, respondents from the EU+EEA+UK and other regions most frequently selected ``institutional guidelines and policies,'' chosen by 60\% and 87.5\% of respondents, respectively. 

\begin{figure}[!htbp]
        \centering
        \includegraphics[width=0.8\textwidth]{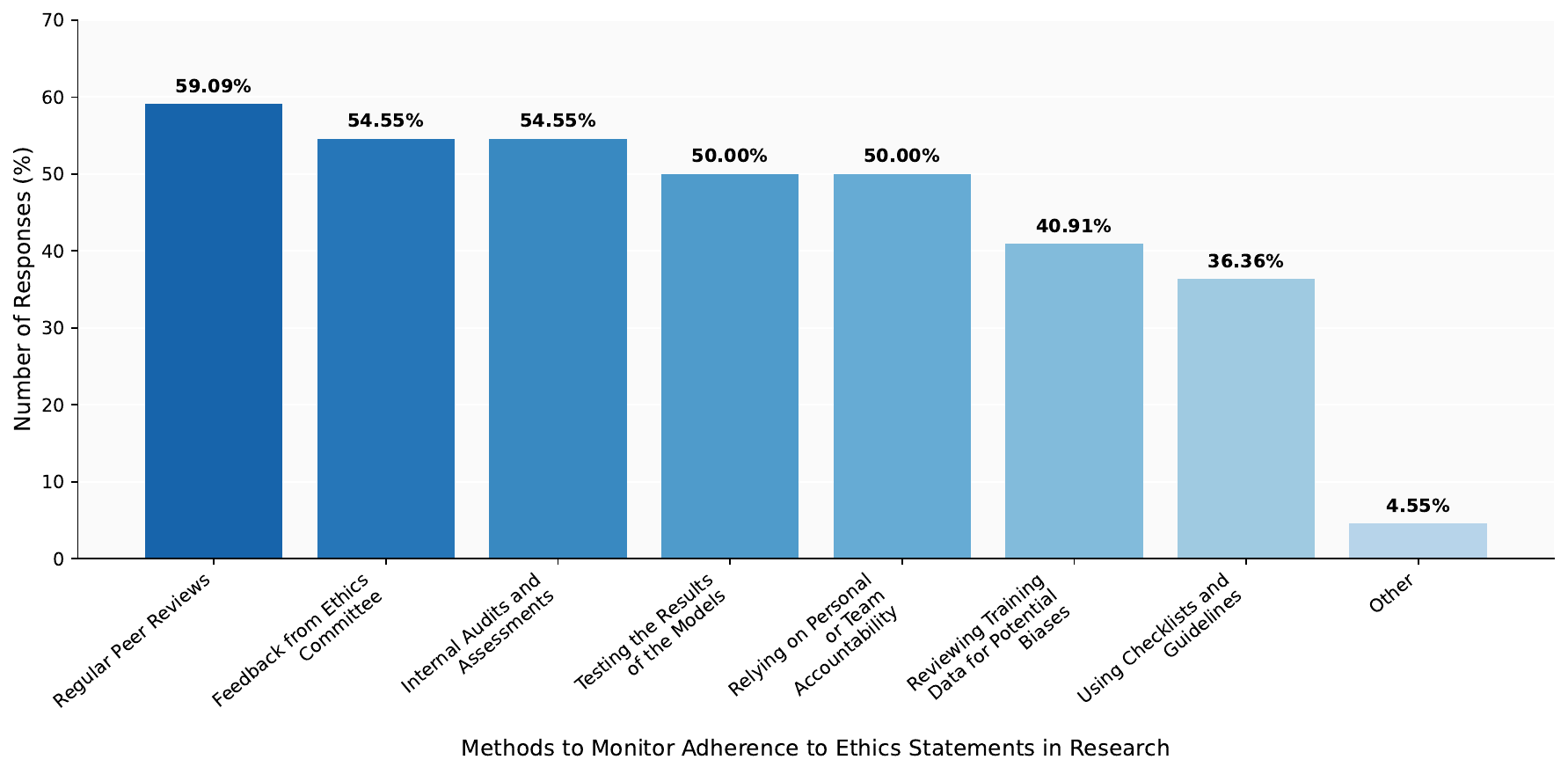} 
        \caption{AR Roles: Adherence to Ethics Statements in Research}
        \label{fig:AR:adherence}
\end{figure}

As for how they monitor their adherence to ethics statements in their research, the most popular strategies are ``regular peer reviews'' (59.05\%), ``feedback from ethics committees'' (54.55\%), and ``internal audits and assessments'' (54.55\%), while ``testing the models' results'' and ``reviewing training data for potential biases'' appear less significant. This contrasts with the practices of AD and QA roles, where peer reviews and feedback are given lower priority than technical aspects (see Figure \ref{fig:AR:adherence}).

Regarding the challenges encountered when writing ethics statements for their research papers, 60\% said, ``difficulty addressing all ethics principles,'' and 50\% said, ``lack of clear guidelines.'' These results highlight the need for clearer guidelines and documentation on ethics statements across academic venues for researchers to follow.  

AI researchers described which elements they consider when writing ethics statements in their papers. Among the twelve answers, some had notable considerations, with one saying: \emph{``Honestly I lie and alway repeat the same 2 sentences that it does not pose any risk whatsoever in different variations. Why nobody is considering the actual risk as the section is an annoyance at best.''} Others mentioned being \emph{``open about the funding received by the research group,''} and \emph{``it is usually includes the intended model use case, data collection details (especially if it includes personal data), etc.''}

We then asked them about strategies to ensure their research does not negatively impact vulnerable or marginalized groups and communities (e.g., ethnic minorities, children, and/or migrants). From the 16 responses, we observed a range of perspectives. One researcher responded: \emph{``Providing support and counselling services to participants during and after data collection.''} Another participant noted: \emph{``It is impossible to fully say that your research doesn't negatively impact marginalized communities, we prioritise the feedback from marginalized communities.''} Other ARs referenced how this is addressed in other research (e.g., \emph{``Read academic journal literature, and some organizations help''} or \emph{``I always check other peoples work and see how they handle it''}), and highlighted the importance of data quality and diversity (e.g., \emph{``Mainly ensuring high-quality diverse data,''} \emph{``Examining the data well, testing the model,''} etc.).

As shown in Table \ref{table:participant_roles}, only three AI ethicists were in Group B. We asked them several questions to explore how their role ensures ethics in AI development. When asked how effectively their company implements AI ethics principles in their project, one responded ``extremely ineffectively,'' while the other two were more positive, mentioning
``moderately effectively.'' This may suggest that AI ethicists see opportunities for improvement in developing ethical AI. Table \ref{tab:AE:actions} shows their company's actions to raise ethical awareness, where the most common actions are ``promote AI ethicists' roles in the company'' and ``use case studies to show ethical dilemmas.'' Regarding the frequency of evaluating AI systems for potential ethical dilemmas, they only mention either ``at a specific project milestone'' or ``after an incident occurs.'' They did not select ``continuously,'' as part of an ongoing process, or ``prior to deploying an AI system.'' Yet, a reliable conclusion cannot be drawn due to the small sample size. 


\begin{figure}[!htbp]
        \centering
        \includegraphics[width=0.8\textwidth]{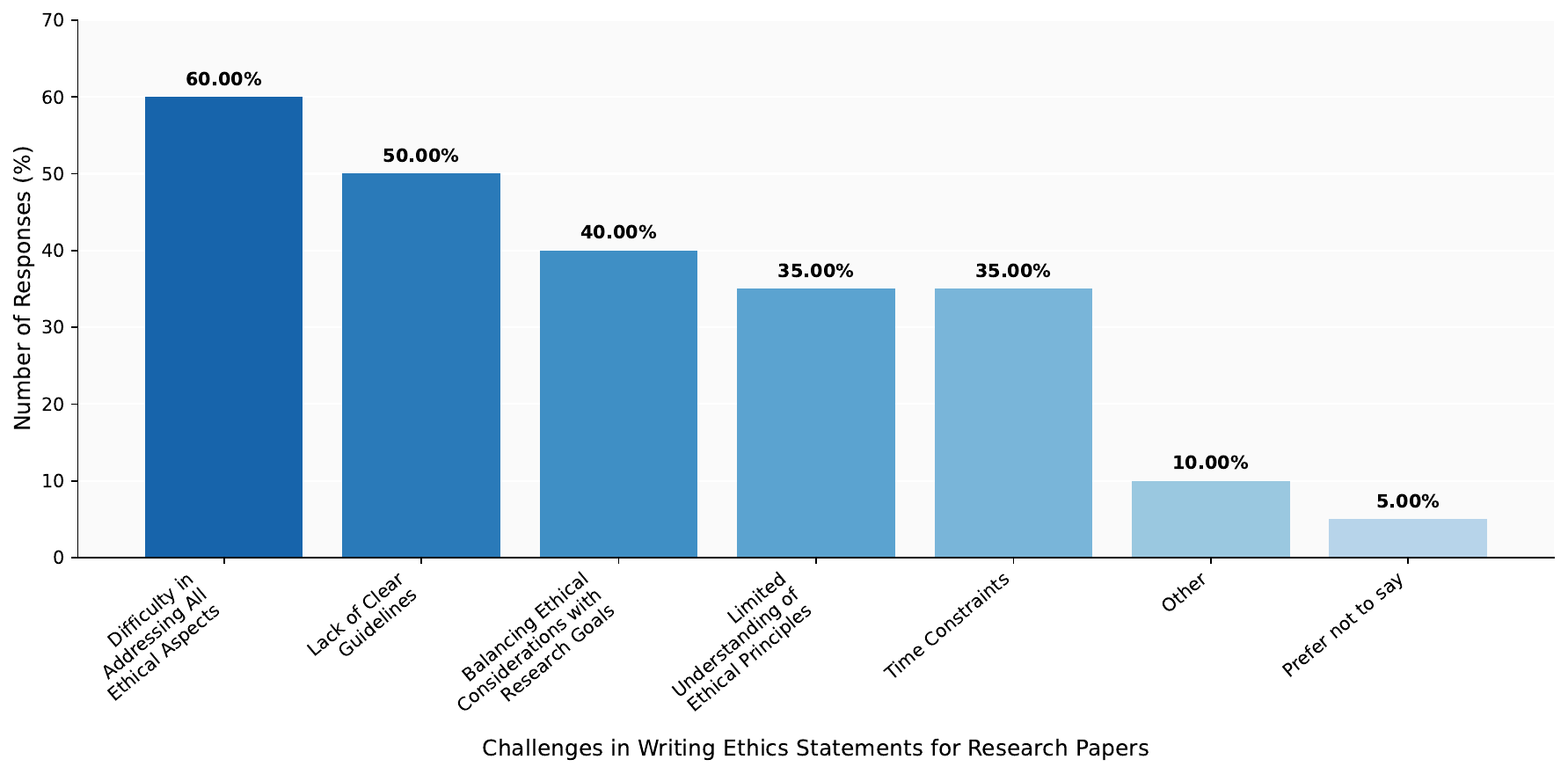} 
        \caption{AR Roles: Challenges when Writing Ethics Statements}
        \label{fig:AR:challenges}
\end{figure}

\begin{table}[ht]
\centering
\caption{AE Roles: Actions Taken by Company to Raise Ethical Awareness}
\label{tab:AE:actions}
\begin{tabular}{lcc}
\toprule
\textbf{Response Option} & \textbf{Count} & \textbf{Percentage} \\ \midrule
Promote AI ethicists' roles in the company         & 2              & 66.67\%           \\
Use case studies to show ethical dilemmas        & 2              & 66.67\%           \\
Create and distribute ethical guidelines        & 1              & 33.33\%           \\
Encourage ongoing education and certification in AI ethics & 1     & 33.33\%           \\
Integrate AI ethics discussions in meetings      & 1              & 33.33\%           \\
Provide regular AI ethics training for employees & 1              & 33.33\%           \\ \bottomrule
\end{tabular}
\end{table}

\begin{tcolorbox}[colback=blue!10,colframe=black!30,boxrule=0.1pt, boxsep=0.1pt] \textbf{Implications.} AI researchers commonly incorporate AI ethics principles into their work, drawing on institutional guidelines, personal judgment, and academic literature. Compliance is typically provided through peer review processes and ethics committees. However, many report challenges in fully addressing all ethical dimensions, due to lack of precise guidance. AI ethicists also identify room for improvement in how companies handle AI ethics during development. They recognize that progress has been made, such as introducing dedicated ethicist roles and using case studies to highlight ethical dilemmas and raise awareness. 
\end{tcolorbox}%

\begin{figure}[!htbp]
        \centering
        \includegraphics[width=0.85\textwidth]{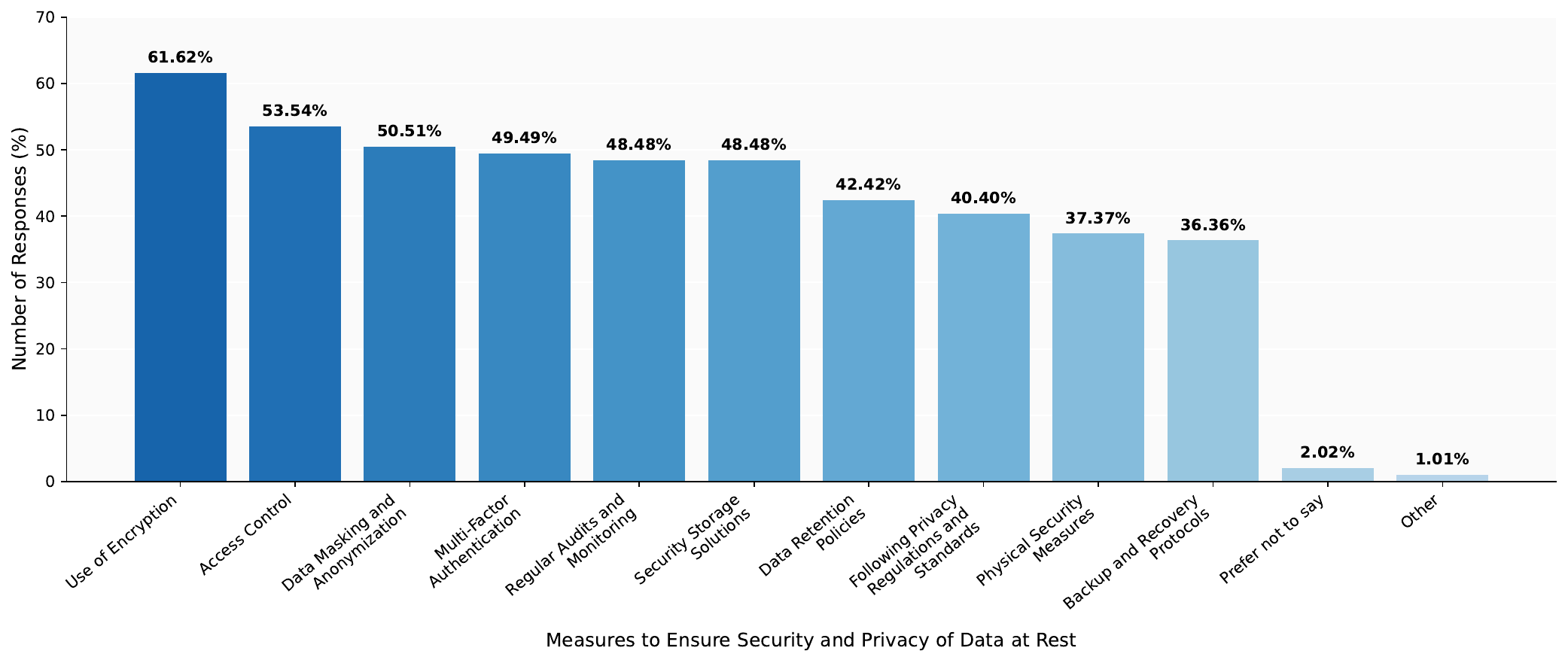} 
        \caption{Group B - Methods For Ensuring Security and Privacy of Data at Rest}
    \label{fig:b_data_security_privacy_atrest}
\end{figure}

\subsubsection{Ethical Considerations in Training AI Models} 
We asked Group B participants questions about the ethical considerations around training their AI models. 77.43\% mentioned that they collect, curate, and manage data for training purposes, while the rest were either ``Unsure'' or said ``No.'' 84.6\% (of those 77.43\%) reported that they train their models in-house, and 62.3\% (of the 77.43\%) mentioned they need ``extensive'' or ``significant resources'' to train these models.

Regarding whether their AI systems use personal data for training or testing, 43.81\% said ``Yes,'' while 35.4\% and 13.72\% said ``No'' or ``Unsure,'' respectively. We asked the 101 participants who answered ``Yes,'' which methods they use to ensure the security and privacy of the data at rest. As Figure \ref{fig:b_data_security_privacy_atrest} shows, the responses varied, but the top two methods cited were security-related (i.e., ``use of encryption'' and  ``access control''), followed by ``data masking and anonymization'' as the third most common approach. Policy-related solutions, such as ``data retention policies'' and ``following privacy regulations and standards,'' were less common. Interestingly, most participants did not consider backup and recovery protocols (36.36\%). 

\begin{figure}[!htbp]
        \centering
        \includegraphics[width=0.85\textwidth]{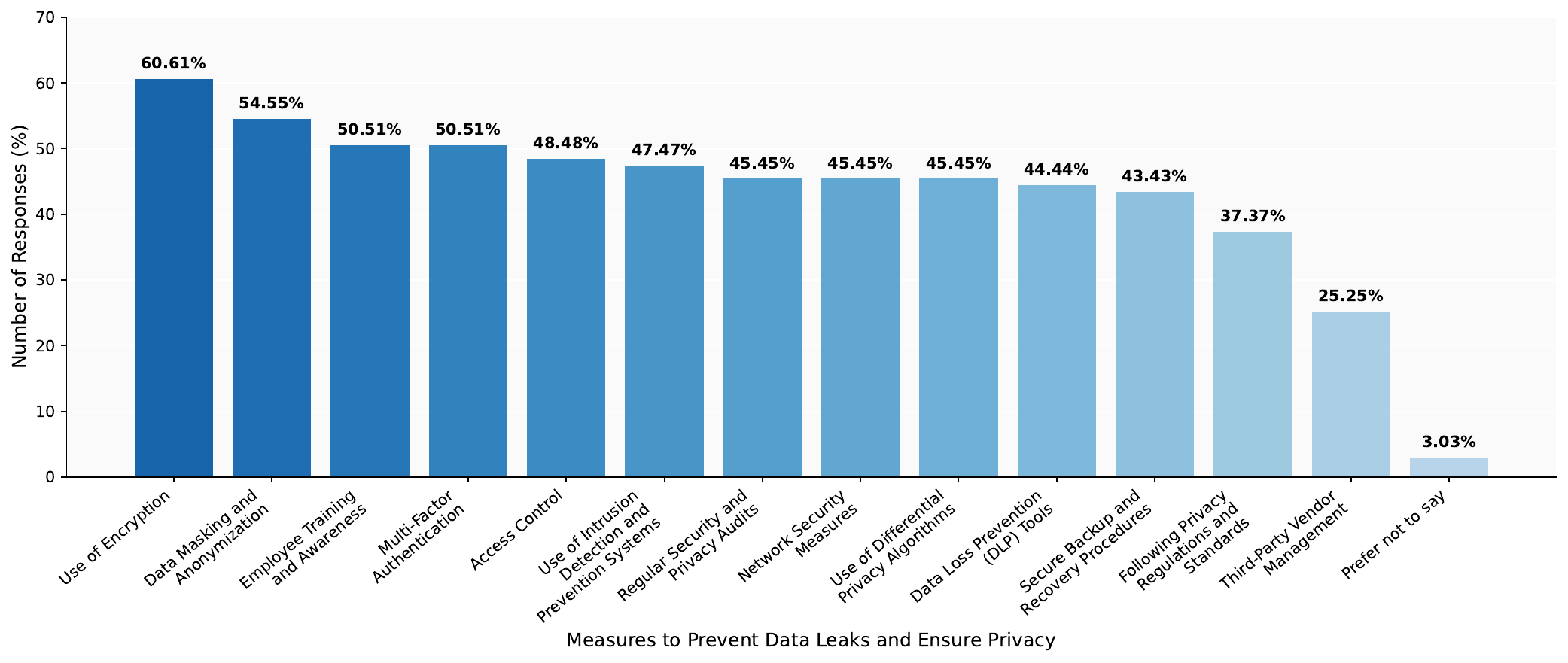} 
        \caption{Group B - Methods For Ensuring Private Data Does Not Get Leaked}
        \label{fig:b_data_privacy_leak}
\end{figure}

We then followed up with the same group to learn about their practices for ensuring data privacy and preventing leaks. We received diverse responses, as shown in Figure \ref{fig:b_data_privacy_leak}; however, ``use of encryption'' remained the most commonly cited method. ``Data masking and minimization'' rose from the third to the second most frequently cited method. One interesting observation is the emphasis on ``employee training and awareness,'' which about half of the participants consider essential. The least popular practice is ``third-party vendor management.'' These findings highlight an area for improvement, given that third-party code is a significant source of data breaches and privacy risks. In contrast, privacy regulations and standards are key in encouraging developers to follow best practices~\cite{prybylo2024evaluating}. Table \ref{tab:rq4_summary} shows the summary of the result.


\begin{tcolorbox}[colback=gray!10,colframe=black!30,boxrule=0.1pt, boxsep=0.1pt]
\textbf{RQ4 Summary - Mitigation Strategies:} Across roles, AI ethics integration varies considerably in approach and depth. AM and RA roles who are more familiar with AI ethics principles are more likely to integrate them during both documentation and design. AD and QA roles employ complementary methods to mitigate biases in models, results, and training data. ADs value collaboration with ethicists and legal experts, while QAs focus on secure testing environments, test data anonymization, and conducting PIAs during testing. ISec roles, by contrast, lean heavily on traditional security measures, such as access control, authentication, and encryption, reflecting a narrower interpretation of ethics. QAs and ADs are generally less involved in ethics impact assessments than those in AM, RA, and ISec roles. AI researchers use institutional guidelines, peer review, and academic literature, yet often cite a lack of precise guidance as a barrier. 
Overall, enhancing communication and collaboration across development teams may support a holistic approach to AI ethics, as suggested by AI ethicists.
\end{tcolorbox}%

\begin{table}[!htbp]
\caption{Summary of Key Risk Mitigation Findings by Role (RQ4)}
\label{tab:rq4_summary}
\small
\begin{tabularx}{\textwidth}{X p{2.5cm} p{1.5cm}}
\toprule
\textbf{Finding} & \textbf{Result} & \textbf{Evidence} \\
\midrule
\textbf{AM Roles (Managers, Administrative)} & & \\
- Top strategies: providing AI ethics training for employees and implementing ethics guidelines for AI development and deployment & 60\% each & H5a: $p=0.008^*$; Table~\ref{tab:am_ai_ethics_actions} \\
- Responsibility is shared between the AI development team (64\%) and upper management (64\%) & & Table~\ref{tab:ai_ethics_responsibility} \\
\midrule
\textbf{RA Roles (Requirements Analysts)} & & \\
- 76\% include ethics in requirements documentation at least ``sometimes''; 53\% report their company conducts ethics-related risk and impact assessments & 76\%; 53\% & H5b: $p=0.013^*$; Table~\ref{tab:ai_ethics_frequency-req} \\
- Prioritization varies by region: N.~America and EU+EEA+UK emphasize data protection (67\%/89\%), while other regions prioritize democracy \& rule of law (63\%) & Region-dependent & Sec.~4.4.2 \\
\midrule
\textbf{AD Roles (Developers, Architects, Data Scientists)} & & \\
- Top strategies focus on technical solutions: evaluate model results (60\%), include diverse training data (49\%), regular audits and testing (44\%), regular data cleaning (43\%) & 60\% top & H5c: $p<0.0001^{**}$; Fig.~\ref{fig:AD:mitigation_strategies} \\
\midrule
\textbf{QA Roles (Quality Assurance, IT Maintenance)} & & \\
- Top strategies complement AD approaches: testing results for vulnerable groups (67\%) and user testing with diverse groups (60\%)---approaches that are among the least commonly selected by AD roles & 67\%; 60\% & H5e: $p=0.827$; Fig.~\ref{fig:QARoles:Techniques} \\
\midrule
\textbf{ISec Roles (Information Security/Privacy)} & & \\
- Rely heavily on traditional security measures: access controls and user authentication (92\%) and encryption (58\%); 50\% implement incident response plans and conduct regular audits & 92\%; 58\% & Fig.~\ref{fig:ISec:measures_mitigating_concerns}; Table~\ref{tab:ISec:AI_privacy_impact} \\
- 46\% report that AI ``somewhat'' or ``significantly'' compromises data privacy within their company & & Table~\ref{tab:ISec:AI_privacy_impact} \\
- Key gap: only 33\% collaborate with AI developers to identify security/privacy risks, and only 33\% integrate privacy-enhancing technologies & 33\% & Fig.~\ref{fig:ISec:measures_mitigating_concerns} \\
\midrule
\textbf{AR+AE Roles (Researchers, Ethicists)} & & \\
- $\sim$85\% incorporate AI ethics principles into their research at least ``sometimes''; top guidance sources: institutional guidelines \& personal judgment & & Fig.~\ref{fig:AR:guidance}; Fig.~\ref{fig:AR:adherence} \\
- Compliance is monitored through peer reviews (59\%), feedback from ethics committees (55\%), and internal audits (55\%) & & Fig.~\ref{fig:AR:adherence} \\
\bottomrule
\end{tabularx}
\begin{tablenotes}
\small
\item $^{*}p < 0.05$; $^{**}p < 0.0019$ (Bonferroni). H5a--H5e test the association between average principle familiarity and role-specific outcomes (see Table~\ref{Table:H4}).
\end{tablenotes}
\end{table}

\section{Discussion} \label{sec:discussion}

\subsection{Summary of Research Findings} 

Our findings on the \textit{general perception of AI (RQ1)} surrounding AI systems indicate that AI is primarily seen as a tool for process automation, performance enhancement, and data synthesis. Those without experience in AI development tend to be more optimistic about emerging generative AI applications but are also concerned about ensuring security and privacy in both the use and development of AI tools. Prioritizing privacy aligns with Vakkuri et al.~\cite{vakkuri2020aiethics}, who found that developers often perceive AI ethics risks related to existing trends in software development (e.g., energy consumption or privacy). On the topic of AI governance initiatives, participants were divided. While the majority saw at least a somewhat positive impact of regulations, some viewed them as hindrances to progress that tend to favor established corporations. 
Our results emphasize the importance of better communicating the AI governance initiatives to developers and for developers to be more proactive in understanding and implementing such initiatives. 
 
We observed considerable variation among AI practitioners in their understanding and implementation of ethical principles and governance initiatives (\textit{familiarity and predicting factors (RQ2/3)}). This aligns with prior research emphasizing the importance of involving diverse stakeholders in ethical decision-making processes (e.g., Pant et al.~\cite{pant2024ethics} and Sanderson et al.~\cite{sanderson2023ai}). Our study reinforces the critical role of explainability and transparency in establishing trust in AI systems, echoing findings from Rui et al.~\cite{rui2023investigating} and others. Notably, AI managers and requirements analysts exhibit a greater awareness of, and emphasis on, integrating ethical guidelines compared to developers, quality assurance personnel, and information security and privacy teams. Individuals working in medium-sized companies, academic institutions, or government agencies tend to report higher familiarity with principles and initiatives than those employed by small businesses, startups, or large multinational corporations. 
Among all ethical principles, data protection and the right to privacy rank highest in familiarity across roles and locations, and are applied the most frequently. 

LASSO identified familiarity with the EU AI Act (+15.39\%) as the strongest predictor of familiarity with AI ethics principles. Mixed-effects models confirmed that while AI development experience significantly predicts familiarity, it accounts for only 6.3\% of the variance. In contrast, for \textit{how often practitioners consider ethics in their work}, familiarity with the AI governance initiatives is the dominant predictor (+0.281), explaining 17.4\% of the variance, whereas experience was not selected. LASSO analysis also revealed that positive perceptions of regulation predict both familiarity with principles (+15.23\%) and the frequency of ethics consideration (+22.6\%). These findings suggest that educating practitioners about AI governance initiatives may effectively enhance both familiarity with and consideration of AI ethics principles, with the EU AI Act showing the strongest association with these benefits. Overall, developers maintain a positive outlook on the role of ethics in the future of AI systems. Although some expressed concerns that ethics and regulatory frameworks could hinder innovation and competitiveness - particularly for small businesses - the majority remain optimistic about integrating ethics principles and governance into AI development practices.

Regarding \textit{practice and experience (RQ2/3)}, our findings align with Pant et al. \cite{pant2024ethicsai}, revealing that roles and demographic factors, such as company size, geographic location, and gender, significantly influence how AI practitioners implement various aspects of AI ethics. North American participants are more comfortable with technical solutions and look to external support for implementing AI ethics, whereas those from the EU+EEA+UK are more open to policy-based solutions rooted in regulations and involving diverse stakeholders in decision-making. 
While these role-based differences exist, random effects analysis revealed that professional role accounts for only 2.9\% of variance in knowledge but 15.5\% in practice frequency, indicating that organizational and role-based norms shape behavior more significantly than knowledge. Participants in AM roles demonstrate the strongest commitment to principles related to respecting human rights, in contrast to those in QA and ISec roles. 
Notably, AM, QA, and ISec roles are critical to ensuring trust, safety, and security in AI systems and are among the most likely to engage with ethical principles in practice. Participants in AR and AE roles, particularly those working in academic or research institutions, emphasize principles such as accountability and responsibility, fairness and justice, and transparency and explainability. Incorporating their perspectives can help development teams address complex or less visible ethical challenges and contribute to designing AI systems that uphold ethical standards. Finally, our findings highlight the importance of gender diversity in AI development teams. Female participants expressed greater concern for ethical principles throughout the AI development process and reported applying them more frequently than their male counterparts.

Concerning \textit{risk mitigation strategies (RQ4)}, we observed various approaches employed by participants across different roles. Overall, those with AI development experience identified key strategies, such as data cleaning to reduce or eliminate bias, ongoing performance monitoring, rigorous testing and validation, and investing in education and training. Role-specific patterns also emerged. Participants in AM roles emphasized the importance of ethics training, the implementation of AI ethics guidelines, and collaboration with external experts. Those in AD and QA roles focused on improving training data, reducing bias and discrimination in outputs, and increasing diversity. Their strategies often included evaluating model results, incorporating representative training data, conducting bias audits, and testing outputs for vulnerable groups. ISec professionals, while generally more cautious about AI’s impact on data privacy, still contributed to integrating privacy and security principles into their organizations’ AI development processes. Ethicists felt their companies could do more to embed and promote AI ethics. Researchers reported regularly considering ethical concerns, often leveraging peer review and ethics committee feedback; however, they also cited challenges, such as unclear guidelines and the tension between ethical goals and research objectives. These findings highlight the need for both technical and organizational strategies to address AI ethics risks and underscore how socio-economic and professional diversity influences the perception and implementation of these principles.

\subsection{Implications for Development Practice}

Our results carry several implications for software engineering practice:

\subparagraph{\textbf{Requirements Engineering}} RAs in North America and the EU+EEA+UK prioritize ``data protection and the right to privacy'' in requirements documentation, reflecting the influence of GDPR and CCPA. But, only 76.5\% of RAs consider ethics at least ``sometimes'' in requirements, suggesting a need for systematic ethics requirements checklists. Particularly for principles like ``democracy and rule of law'' that are less frequently considered. The EU AI Act's requirements for risk classification (Article 6), technical documentation (Annex IV), and conformity assessment (Article 43) provide a concrete framework that could be translated into requirements templates.

\subparagraph{\textbf{Design and Development}} AD roles predominantly use technical mitigation strategies (evaluating model results, diverse training data, regular audits). The relatively low adoption of policy-related solutions such as ``peer-review and collaborative development'' (34.06\%) and ``ethics impact assessments'' (33.33\%) suggests that development processes would benefit from integrating structured ethical review points---analogous to security review
gates---at key design milestones.

\subparagraph{\textbf{Testing and Quality Assurance}} QA roles complement AD roles by focusing on ``identifying and testing results for vulnerable groups'' (66.67\%) and ``user testing with diverse groups'' (60\%)---strategies less commonly selected by AD roles. This complementarity underscores the value of cross-role collaboration during testing, yet QA professionals are generally less involved in ethics-related impact assessments than those in AM and RA roles, representing an opportunity to expand their mandate.

\subparagraph{\textbf{Governance and Compliance}} Only $\sim$50\% of AM roles conduct regular ethics audits, despite this being a recommended practice in OWASP Top 10 for LLMs~\cite{owasp2025llm}. The NIST AI Risk Management Framework's ``Govern, Map, Measure, Manage'' functions map to several of our observed mitigation strategies: ``clean data for biases'' and ``monitor AI system performance'' correspond to Measure and Manage, while ``establish AI ethics guidelines'' and ``create contingency plans'' map to Govern. Organizations could use this mapping to assess coverage gaps.

\subsection{Research Directions} 

Building on our findings and the gaps identified in related work, future research should explore interdisciplinary frameworks that bridge technical solutions with organizational processes. We propose five key research directions: 

(1) Development of standardized threat modeling for ethics in AI systems. Software engineering and privacy research have presented threat modeling frameworks, such as LINDDUN~\cite{deng2011privacy}, which can capture privacy risks and identify countermeasures early in development. Similar frameworks can be applied to measure and mitigate ethical and regulatory risks in AI systems, enabling developers to understand the risks and threats posed by their applications and providing actionable mitigation methods. 

(2) Development of practical tools and automated frameworks to support ``ethics by design'' in AI development. This addresses Pant et al.~\cite{pant2024ethicsai}'s finding that practitioners cite lack of practical tools as a major barrier, and Khan et al.~\cite{khan2023aiethics}'s observation that vague principles are difficult to operationalize. 

(3) Development of tools and methods for the automated translation of AI ethics regulations into software requirements. 
This includes approaches for extracting legal requirements, automating compliance analysis to detect potential violations, and encouraging practitioners to select ethical design solutions. As Vakkuri et al.~\cite{vakkuri2020aiethics} identified, AI features tend to be treated like regular software features. Translating policy into documentation that aligns with development workflows, such as requirements and user stories~\cite{Baldwin2025}, can help bridge the gap in developer understanding. 

(4) Investigation and development of nudging tools embedded within development environments. Drawing on privacy nudging approaches~\cite{balebako2014improving}, future research could focus on developing tool-supported methods and approaches that could detect potential AI ethics risks in real-time and prompt consideration of specific mitigation mechanisms. 

(5) Tools that enable automated testing of ethics requirements in AI applications. By leveraging recent advancements in LLM-based code analysis, new tools may automatically detect risks, such as biases in training pipelines, explainability in documentation or pull requests, and prompt injection risks in user interfaces. Extending such tools with information retrieval opens the door to analyzing policy alongside applications and providing individualized solutions. 

(6) Investigation of regulatory and cultural influences. Building on prior work~\cite{rakova2021where,vakkuri2022how}, future work could explore how different regulatory environments and organizational cultures shape ethical decision-making. This could inform best practices, such as those outlined by OWASP \cite{owasp2025llm}, help create supplementary documentation (similar to that of the EDPB \cite{EDPB_website}), and enable context-specific interventions.


\subsection{Educational Takeaways} 

Our work underscores the importance of incorporating ethics education for all roles within AI development teams. Educational programs should move beyond traditional technical training to include modules on ethical risk assessment, transparency, communication, and the practical implementation of governance frameworks. Building on prior studies that highlight the lack of formal training in ethical AI practices~\cite{griffin2024ethical,agbese2023implementing}, our findings suggest that curriculum designers and organizational leaders should develop role-specific training programs that integrate both technical measures (e.g., bias audits and explainable AI techniques) and collaborative, interdisciplinary approaches to ethical decision-making. Given the evolving nature of AI ethics, it is essential to equip AI practitioners with educational modules that promote lifelong learning. As Balebako et al. \cite{balebako2014improving} suggest, with appropriate guidance and well-designed educational tools, AI practitioners will be better prepared to navigate the complex trade-offs among innovation, trust, and compliance. 

\subsection{Limitations}
\label{sec:limitations}

Similar to any survey study, our analysis relies on participants' self-reported data. Thus, it may be affected by \emph{self-report bias}, \emph{recall bias}, and/or \emph{social desirability bias}. During the consent process, we informed the participants that the survey pertained to AI practices, but we intentionally did not mention ethics to minimize the impact of self-selection biases. 
The Prolific user base may not fully represent the diverse population of AI practitioners; thus, the survey may face \emph{recruitment bias}. We diversified our sample by including participants from various AI-related forums, as well as those from Prolific. We used multiple screening questions to ensure that recruited participants have experience in AI development activities (Section \ref{sec:recruitment-filtering}). We adopted a conservative process to remove participants who appeared to provide \emph{AI-generated answers} by carefully scrutinizing the write-in responses and investigating semantic qualities of LLM vs. human responses \cite{tang2024science,munoz2024contrasting,guo2023how}. However, AI-generated responses may have influenced the study's outcomes, and we may have excluded some professionals. 
We carefully framed our questions to avoid prompting biased responses and began the survey with general questions about AI practices to minimize the \emph{framing bias}. 
Although we aimed to minimize the risks associated with answering follow-up write-in questions, some biases may still exist in those questions. We employed statistical analyses to ensure the broad applicability of our findings. To control for Type I errors in the presence of multiple hypothesis tests, we reported our results after employing the Bonferroni correction. Finally, we frame our threats to validity using the classification of Wohlin et al.~\cite{Wohlin2012ExperimentationIS}, covering construct, internal, external, and conclusion validity. 

\subparagraph{\textbf{Construct Validity}} To mitigate \textit{construct validity} threats, such as variable definitions that may be interpreted differently across demographics, and social desirability bias, we employed validated taxonomies (e.g., NIST Human-AI Taxonomy, AI ethics principles defined in~\cite{worsdorfer2023aibill}) and avoided mentioning ethics during recruitment. However, our measures rely on \emph{self-reported familiarity} rather than objective knowledge tests, which may overestimate actual understanding. The internal consistency of our Likert-scale instruments is supported by Cronbach's alpha values of 0.916 (principle familiarity) and 0.950 (governance initiative familiarity); see Section~\ref{sec:method-validation}.

\subparagraph{\textbf{Internal Validity}} To reduce \textit{internal validity} threats, including confounding variables, we employed statistical methods such as mixed-effects models and LASSO regression to isolate predictive factors across demographics. As a cross-sectional survey, our study cannot establish causation, only associations. The question ordering (general AI practices before ethics-specific questions) may have influenced later responses through priming effects. We adopted a conservative process to remove participants who appeared to provide \emph{AI-generated answers} by carefully scrutinizing the write-in responses and investigating semantic qualities of LLM vs.\ human responses~\cite{tang2024science,munoz2024contrasting,guo2023how}. However, AI-generated responses may have influenced the study's outcomes. Additionally, our AI-generated response filtering, while conservative, may have introduced bias by excluding some genuine responses or retaining some AI-generated ones.

\subparagraph{\textbf{External Validity}} The \textit{external validity} was strengthened by recruiting diverse participants from multiple sources, including Prolific and online forums across 43 countries, and filtering for relevant AI development experience via Prolific. The Prolific user base may not fully represent the diverse population of AI development practitioners; thus, the survey may be subject to \emph{recruitment bias}. We used multiple screening questions to ensure that recruited participants have experience in AI development activities (Section~\ref{sec:recruitment-filtering}). Regarding dropout, 120 of 450 Prolific starters ($\sim$27\%) and 154 of 356 forum starters did not complete the survey. Those who dropped out may be inherently less engaged with ethics topics than completers, potentially skewing our results toward more positive perceptions of ethical practices. 
Not framing the survey as ethics-related during recruitment limits self-selection into the study, but it does not address self-selection at the completion stage, and we did not collect ethics-attitude data from non-completers that would allow us to correct for it. The most likely effect is a completer sample skewed toward more ethics-aware practitioners, which would shift the reported perceptions, familiarity, and practice frequencies in a more positive direction. We note, however, that our central conclusions are comparative, differences across roles, demographics, and levels of governance familiarity, and a uniform positivity shift among completers would not readily explain the role- and demographic-specific gaps we observe.
Notable regional imbalances limit generalizability: 48.9\% of participants are from North America, while Central/South America (6.1\%), Asia (3.9\%), and the Middle East (0.7\%) are underrepresented. The Prolific platform skews toward English-speaking participants with programming skills. Small subgroup sizes for certain roles (AE: $n=5$, ISec: $n=19$) limit the reliability of role-specific conclusions for these groups.

\subparagraph{\textbf{Conclusion Validity}} To enhance \textit{conclusion validity}, we filtered out low-quality or AI-generated responses and applied the Bonferroni correction to control for Type I errors across 16 hypothesis tests, with an adjusted significance threshold of $\alpha = 0.0019$. This conservative approach increases the risk of Type II errors, particularly for smaller subgroups. We report both corrected and uncorrected p-values to enable readers to assess the robustness of findings. We employed statistical analyses to ensure the broad applicability of our findings. Sample sizes per sub-analysis vary considerably (e.g., AE: $n=3$ in Group B), and conclusions for small subgroups should be interpreted with appropriate caution.
\section{Conclusion} 
\label{sec:conclusions}

This paper presented a large-scale mixed-method study examining the perceptions, practices, and risk mitigation strategies related to AI ethics across diverse roles in AI development teams. Our findings reveal that explainability, transparency, and role-specific familiarity with ethical guidelines foster trust and accountability in AI systems, promoting the consideration and confidence in ethical principles during development. We observed differences in how various groups, from AM and RA to AD, QA, and ISec teams, understand and implement AI ethics principles. These disparities, which are further influenced by organizational size, geographic region, and gender, underscore the need for education and enhanced interdisciplinary collaboration in the development of ethical AI. While our study advances the current understanding by highlighting both technical and organizational challenges, it also points to several avenues for future work, including longitudinal analyses, deeper investigations into the regulatory impacts on AI practices, and the development of frameworks for collaboration among the various demographics that comprise AI development teams.

\section{ Declarations}

\subsection{Funding} 

This research was supported by NSF Award \#2238047 and a Maine Business School Research Award. 

\subsection{Ethical Approval} 

This research adheres to our university's ethical guidelines and was approved by our Institutional Review Board (IRB) with the approval number `IRB 2024-07-21'.

\subsection{Informed Consent}

All participants agreed to a thorough consent form that included information about the investigators, risks, benefits, compensation, and confidentiality. All participants were informed about their voluntary participation, with the emphasis on their right to withdraw at any time. No personally identifiable information was collected, and measures were in place to ensure the anonymity, confidentiality, and security of responses. The contact information of all investigators and the IRB team was also included. No participants contacted the investigators or the IRB about the study or the compensation. 
The informed consent forms can be found here: https://tinyurl.com/3b52cfux.

\subsection{Author Contributions} 
The first author conducted the study and the analysis of the results under the supervision of the PIs. The second and the third authors are the PIs of the project with equal contributions to the project.

\subsection{Data Availability Statement}
We have published a replication package, which contains all responses and the code for any analysis presented in this paper. The replication package is available at: https://zenodo.org/records/17905482.

\subsection{Conflict of Interest} There is no conflict of interest for this project.

\subsection{Clinical Trial Number} `Clinical trial number: not applicable.’


\backmatter





\bmhead{Acknowledgements}

We would like to thank Ersilda Cako, an undergraduate student at the University of Maine, for her valuable assistance in developing the survey and conducting the open coding analysis of the qualitative data.

\bibliography{references}

\begin{appendices}



\section{Definitions, Taxonomy, and Codebook}
\label{Appendix:Defs}

\begin{table}[h]
    \centering
    \caption{Definition of AI Ethics Principles}
    \begin{tabular}{p{0.3\textwidth}p{0.7\textwidth}}
        \toprule
        \textbf{AI Ethics Principles} & \textbf{Definition} \\
       \midrule
        Respect for human rights & A human-centered approach to AI that preserves individual agency, control, and oversight, ensuring ethical business practices, human review of automated decisions, and promoting societal well-being. \\
        Data protection \& privacy & Involves safeguarding human dignity, autonomy, and anonymity by ensuring control over data, access limitations, communication freedoms, and the right to data stewardship, minimization, correction, erasure, and privacy by design/default, supported by strong privacy laws. \\
        Harm prevention \& beneficence & Includes the robustness, safety, and security of AI systems, preventing misuse, ensuring the reliability and reproducibility of research methods and applications, addressing unintended consequences, and establishing mandatory safety standards, audits, and certifications. \\
        Non-discrimination \& freedom of privileges & Implies preventing all forms of discrimination, manipulation, negative profiling, and minimizing algorithmic biases. \\
        Fairness \& justice & Emphasizes four types—data, design, outcome, and implementation—and promotes open innovation, accessibility, inclusion, and competitive market practices to prevent discriminatory business behaviors. \\
        Transparency and explainability of AI systems & Encompasses explainability, open-source data and algorithms, the right to information, notification of AI decision making and human interaction, and regular reporting. \\
        Accountability \& responsibility & Includes verifiability, replicability, evaluation and assessment, creation of oversight bodies, the ability to appeal, remedies for automated decisions, legal responsibility, and accountability-by-design. \\
        Democracy \& rule of law & Involves embedding technologies within democratic oversight, judicial governance, and public engagement, ensuring inclusion, stakeholder dialogue, and a “community-in-the-loop” approach. \\
        Environmental \& social responsibility & Implies reducing ecological impacts such as carbon emissions and electronic waste, supporting sustainable AI, and ensuring social sustainability through human rights due diligence, stakeholder impact assessments, and workforce education and training. \\
        \bottomrule
        \label{tab:AIEthics-Definition}
    \end{tabular}
\end{table}

\begin{table}[h]
\caption{Taxonomy of Human-AI Activities}
\label{Table:NIST_taxonomy}
\centering
\begin{tabular}{lp{6cm}p{4.5cm}}
  \toprule
  \textbf{Human-AI Activity} &  \textbf{Description} &  \textbf{Example AI Outcomes} \\
  \midrule
  Content creation & Generating new artifacts such as video, narrative, software, synthetic data. & subtitle creation, text-to-image  \\ 
  Content synthesis & Combining and/or summarizing parts, elements, or concepts into a coherent whole. & converting doctors' unstructured notes; summarizing a book  \\ 
  Decision making & Selecting a course of action from among possible alternatives in order to arrive at a solution. & buy/sell financial decisions  \\ 
  Detection & Identifying, by careful search, examination, or probing, the existence or presence of (something). & detect cybersecurity threats  \\ 
  Digital assistance & Acting as a personal agent for understanding and responding to commands and questions, and carrying out requested tasks in a conversational manner. & reminders from smart assistants (e.g., Siri, Amazon Echo, Google Assistant, Alexa) \\ 
  Discovery & Finding, recognizing, or unearthing something for the first time. & drug discovery and production  \\ 
  Image analysis & Recognizing attributes within digital images to extract meaningful information. & medical diagnostics  \\ 
  Information retrieval & Finding information about specific topics of interest. & speed the search for stable proteins used in drug development  \\ 
  Monitoring & Observing, checking, and watching over the process, quality, or state of (something) over time to gain insights into how (something) is behaving or performing. & wildlife monitoring  \\ 
  Performance improvement & Improving the quality and efficiency of the intended outcomes. & graph analytics; increasing efficiency and scalability for graph computing \\ 
  Personalization & Designing and tailoring (something) to meet an individual's characteristics, preferences, or behaviors. & sales content personalization and analytics  \\ 
  Prediction & Forecasting the likelihood of a future outcome. & sales forecasting; weather forecasting  \\ 
  Process automation & Performing repetitive tasks, removing bottlenecks, reducing errors and loss of data, and increasing efficiency of a process. & automating administrative tasks  \\ 
  Recommendation & Suggesting or proposing a manageable set of viable options to aid decision-making. & customer service response suggestions; purchase recommendations; content recommendations  \\ 
  Robotic automation & Using physical machines to automate, improve, and/or optimize a variety of tasks. & intelligent robots in surgery  \\ 
  Vehicular automation & Automating physical transportation of goods, instrumentation and/or people. & self-driving cars/trucks/trains; drones; spacecraft; airplanes  \\ 
  \bottomrule
\end{tabular}%
\end{table}

\begin{table}[!htbp]
\centering
\caption{AI Ethics Principles Risk and Mitigation Codebook}
\label{Table:risk_mitigation_codebook}
\begin{tabularx}{\textwidth}{p{0.2\textwidth}p{0.32\textwidth}X}
\toprule
\textbf{Theme} & \textbf{Codes} & \textbf{Example} \\ 
\midrule
Bias and Discrimination & Data Diversity, Non-Discrimination, Bias Mitigation, Bias Detection & ``Lack of Diverse Representation: Insufficient diversity in the development team or in data can result in overlooking fairness issues." \\ \hline
Harm Prevention and Beneficence & AI Replacing Jobs & ``While AI systems in finance are intended to optimize productivity and profitability, if they are not properly controlled, they might cause harm."\\ \hline
Data Security and Privacy & Technical Security, Data Leak, User Consent,  Transparency & ``AI powered mass surveillance infringing on peoples rights of privacy." \\ \hline
Transparency and Explainability & Open Documentation, Explainability Features, Transparency, Unclear Data Collection & "Clear implementation of documents AI algorithms." \\ \hline
Environmental and Social Responsibility & Energy Efficiency, Resource Management & ``Using unnecessary resources that arent sustainable and do not help the environment. Or even using materials that are harmful to the environment." \\ \hline
Fairness and Justice & - & ``We audit our data collection and cleaning methodology to ensure fairness when creating models." \\ \hline
Accountability and Responsibility & Lack of Accountability, Quality Assurance, Regular Audits, Defining Roles and Responsibilities & ``At the outset, I define clear roles and responsibilities for all team members involved in the AI project. This includes identifying who is responsible for data collection, model training, validation, and deployment, ensuring that everyone understands their accountability in the process." \\ \hline
Respect for Human Rights & -  & ``Protect the rights of research participants." \\ \hline
Democracy and Rule of Law & - & ``To ensure the Democracy and Rule of Law principle in AI systems, I engage diverse stakeholders, adhere to ethical frameworks, ensure transparency, comply with laws, and establish monitoring and redress mechanisms." \\ \hline
Ethical Guidelines & Establishing AI Guidelines, Following Regulations, Ethical Reviews, System Monitoring, Conducting Impact Assessments & ``It is important that intelligent systems should have some railguards when it comes to Human-World-models. Need to encode certain laws regarding democratic solutions." \\ \hline
Stakeholder Involvement & User Feedback, Engage Diverse Stakeholders & ``Lack of Diverse Representation: Insufficient diversity in the development team or in data can result in overlooking fairness issues." \\ \hline
AI Design & Non-Transparent AI, Using Controlled AI Models, Monitor System Outputs, Value Misalignment, AI Misuse & ``Develop interpretable AI models to ensure that key decision-making processes can be scrutinized and understood." \\ 
\bottomrule
\end{tabularx}
\label{Table:risk_mitigation_codebook}
\end{table}


\section{Participants Demographcis}
\label{appendix:dem}

\begin{sidewaystable}
    \centering
    \caption{Breakdown of Participants' Demographics}
    \label{Table:Demographics}
        \begin{tabular*}{\textheight}{@{\extracolsep\fill}|l|c|c|c|c|c|} 
            \toprule
            \textbf{Gender} & Female (31.2\%) & Male (67.6\%) & Non-Binary (0.7\%) & Other (0.0\%) & PnS (0.5\%)\\
           \hline
            \textbf{Age} & 18-25 (13.0\%) & 26-35 (50.5\%) & 36-45 (24.4\%) & 46+ (11.8\%) & PnS (0.3\%)\\
            \hline
            \textbf{Education} & High School (8.7\%) & BSc. (40.0\%) & MSc./Graduate Cert. (41.4\%) & Ph.D. (8.0\%) & Other (1.9\%)\\
           \hline
            \textbf{Degree Field} & CS/ECE (31.4\%) & SWE/DS (21.2\%) & IT/Info Sec/Privacy (24.1\%) & Business (13.1\%) & Other (10.2\%)\\
            \hline
            \textbf{Company Size} & 1-5 (9.2\%) & 6-20 (12.9\%) & 21-50 (17.5\%) & 51-100 (14.5\%) & 100+ (45.9\%)\\
            \hline
            \textbf{Company Type} & Multi-national (31.2\%) & Startup/Small (32.7\%) & Academic/Research (13.3\%) & Government (8.0\%) & Other (14.8\%)\\
            \hline
            \textbf{Location} & N. America (48.9\%) & EU/UK/EEA (25.8\%) & C/S America (6.1\%) & World (19.0\%) & Other (0.2\%)\\
           \hline
            \textbf{AI Dev. Exp.} & None (10.1\%) & 1-2 yrs (30.0\%) & 2-5 yrs (36.0\%) & 5-10 yrs (15.7\%) & 10+ yrs (8.2\%)\\
            \hline
            \textbf{Company Exp.} & - & 0-2 yrs (11\%) & 2-5 yrs (49.4\%) & 5-10 yrs (23.0\%) & 10+ yrs (16.6\%)\\
            \bottomrule
        \end{tabular*}
\end{sidewaystable}

\newpage


\section{Statistical Tests and LASSO Analysis}
\label{Appendix:stats}

\begin{table}[!htbp]
\centering
\caption{LASSO Regression Predictor Categories by Participant Group}
\small
\begin{tabular}{@{}lcc@{}}
\toprule
\textbf{Predictor Category} & \textbf{Group A} & \textbf{Group B} \\
\midrule
\multicolumn{3}{@{}l}{\textit{Demographics}} \\
\quad Age & $\bullet$ & $\bullet$ \\
\quad Gender & $\bullet$ & $\bullet$ \\
\quad Education Level & $\bullet$ & $\bullet$ \\
\quad Degree Field & $\bullet$ & $\bullet$ \\
\quad Company Size & $\bullet$ & $\bullet$ \\
\quad Company Type & $\bullet$ & $\bullet$ \\
\quad Location & $\bullet$ & $\bullet$ \\
\quad Role & $\bullet$ & $\bullet$ \\
\quad Current Job Experience & $\bullet$ & $\bullet$ \\
\quad AI Development Experience & -- & $\bullet$ \\
\midrule
\multicolumn{3}{@{}l}{\textit{Familiarity}} \\
\quad AI Ethics Principles (individual + average) & $\bullet^{c}$ & $\bullet^{c}$ \\
\quad Governance Initiatives (individual + average) & $\bullet^{c}$ & $\bullet^{c}$ \\
\midrule
\multicolumn{3}{@{}l}{\textit{AI Practice}} \\
\quad Company Builds/Deploys AI & $\bullet$ & $\bullet$ \\
\quad Currently Developing AI & $\bullet$ & $\bullet$ \\
\quad AI Usage Frequency & $\bullet$ & $\bullet$ \\
\quad AI Applications (multi-label) & $\bullet^{a}$ & $\bullet$ \\
\quad AI Effects/Impacts (multi-label) & $\bullet^{a}$ & $\bullet$ \\
\quad AI Tools/Technologies & -- & $\bullet$ \\
\quad Ethics Consideration Frequency & -- & $\bullet$ \\
\quad Mitigation Strategies & -- & $\bullet$ \\
\midrule
\multicolumn{3}{@{}l}{\textit{Perception}} \\
\quad Regulation Impact & $\bullet$ & $\bullet$ \\
\quad Regulation Negative Impact & $\bullet$ & $\bullet$ \\
\quad Principles Considered & -- & $\bullet$ \\
\quad Principles at Risk & $\bullet^{a}$ & $\bullet$ \\
\midrule
\textbf{Total Features} & \textbf{68--78}$^{b}$ & \textbf{141--142} \\
\textbf{Sample Size (N)} & \textbf{76--90} & \textbf{141--178} \\
\bottomrule
\multicolumn{3}{@{}p{0.85\linewidth}@{}}{\footnotesize
$^{a}$Included in Track A but excluded in some models to prevent overfitting.
$^{b}$Feature count varies; multi-label questions are one-hot encoded.
$^{c}$Included only when not the target variable and when circularity is not a concern (see Sec.~3.2).} \\
\end{tabular}
\label{tab:lasso-predictors}
\end{table}

\begin{table}[htbp]
\centering
\caption{Statistical Tests for Group and Demographic Comparisons}
\label{tab:new_stat_tests}
\small
\begin{tabularx}{\textwidth}{X l l l l}
\toprule
\textbf{Comparison} & \textbf{Test} & \textbf{Statistic} & \textbf{$p$-value} & \textbf{Effect} \\
\midrule
\textbf{Group A vs Group B} & & & & \\
AI usage frequency distribution & Chi-square & $\chi^2(4) = 32.81$ & $< 0.001$ & $V = 0.29$ \\
Cost reduction as AI impact & Chi-square & $\chi^2(1) = 14.89$ & $< 0.001$ & $V = 0.19$ \\
Mean principle familiarity & Mann-Whitney & $U = 9684.5$ & $< 0.001$ & $r = 0.31$ \\
\midrule
\textbf{Gender (Group B)} & & & & \\
Ethics consideration frequency & Mann-Whitney & $U = 6264.0$ & $< 0.001$ & $r = 0.25$ \\
\# of principles considered & Mann-Whitney & $U = 7013.0$ & $0.011$ & $r = 0.15$ \\
\midrule
\textbf{Role (Group B)} & & & & \\
Ethics consideration frequency & Kruskal-Wallis & $H(6) = 12.34$ & $0.055$ & $\epsilon^2 = 0.04$ \\
\midrule
\textbf{Education (Group B)} & & & & \\
Ethics consideration frequency  & Mann-Whitney & $U = 3876.0$ & $0.815$ & $r = 0.01$ \\
\bottomrule
\end{tabularx}
\begin{tablenotes}
\small
\item Effect sizes: $V$ = Cram\'{e}r's V, $r = Z/\sqrt{N}$, $\epsilon^2$ = epsilon-squared. These tests complement the hypothesis tests in Tables~\ref{table:demographics-familiarity} and~\ref{table:demographics-familiarity-ai-initiatives}.
\end{tablenotes}
\end{table}

\begin{table}[ht]
\centering
\caption{LASSO Results: Predictors of AI Ethics Principle Familiarity (Gr. B)}
\label{tab:lasso_trackb_principles}
\begin{tabular}{lcc}
\hline
\textbf{Predictor} & \textbf{Coefficient} & \textbf{Std. Error} \\
\hline
\multicolumn{3}{l}{\textit{Familiarity with Regulations}} \\
\quad EU AI Act & $0.154^{*}$ & $(0.074)$ \\
\quad NIST Risk Management Framework & $0.092$ & $(0.106)$ \\
\quad Average regulations familiarity & $0.032$ & $(0.129)$ \\
\\[-6pt]
\multicolumn{3}{l}{\textit{Perception}} \\
\quad Positive perception of regulatory impact & $0.152^{**}$ & $(0.054)$ \\
\quad Considers environment/social responsibility & $0.140^{*}$ & $(0.054)$ \\
\\[-6pt]
\multicolumn{3}{l}{\textit{AI Practice}} \\
\quad AI used for cost reduction & $0.124^{*}$ & $(0.052)$ \\
\quad Frequency of considering ethics & $0.111$ & $(0.062)$ \\
\quad Conducts regular audits/assessments & $0.101$ & $(0.053)$ \\
\quad Adopts AI governance frameworks & $0.072$ & $(0.056)$ \\
\hline
Overall Mean & \multicolumn{2}{c}{3.88} \\
Observations & \multicolumn{2}{c}{142} \\
R² (test) & \multicolumn{2}{c}{0.56} \\
Features Selected & \multicolumn{2}{c}{9 / 141} \\
\hline
\multicolumn{3}{p{12cm}}{\footnotesize \textit{Note:} $^*p < 0.05$; $^{**}p < 0.01$; $^{***}p < 0.001$. Dependent variable: Average familiarity with 9 AI ethics principles (1--5 scale). Coefficients are standardized (per 1-SD change in predictor).} \\
\end{tabular}
\end{table}

\begin{table}[ht]
\centering
\caption{LASSO Results: Predictors of Ethics Consideration Frequency (Gr. B)}
\begin{tabular}{lcc}
\hline
\textbf{Predictor} & \textbf{Coefficient} & \textbf{Std. Error} \\
\hline
\multicolumn{3}{l}{\textit{Familiarity with AI Ethics Principles}} \\
\quad Average principle familiarity & $0.281^{**}$ & $(0.088)$ \\
\\[-6pt]
\multicolumn{3}{l}{\textit{Perception}} \\
\quad Positive perception of regulatory impact & $0.226^{**}$ & $(0.076)$ \\
\\[-6pt]
\multicolumn{3}{l}{\textit{Familiarity with Regulations}} \\
\quad US Executive Order on Safe AI & $0.210^{*}$ & $(0.083)$ \\
\hline
Overall Mean & \multicolumn{2}{c}{3.73} \\
Observations & \multicolumn{2}{c}{137} \\
R² (test) & \multicolumn{2}{c}{0.40} \\
Features Selected & \multicolumn{2}{c}{3 / 150} \\
\hline
\multicolumn{3}{p{12cm}}{\footnotesize \textit{Note:} $^*p < 0.05$; $^{**}p < 0.01$; $^{***}p < 0.001$. Dependent variable: Frequency of considering AI ethics principles in development (1--5 scale, from never to always). Coefficients are standardized (per 1-SD change in predictor).} \\
\end{tabular}
\label{table:lasso_ethics_frequency}
\end{table}

\begin{table}[ht]
\centering
\caption{LASSO Results: Predictors of AI Governance Initiative Familiarity}
\begin{tabular}{lcc}
\hline
\textbf{Predictor} & \textbf{Coefficient} & \textbf{Std. Error} \\
\hline
\multicolumn{3}{l}{\textit{AI Tools/Technologies}} \\
\quad Google Cloud AI & $0.251^{**}$ & $(0.090)$ \\
\quad LangChain & $0.240^{**}$ & $(0.091)$ \\
\quad Neural Auto Encoder & $0.192^{*}$ & $(0.084)$ \\
\\[-6pt]
\multicolumn{3}{l}{\textit{Familiarity with AI Ethics Principles}} \\
\quad Harm Prevention and Beneficence & $0.221$ & $(0.113)$ \\
\quad Accountability and Responsibility & $0.178$ & $(0.135)$ \\
\quad Environment and Social Responsibility & $0.175$ & $(0.139)$ \\
\quad Average principle familiarity & $0.056$ & $(0.212)$ \\
\\[-6pt]
\multicolumn{3}{l}{\textit{Mitigation Strategies}} \\
\quad Create contingency plans & $0.181^{*}$ & $(0.090)$ \\
\quad Implement transparent/explainable approaches & $0.155$ & $(0.087)$ \\
\\[-6pt]
\multicolumn{3}{l}{\textit{Demographics}} \\
\quad High school degree (vs. higher education) & $-0.213^{*}$ & $(0.085)$ \\
\hline
Overall Mean & \multicolumn{2}{c}{2.86} \\
Observations & \multicolumn{2}{c}{141} \\
R² (test) & \multicolumn{2}{c}{0.41} \\
Features Selected & \multicolumn{2}{c}{10 / 142} \\
\hline
\multicolumn{3}{p{12cm}}{\footnotesize \textit{Note:} $^*p < 0.05$; $^{**}p < 0.01$; $^{***}p < 0.001$. Dependent variable: Average familiarity with 8 AI governance initiatives (1--5 scale). Coefficients are standardized (per 1-SD change in predictor).} \\
\end{tabular}
\label{table:lasso_regulation_familiarity_b}
\end{table}

\begin{table}[h]
\centering
\caption{Model Comparison via Likelihood Ratio Tests}
\label{table:model_comparison_interactions}
\begin{tabular}{llcccc}
\toprule
\textbf{Model} & \textbf{Fixed Effects} & \textbf{AIC} & \textbf{$\Delta$AIC} & \textbf{LRT $\chi^2$} & \textbf{$p$-value} \\
\midrule
\multicolumn{6}{l}{\textit{Principle Familiarity Models ($n=271$)}} \\
Null & Intercept only & 717.60 & --- & --- & --- \\
M1a & Experience & 705.48 & $-12.12$ & 12.04 & 0.001 \\
M1b & Experience + Develops AI & 706.27 & $+0.79$ & 1.20 & 0.271 \\
M1c & Experience $\times$ Location + Develops AI & 698.03 & $-8.24$ & $-2.45$ & 1.000 \\
M1d & Experience $\times$ Develops AI + Location & 689.99 & $-8.04$ & 3.60 & 0.309 \\
\\[-6pt]
\multicolumn{6}{l}{\textit{Ethics Frequency Models ($n=272$)}} \\
Null & Intercept only & 788.82 & --- & --- & --- \\
M2a & Governance Initiative familiarity & 736.36 & $-52.46$ & 52.03 & $<0.001$ \\
M2b & Reg. fam. + Experience & 737.13 & $+0.77$ & 1.24 & 0.265 \\
M2c & Reg. fam. $\times$ Location + Experience & 734.38 & $-2.75$ & $-5.62$ & 1.000 \\
M2d & Reg. fam. $\times$ Experience + Location & 727.78 & $-6.60$ & $-1.02$ & 1.000 \\
\bottomrule
\end{tabular}
\begin{tablenotes}
\small
\item Note: $\Delta$AIC is relative to previous model. LRT compares to baseline (M1b or M2b). Negative $\Delta$AIC indicates a worse fit.
\end{tablenotes}
\end{table}






\end{appendices}



\end{document}